%% file: main.tex
\documentclass[12pt]{article}
\usepackage{graphicx, color}
\usepackage[hidelinks]{hyperref}
\usepackage[dvipsnames,table]{xcolor}
\usepackage[section]{placeins}
\definecolor{safegreen}{RGB}{223,240,216}
\definecolor{biasred}{RGB}{248,215,218}

\usepackage[top=1.1in, left=1in, right=1in, bottom=1.1in]{geometry}
\usepackage[parfill]{parskip}	
\usepackage{caption}
\usepackage{subcaption}
\usepackage[export]{adjustbox}
\usepackage{longtable}
\usepackage{amssymb}
\usepackage{bbm}
\usepackage{bm}
\usepackage{amsmath}
\usepackage{amsfonts}
\newcommand{\indicator}[1]{\mathbbm{1}_{ {#1} }}
\usepackage{enumitem}
\usepackage{setspace}
\usepackage{booktabs}
\usepackage[compact]{titlesec}

\newcommand*{\dbar}[1]{\overline{\overline{#1}}}
\usepackage{lscape}
\usepackage{floatrow}
\usepackage{accents}

\usepackage[nocitation]{apacite}
\usepackage{natbib}
\bibpunct{(}{)}{;}{a}{}{,}
\usepackage{titling}
\usepackage{pifont}
\usepackage{amsthm}
\usepackage{multirow}
\usepackage{multicol}
\usepackage{mathrsfs}

\usepackage[displaymath, mathlines, pagewise]{lineno}

\usepackage{tikz}
\usepackage[scaled=.90]{helvet}
\usetikzlibrary{arrows.meta,calc}

  {\end{smallmatrix}\right)}

\usepackage{epstopdf}
\renewcommand{\indicator}[1]{\mathbbm{1}_{ {#1} }}
\usepackage{rotating}
\usepackage{threeparttable}
\usepackage{lmodern}
\usepackage{comment}
\usepackage{color}

\usepackage{array}
\newcolumntype{C}[1]{>{\centering\arraybackslash}m{#1}}
\renewcommand*{\dbar}[1]{\overline{\overline{#1}}}

\input{macros}

\begin{document}
\pagestyle{plain}

\renewcommand{\arraystretch}{1.3}

\newcommand{\blind}{0}
\newcommand{\tit}{\Large An Anatomy of Event Studies: Hypothetical Experiments, Exact Decomposition, and Weighting Diagnostics}

\if0\blind
{\title{\tit
\thanks{The authors thank Dmitry Arkhangelsky, Francesca Dominici, Avi Feller, Phillip Heiler, Miguel Hern\'{a}n, Guido Imbens, Andrea Rotnitzky, Pedro Sant'Anna, Zach Shahn, Tymon S\l{}oczy\'nski, Dylan Small, Anton Strezhnev, Davide Viviano, Ting Ye, and three anonymous reviewers for helpful comments. \\
\indent \indent \emph{Funding and support}: This research was supported by award ME-2022C1-25648 from the Patient-Centered Outcomes Research Institute and grant P01 AG032952 from the National Institute on Aging. 
}\vspace*{.3in}}
\author{\normalsize Zhu Shen\thanks{Stanford Institute for Human-Centered AI \& Stanford Data Science, Stanford University.} \and \normalsize Ambarish Chattopadhyay\thanks{Applied Statistics Unit, Indian Statistical Institute, Kolkata.}\and \normalsize Yuzhou Lin\thanks{Department of Statistics, Harvard University.} \and  \normalsize Jos\'{e} R. Zubizarreta\thanks{Departments of Health Care Policy, Biostatistics, and Statistics, Harvard University, 180 Longwood Avenue, Office 215-A, Boston, MA 02115; email: \url{zubizarreta@hms.harvard.edu}.}}

\date{} 

\maketitle
}\fi

\if1\blind
\title{\tit}
\date{} 
\maketitle
\fi

\begin{abstract}
\singlespacing
\input{sec0}
\end{abstract}

\begin{center}

\end{center}
\clearpage

\tableofcontents
\clearpage

\doublespacing

\input{sec1}

\input{sec2}

\input{sec3}

\input{sec4}

\input{sec5}

\input{sec7}

\bibliographystyle{abbrvnat}
\bibliography{bib2022}



\input{sec_supp}

\end{document}

%% file: macros.tex
\usepackage{amsthm}

\newif\ifsuppstandalone
\suppstandalonefalse

\theoremstyle{plain}
\newtheorem{theorem}{Theorem}
\newtheorem{lemma}[theorem]{Lemma}
\newtheorem{proposition}[theorem]{Proposition}

\theoremstyle{definition}
\newtheorem{assumption}{Assumption}

\providecommand{\indicator}[1]{}\renewcommand{\indicator}[1]{\mathbbm{1}_{ {#1} }}
\providecommand{\dbar}[1]{}\renewcommand{\dbar}[1]{\overline{\overline{#1}}}
\providecommand{\argmin}{}\renewcommand{\argmin}{\mathop{\rm arg\,min}}

\DeclareMathOperator*{\Var}{{\rm Var}}

\renewcommand{\hat}{\widehat}
\renewcommand{\bar}{\overline}

\newcommand{\indep}{\perp\!\!\!\perp}

\newcommand{\Ccal}{\mathcal{C}}
\newcommand{\Ical}{\mathcal{I}}

\newcommand{\Pcal}{\mathcal{P}}
\newcommand{\Scal}{\mathcal{S}}
\newcommand{\Tcal}{\mathcal{T}}

\newcommand{\Fcal}{\mathcal{F}}

\newcommand{\Mcal}{\mathcal{M}}
\newcommand{\Gcal}{\mathcal{G}}

\newcommand{\ate}{\text{ATE}}

\newcommand{\obs}{\text{obs}}

\newcommand{\bU}{\mathbf{U}}

\newcommand{\bX}{\mathbf{X}}
\newcommand{\bZ}{\mathbf{Z}}
\newcommand{\bY}{\mathbf{Y}}

\newcommand{\E}{\mathbb{E}}
\renewcommand{\P}{\mathbb{P}}
\renewcommand{\S}{\mathbb{S}}
\newcommand{\R}{\mathbb{R}}

\DeclareMathAlphabet{\mathbbmsl}{U}{bbm}{m}{sl}

%% file: sec0.tex
\indent \hspace{.5cm} 

In recent decades, event studies have become widely used in health and social research to estimate the dynamic effects of staggered interventions. In this paper, we develop a framework grounded in experimental design principles for observational studies that encompasses common event study estimators and makes explicit how they borrow information across units and time periods. We consider two routes to identification, unconfoundedness in treatment initiation and parallel trends with limited anticipation, and construct weighted contrasts from outcome levels under the first and within-unit differences under the second. We propose a robust weighting estimator that draws on progressively larger sets of observations as stronger assumptions on treatment assignment and potential outcomes are imposed. We derive exact finite-sample decompositions of dynamic two-way fixed effects and interaction-weighted estimators, and of doubly robust estimators, into weighted contrasts of within-unit outcome differences, revealing the hypothetical experiment each approximates. These decompositions show which unit-period observations enter each contrast and how they are weighted, clarifying and supplementing the notion of forbidden comparisons. The implied weights yield diagnostics for covariate balance, sign reversal, effective sample size, information shares across observation groups, and each observation’s influence on the estimate. We illustrate these diagnostics in a study of the effects of Castle Doctrine laws on homicide rates.

%% file: sec1.tex
\section{Introduction}
\label{section_intro}

Event studies are widely used in the social sciences and policy research to analyze the effects of staggered interventions. Applications range from the effects of bank branching deregulation on income distribution \citep{beck2010big}, to the role of health insurance coverage in treating opioid use disorder \citep{meinhofer2018role}, to the effects of Castle Doctrine laws on homicide rates \citep{cheng2013does}. Across these applications, the common structure is a panel in which units, typically states or counties, adopt a policy at different calendar times and outcomes are measured before and after adoption. Interest lies in how the effect of the policy evolves with the duration of exposure. 

A substantial methodological literature has clarified the properties of common event study estimators. Conventional two-way fixed effects (TWFE) regression can assign negative weights to some comparisons and admits contrasts in which already treated units serve as controls, so that it can be biased when treatment effects vary across adoption cohorts or across event times \citep{baker2022much, borusyak2024revisiting, de2020two, goodman2021difference, sun2021estimating}. A family of estimators addresses these problems by specifying the effects of interest and the comparisons used to estimate them, rather than leaving these choices implicit in a restrictive regression specification. These include the group-time average treatment effect estimators of \citet{callaway2021difference}, the interaction-weighted estimator of \citet{sun2021estimating}, the imputation estimator of \citet{borusyak2024revisiting}, and the two-stage and Mundlak formulations of \citet{gardner2024two} and \citet{wooldridge2025two}; see \citet{de2023twosurvey} and \citet{roth2023s} for surveys. A complementary line of work examines the same designs from the standpoint of treatment assignment, describing how variation in adoption timing arose and how that assignment process can be used to identify and estimate effects \citep{arkhangelsky2022doubly, arkhangelsky2024double, athey2022design, roth2023efficient}.

Alongside these advances, three limitations remain, each regarding how an event study estimate is formed and how it can be compared with others. First, existing decompositions \citep{de2023two, goodman2021difference, li2024guide, sun2021estimating} operate on cohorts, on $2 \times 2$ blocks, or on aggregated groups (the ``molecules’’ of a panel dataset), so that for a given point estimate they do not directly show which individual observations (the ``atoms”) produced it and with what weights. Without this fine-grained decomposition, it is difficult to clearly see the treatment and control group contrasts actually formed by the estimator and assess whether they form the intended causal comparison in connection to a hypothetical experiment. Second, how an estimator borrows strength across the panel is often an implication of the model specification rather than a design choice made in advance and justified through assumptions about treatment assignment and restrictions on effect dynamics. Third, estimators that draw on different observations under different assumptions take quite different forms, which makes them difficult to compare side by side, and a comparison of their standard errors is not especially informative when they use different information to target different quantities.

This paper takes up these limitations by separating three elements that an event study estimator often bundles together. Based on suitable identification assumptions, an event study estimator is a weighted contrast of transformed outcomes $Y^*$ over an eligible set of observations. Based on suitable borrowing information assumptions, the set of eligible observations can be expanded. Then the causal contrast is built using weights that take into account design and estimation considerations, such as covariate balance and sample bounded estimation to avoid sign reversal \citep{robins2007comment, small2017instrumental}. Keeping these apart allows one to state a design and construct weights for it, to take an estimator as given and recover in closed form the design it implies, and to compare several estimators on common terms. Figure~\ref{fig:roadmap} summarizes the framework and the argument of the paper: what is estimated, how it is identified, which observations enter, how the weights are built, and what the weights say. Columns three and four get to the main point, namely to describe event study estimators by a pair consisting of the set of unit period observations it is eligible to use and the weights it places on them.

\begin{figure}[t]
\centering
\begingroup
\renewcommand{\sfdefault}{phv}
\definecolor{RouteU}{HTML}{6B8C21}
\definecolor{RouteP}{HTML}{4298B5}   
\resizebox{\linewidth}{!}{%
\begin{tikzpicture}[
  x=1cm,
  y=1cm,
  font=\sffamily\fontsize{5.0}{5.8}\selectfont,
  common/.style={
    draw=black!40,
    fill=black!2,
    rounded corners=2.5pt,
    line width=.45pt,
    align=center,
    inner xsep=2pt,
    inner ysep=4pt,
    anchor=center,
    execute at begin node={\hyphenpenalty=10000\exhyphenpenalty=10000}
  },
  routeU/.style={common,draw=RouteU,fill=RouteU!12},
  routeP/.style={common,draw=RouteP,fill=RouteP!12},
  question/.style={
    font=\sffamily\fontsize{5.4}{6.2}\selectfont\bfseries,
    text=black!74,
    draw=black!22,
    fill=black!4,
    rounded corners=2.5pt,
    line width=.35pt,
    align=center,
    anchor=south,
    inner xsep=4pt,
    inner ysep=2pt
  },
  sectiontag/.style={
    font=\sffamily\fontsize{5.0}{5.8}\selectfont,
    text=black!58,
    anchor=south east,
    inner sep=0pt
  },
  flow/.style={
    -{Latex[length=2.4pt,width=2.0pt]},
    draw=black!52,
    line width=.45pt,
    line cap=round,
    line join=round
  }
]

\pgfmathsetmacro{\cA}{0}
\pgfmathsetmacro{\cB}{2.92}
\pgfmathsetmacro{\cC}{5.81}
\pgfmathsetmacro{\cD}{8.67}
\pgfmathsetmacro{\cE}{11.11}
\pgfmathsetmacro{\cF}{13.83}
\pgfmathsetmacro{\cDE}{0.5*(\cD+\cE)}

\node[question] at (\cA,1.70) {What is estimated?};
\node[question] at (\cB,1.70) {How is it identified?};
\node[question] at (\cC,1.70) {Which observations enter?};
\node[question,minimum width=4.55cm] at (\cDE,1.70) {How are weights built?};
\node[question] at (\cF,1.70) {What do the weights say?};

\node[common,text width=3.00cm,minimum height=2.82cm] (target) at (\cA,0) {%
  {\sffamily\fontsize{5.4}{6.2}\selectfont\bfseries Building-block estimand}\\[4pt]
  $\mathrm{ATE}^{\mathcal P}_{t_y}(t_1,\infty)$\\[4pt]
  Treatment initiation: $t_1$\\[1pt]
  Outcome measurement: $t_y$\\[1pt]
  Target population: $\mathcal P$\\[3pt]
  \rule{2.40cm}{.25pt}\\[2pt]
  {\bfseries Aggregate event study estimand}\\[1pt]
  $\mathrm{ATE}^{\mathcal S_t}_{\mathrm{agg}}(l)$
};

\node[routeU,text width=1.95cm,minimum height=1.20cm] (unc) at (\cB,.79) {%
  {\sffamily\fontsize{5.4}{6.2}\selectfont\bfseries Route A}\\[-1pt]
  unconfoundedness\\[-1pt]
  A1--A3, A4\\[-1pt]
  $Y^*=Y$
};
\node[routeP,text width=1.95cm,minimum height=1.20cm] (pt) at (\cB,-.79) {%
  {\sffamily\fontsize{5.4}{6.2}\selectfont\bfseries Route B}\\[-1pt]
  parallel trends\\[-1pt]
  A1--A3, A5--A6\\[-1pt]
  $Y^*=\Delta^lY$
};

\node[common,text width=2.50cm,minimum height=2.82cm] (experiment) at (\cC,0) {%
  \mbox{\sffamily\fontsize{5.4}{6.2}\selectfont\bfseries Information-borrowing}\\[-1pt]
  {\sffamily\fontsize{5.4}{6.2}\selectfont\bfseries assumptions}\\[5pt]
  Hypothetical experiment based on an information set
};

\node[common,text width=1.90cm,minimum height=2.82cm] (build) at (\cD,0) {%
  {\sffamily\fontsize{5.4}{6.2}\selectfont\bfseries From first principles}\\[5pt]
  Construction of weighted contrasts with minimal weights
};

\node[common,text width=2.10cm,minimum height=2.82cm] (decomposition) at (\cE,0) {%
  {\sffamily\fontsize{5.4}{6.2}\selectfont\bfseries Implied, from\\ existing estimators}\\[5pt]
  Exact decomposition of TWFE, IW, and DR estimators
};

\node[common,text width=2.45cm,minimum height=2.82cm] (diagnostics) at (\cF,0) {%
  \mbox{\sffamily\fontsize{5.4}{6.2}\selectfont\bfseries Weight-based diagnostics}\\[5pt]
  Information, Influence, Balance,\\
  Profile, and Extrapolation
};

\node[sectiontag] at ($(target.north east)+(0,1.2pt)$) {Sec. 2.2};
\node[sectiontag] at ($(unc.north east)+(0,1.2pt)$) {Sec. 2.3.1};
\node[sectiontag] at ($(pt.north east)+(0,1.2pt)$) {Sec. 2.3.2};
\node[sectiontag] at ($(experiment.north east)+(0,1.2pt)$) {Sec. 2.4--2.5};
\node[sectiontag] at ($(build.north east)+(0,1.2pt)$) {Sec. 3.1--3.2};
\node[sectiontag] at ($(decomposition.north east)+(0,1.2pt)$) {Sec. 4.1};
\node[sectiontag] at ($(diagnostics.north east)+(0,1.2pt)$) {Sec. 4.2};

\coordinate (split-a) at ($(target.east)!0.5!(unc.west |- target.east)$);
\coordinate (merge-a) at ($(unc.east |- experiment.west)!0.5!(experiment.west)$);
\draw[flow] (target.east) -- (split-a) |- (unc.west);
\draw[flow] (target.east) -- (split-a) |- (pt.west);
\draw[flow] (unc.east) -| (merge-a) -- (experiment.west);
\draw[flow] (pt.east) -| (merge-a) -- (experiment.west);
\draw[flow] (experiment.east) -- (build.west);
\draw[flow] (build.east) -- (decomposition.west);
\draw[flow] (decomposition.east) -- (diagnostics.west);

\end{tikzpicture}%
}
\endgroup
\caption{Roadmap of the paper. Building-block estimands can aggregate to the event study estimand and are identified under either unconfoundedness or parallel trends. Each route yields a common weighted contrast in $Y^*$ based on an information set. Section~\ref{sec3:weight_estimator} constructs weights from first principles; Section~\ref{sec:diagnostics} recovers the weights implied by existing estimators and evaluates them with weight-based diagnostics.}
\label{fig:roadmap}
\end{figure}

From this description, the paper offers four contributions. (i) It provides a framework that progressively maps assumptions to eligible observations along two routes to identification: unconfoundedness in treatment initiation, and parallel trends with limited anticipation \citep{callaway2021difference, malani2015interpreting, sun2021estimating}. Each additional assumption about assignment or effect dynamics admits a specified group of observations into the causal contrast (columns two and three of Figure~\ref{fig:roadmap}). (ii) It introduces a robust weighting estimator that constructs the contrast implied by stated identification and information assumptions by minimizing weight dispersion subject to approximate balance. The estimator is consistent under any of four combinations of conditions on the treatment initiation and potential outcome models, asymptotically Normal, and semiparametrically efficient. (iii) It derives exact, closed-form weighting representations at the level of the individual unit period observation. These representations encompass the dynamic TWFE, interaction-weighted \citep{sun2021estimating}, and doubly robust estimators \citep{callaway2021difference, sant2020doubly}. They reveal the hypothetical experiment that each estimator approximates and identify precisely which observations in a given panel carry the estimate, and by how much. (iv) It develops diagnostics that follow directly from the weights. These report the effective sample size and information share of each observation group, the influence of each observation on the point estimate, the covariate balance achieved, any reversal in the sign of a weight, and the target profile implied by the adjustment. Because the diagnostics require only a linear representation in the observed outcomes, they apply both to the estimators studied here and to others that admit such a representation \citep{knaus2024treatment}.

In an analysis of the effects of Castle Doctrine laws on homicide rates \citep{cheng2013does}, these diagnostics show how much of a conventional event study estimate rests on observations that enter only under additional assumptions about effect dynamics, and what covariate adjustment costs in effective sample size. Figure~\ref{fig:map_agg} reports the exact weight that commonly used estimators assign to each state in each year, thereby permitting a direct comparison of how they use each of the observations in the panel dataset. For instance, for one of the point estimates the observations that TWFE draws in under assumptions on delayed onset and effect dissipation supply 12.9\% of the information share, against 8.5\% for the treated group of the ideal experiment, so that the two groups whose inclusion rests on the strongest assumptions together contribute more information than the group the design was meant to isolate. Without covariate adjustment, the interaction-weighted, doubly robust, and robust weighting estimators coincide exactly, while after adjustment the effective size of the control group falls to 32\% of its unadjusted value under robust weighting and to 17\% under doubly robust estimation. 

The paper draws on several strands of the weighting literature for observational studies with time invariant treatments. It extends the optimization approach of \citet{wang2020minimal} that finds the weights, the machinery of \citet{chattopadhyay2023implied} for recovering weights implied by a linear estimator, and the one-step construction of \citet{chattopadhyay2024one} that targets a population distinct from the study sample. What is new is the framework that imposes assumptions progressively and places existing estimators within it, the unit period resolution of the decomposition, the extension of the implied weight machinery to the interaction-weighted and doubly robust estimators, and the diagnostics that follow. Among related approaches, \citet{chen2024potential} asks which implicit design justifies a linear regression estimand, whereas we ask, for a stated design, which observations in the panel may form a contrast, and we work with dynamic causal estimands. \citet{arkhangelsky2024double} build estimators from restrictions on the assignment process, and \citet{roth2023efficient} study efficiency under randomized adoption timing, whereas our concern is the eligibility of observations and the weights placed on them under either route to identification. \citet{ghanem2022selection} examine how selection and the outcome process trade off to justify parallel trends, a question precedes the one considered here.

The paper is organized as basically described in Figure~\ref{fig:roadmap}. Section~\ref{sec2} defines what is estimated and how it is identified, describes the hypothetical experiment behind each route, and states the assumptions that determine which observations enter. Section~\ref{sec3:weight_estimator} builds the weights from first principles and gives the asymptotic properties of the resulting estimator. Section~\ref{sec:diagnostics} recovers the weights implied by the dynamic TWFE, interaction-weighted, and doubly robust estimators, and develops diagnostics from them. Section~\ref{sec:case_study} applies these to the effect of Castle Doctrine laws on homicide rates. Section~\ref{sec:conclusion} closes with remarks on extensions to other panel data methods. 


%% file: sec2.tex
\section{An experimental perspective of event study estimators}
\label{sec2}

This section connects the causal effect of interest to the observations that can be used to estimate it. Following the first three questions in Figure~\ref{fig:roadmap}, we begin by specifying when treatment starts, when outcomes are measured, and the target population. We then consider two routes to identification: unconfoundedness in treatment initiation and parallel trends with limited anticipation. These routes lead to comparisons of outcome levels and within-unit outcome differences, respectively. For each route, we describe the treatment and control components of a hypothetical experiment and the observations they use. We then introduce additional assumptions that allow this baseline contrast to draw on more of the panel. The aim is to establish which observations can enter the contrast and why, before constructing their weights in Section~\ref{sec3:weight_estimator}.

\subsection{Setup}
\label{sec2:setup}
Consider a sample of $n$ units randomly drawn from a study population and observed at $T$ regular time points. 
We assume there is perfect compliance and no loss to follow-up. 
At each time point $t \in \{1, 2, ..., T\} = \Tcal$, the policy or treatment for each unit $i = 1, ..., n$ is assigned before the outcome is measured. Baseline covariates $\bX_{i} \in \R^k$ are measured at the start of the follow-up period. 
We denote $Z_{it}$ for the treatment assignment indicator with $Z_{it} = 1$ if unit $i$ is assigned to treatment at time $t$ and $Z_{it} = 0$ otherwise.
We write $Y^{\text{obs}}_{it}$ for the observed outcome. We collect the treatment and outcome histories for unit $i$ in the vectors $\bZ_i=(Z_{i1},\ldots,Z_{iT})^\top \in \{0,1\}^T$ and $\bY_i^{\obs}=(Y^{\obs}_{i1},\ldots,Y^{\obs}_{iT})^\top \in \mathbb{R}^T$, respectively. 
Throughout, we refer to a unit-time pair as an observation. 
Thus, the pair $(i,t)$ is called a treated observation if $Z_{it} = 1$ and a control observation if $Z_{it} = 0$.

We focus on staggered treatment adoption: once a unit begins treatment, it remains treated. We can therefore summarize the treatment path by the treatment-initiation time. Let $G_i \in \Tcal^+ \equiv \Tcal \cup \{\infty\}$ denote the first time at which unit $i$ receives treatment. If a unit remains untreated throughout follow-up, then $G_i = \infty$. Thus, $Z_{it}=\mathbbm{1}\{G_i \leq t\}$. For each $g \in \Tcal^+$, let $Y_{it}(g)$ denote the potential outcome for unit $i$ at time $t$ under treatment initiation at $g$. We assume no interference across units, as formalized in Section~\ref{sec_causal_identification}. 

For conciseness, let $\Scal$ and $\Pcal$ denote the study and target populations, with probability measures $\mathbb{S}$ and $\mathbb{P}$, respectively. We assume no particular relationship between them, beyond the transportability assumptions stated in Supplement Section~\ref{sec:app_identification_proof}. Let $\Scal_t$ and $\Pcal_t$ denote the subpopulations that initiate treatment at time $t \in \Tcal$, with probability measures $\mathbb{S}_t$ and $\mathbb{P}_t$. In the main paper, we take $\Pcal$ to be the study population $\Scal$ or one of its treatment cohorts $\Scal_{t}$, while the extension to a target population distinct from $\Scal$ is developed in Supplement Section~\ref{sec:app_robust_generalization}.

\subsection{Target estimand}
\label{sec_estimand}
Our causal estimand is the average treatment effect on a specified target population $\Pcal$: $\ate^\Pcal_{t_y} (t_1,t_0)$. It compares outcomes measured at time ${t_y} \in \Tcal$ under two treatment regimes: initiation at $t_1$ and initiation at $t_0$, where $t_1,t_0 \in \Tcal^+$:
\begin{equation}\label{equation_estimand_particular}
    \begin{aligned}[b]
    \ate^\Pcal_{t_y}(t_1, t_0) 
    &= \E_{\mathbb{P}}\{ Y_{i{t_y}}(t_1) - Y_{i{t_y}}(t_0) \}.
    \end{aligned}
\end{equation}
More generally, we can specify a stochastic reference regime with initiation $t_\mathbf{p}$ with varying probabilities at times $t \in \Tcal^{+}$, such that 
\vspace{-.2cm}
\begin{equation}
    \label{equation_estimand_general}
    \begin{aligned}[b]
    \ate^\Pcal_{t_y}(t_1, t_\mathbf{p}) 
    = \E_{\mathbb{P}} \left\{Y_{i{t_y}}(t_1) - \sum_{t \in \Tcal^{+}} p_{t} Y_{i{t_y}}(t) \right\} 
    = \sum_{t \in \Tcal^{+} } p_{t} \E_{\mathbb{P}} \{ Y_{i{t_y}}(t_1) - Y_{i{t_y}}(t) \} 
    = \sum_{t \in \Tcal^{+} } p_{t} \ate^\Pcal_{t_y}(t_1, t), 
    \end{aligned}
\end{equation}
where $p_t \in \left[0, 1\right]$ and $\sum_{t \in \Tcal^{+}} p_t = 1$. Estimand~\eqref{equation_estimand_general} is a convex combination of average treatment effects in Equation~\eqref{equation_estimand_particular}. It includes several familiar estimands as special cases. For example, setting $p_\infty = 1$ recovers the conventional comparison with never-treated units, $\ate^\Pcal_{t_y}(t_1, \infty)$. However, in some applications, such as evaluations of active labor market programs, the relevant comparison is between earlier and later participation rather than between participation and never participation, which in turn makes the conventional never-treated estimand less appropriate as a target of interest \citep{sianesi2004evaluation}.

We use $\ate^\Pcal_{t_y}(t_1,t_0)$ as a building-block estimand indexed by treatment initiation, outcome measurement time, and the target population. When $\Pcal=\Scal_{t_1}$, $t_0=\infty$, and $t_y=t_1+l$, it coincides with the group--time average treatment effect of \citet{callaway2021difference} and the cohort-specific effect of \citet{sun2021estimating} at event time $l$. Aggregating these building blocks across cohorts observed at event time $l$ yields the conventional event study parameter. Our formulation adds flexibility in two ways: the target population $\Pcal$ need not be a treated population in the study, and the reference regime $t_{\mathbf{p}}$ may be a mixture of treatment initiation times. To facilitate comparison with existing approaches, we focus on the case $p_\infty = 1$, in which the counterfactual regime is never treatment, while the more general formulation is discussed in Supplement Section~\ref{sec:app_genestimand}.

\subsection{Causal identification}
\label{sec_causal_identification}

We consider two routes to identification: unconfoundedness in treatment initiation and parallel trends with limited anticipation. We first state three assumptions common to both routes and then introduce the assumptions specific to each. Let $\Fcal_i$ denote the adjustment set, which contains unit-level information used to characterize treatment assignment for unit $i$.

\begin{assumption}[Consistency and no interference across units]
\label{assumption_consistency}
$Y^\obs_{it} = Y_{it}(G_i)$.
\end{assumption}

\begin{assumption}[Random sampling]
\label{assumption_random_sample}
The study sample $\{\mathbf{\mathcal{F}}_i, \bZ_i, \bY_i^{\obs} \}_{i = 1}^n$ is independently and identically distributed from the study population $\Scal$.
\end{assumption}

\begin{assumption}[Positivity of treatment]
\label{assumption_positive_treatment_initiation}
$\Pr(G_i = t \mid \bX_i) > 0$ for all $i = 1, ..., n$ and every $t \in \Tcal^{+}$.
\end{assumption}

\subsubsection{Identification under unconfoundedness} The first route adds conditional exchangeability of treatment initiation given an observed adjustment set.
\begin{assumption}[Conditional exchangeability of treatment initiation]
\label{assumption_random_treatment_initiation}
$\mathbbm{1}(G_i = t) \indep Y_{it_y}(t) \mid \mathcal{F}_i$ for all $i = 1, ..., n$ and every $t \in \Tcal^{+}$.
\end{assumption}

Assumption~\ref{assumption_random_treatment_initiation} characterizes the treatment assignment. The information included in $\Fcal_i$ determines how exchangeability is justified. Taking $\mathcal{F}_i = \emptyset$ corresponds to unconditionally as-if-random assignment of adoption time, as outlined in Section~\ref{sec_hypothetical_experiment}; taking $\mathcal{F}_i = \mathbf{X}_i$ gives a baseline conditionally randomized design; and allowing $\mathcal{F}_i$ to include past outcomes, treatments, and covariates corresponds to sequential randomization, as in \citet{bojinov2021panel, robins1986new}.

\subsubsection{Identification under parallel trends} 

Of particular relevance to event studies is the case in which $\Fcal_i$ includes time-invariant, unit-level unobserved factors $\mathbf{U}_i$ that may affect both treatment initiation and outcomes. Assumption~\ref{assumption_random_treatment_initiation} then conditions on unobserved information and does not, by itself, identify the estimand from observed data. This motivates the second identification route, based on limited anticipation and parallel trends conditional on observed covariates $\mathbf{X}_i$.
\begin{assumption}[Limited/No anticipation]
\label{assumption_limited_anticipation}
There is a known $\kappa \geq 0$ such that, for any $\kappa^\prime > \kappa$ and any $t \in \Tcal$, $\E_{\mathbb{S}}\left\{Y_{it}(t + \kappa^\prime) - Y_{it}(\infty) \mid \bX_i\right\} = 0$.
\end{assumption}

\begin{assumption}[Double parallel trend]
\label{assumption_two_way_parallel}
For any relative time period $l \in \{1, ..., t_y - 1\}$ and $t \in \Tcal^{+}$,
$\E_{\mathbb{S}}\left\{Y_{it_y}(t) - Y_{i(t_y-l)}(t) \mid \bX_i, G_i\right\} = \E_{\mathbb{S}}\left\{Y_{it_y}(t) - Y_{i(t_y-l)}(t) \mid \bX_i\right\}$.
\end{assumption}

When $\kappa=0$, Assumption~\ref{assumption_limited_anticipation} reduces to conditional no anticipation; when $\kappa>0$, it allows a known anticipation window of up to $\kappa$ periods. This assumption is standard in the literature \citep{callaway2021difference, malani2015interpreting, sun2021estimating}.
Assumption~\ref{assumption_two_way_parallel} extends conventional parallel trends to potential outcomes under every treatment initiation time. While conventional parallel trends is sufficient for identifying effects in the ever-treated population by comparing the observed treatment regime to the counterfactual never-treated regime, this stronger condition accommodates other target populations. Assumption~\ref{assumption_two_way_parallel} gives conditional mean exchangeability of within-unit potential outcome differences across treatment cohorts.

The two routes lead to contrasts in different outcomes: levels under unconfoundedness and within-unit differences under parallel trends with limited anticipation. We use each route to define a hypothetical experiment and then state the additional assumptions that allow its information set to expand.

\subsection{Hypothetical experiment}
\label{sec_hypothetical_experiment}
We consider a target estimand, $\ate^\Pcal_{t_y}(t_1, \infty)$, defined as a population-level contrast of potential outcomes at time $t_y$, and examine how it can be identified and estimated in a hypothetical randomized experiment. We consider the case where $\Pcal=\Scal$ and $\Fcal_i=\emptyset$, corresponding to a baseline unconditional randomized staggered rollout design. Extensions to study subpopulations and to settings in which $\Fcal_i=\bX_i$ are straightforward by standardization over $\bX_i$. In this hypothetical experiment, $n$ study units are randomly assigned to $T + 1$ treatment groups indexed by $\{1, \ldots, T, \infty\}$, with corresponding sample sizes $n_1, \ldots, n_T$, and $n_\infty$. Units assigned to group $t$ initiate treatment at time $t$, whereas those assigned to group $\infty$ never receive treatment. Outcomes $Y_{it}^{\obs}$ are measured for each unit $i$ at each time point $t$. We estimate $\E_{\mathbb{S}}\{ Y_{it_y}(t_1) \}$ and $\E_{\mathbb{S}}\{ Y_{it_y}(\infty) \}$ with their sample analogs, and derive an unbiased estimator for the target estimand:
   \begin{equation}\label{equation_idealexperiment_standard}
    \begin{aligned}[b]
    \hat{\ate}^{\Scal}_{t_y}(t_1, \infty)
    & = \hat{\E}_{\mathbb{S}}\{ Y_{i{t_y}}(t_1) \} - \hat{\E}_{\mathbb{S}}\{ Y_{i{t_y}}(\infty) \} \\ 
    & = \underbrace{\frac{1}{n_{t_1}} \sum_{i = 1}^{n} \mathbbm{1}(G_i = t_1) Y^{\obs}_{it_y}}_{\text{treatment component}} - \underbrace{\frac{1}{n_{\infty}} \sum_{i = 1}^{n} \mathbbm{1}(G_i = \infty) Y^{\obs}_{it_y}}_{\text{control component}}. 
    \end{aligned}
    \end{equation}
    The treatment component comprises outcomes at time $t_y$ from units assigned to initiate treatment at $t_1$, whereas the control component comprises outcomes at the same time from units assigned to never receive treatment. The contrast therefore involves only the two treatment arms $t_1$ and $\infty$; units assigned to other initiation times do not enter this contrast. Each unit in these two arms contributes exactly one observation at time $t_y$, either to the treatment component or to the control component. We refer to these observations as Ideal Experiment observations. This is closely related to target trial emulation \citep{ben2021trial, hernan2016}, where the design of the hypothetical experiment is specified before the analysis.

\begin{figure}[h!tbp]
        \centering
        \begin{subfigure}[b]{0.45\textwidth}
        \centering
        \includegraphics[width=\textwidth]{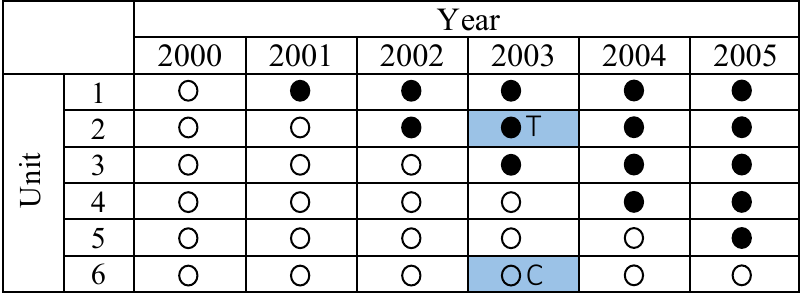}
        \caption{Assumptions \ref{assumption_consistency}-\ref{assumption_random_treatment_initiation}} 
        \label{fig:decomp_usual_a} 
        \end{subfigure}
        \hfill
        \begin{subfigure}[b]{0.45\textwidth}
        \centering
        \includegraphics[width=\textwidth]{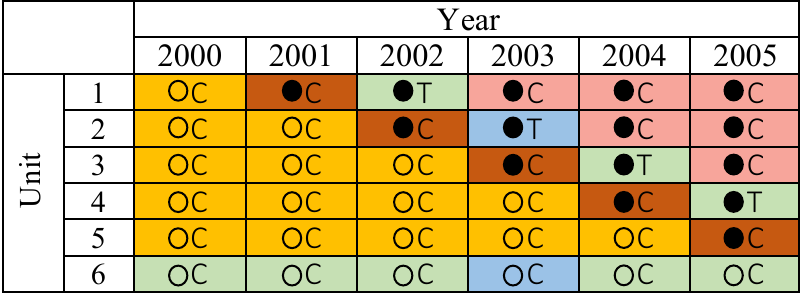}
        \caption{Assumptions \ref{assumption_consistency}-\ref{assumption_limited_anticipation} and  \ref{assumption_invariance_time_shifts}-\ref{assumption_washout}} 
        \label{fig:decomp_usual_b} 
        \end{subfigure}
    \caption{Observations used to estimate $\ate^\Scal_{2003}(2002, \infty)$ under (a) Assumptions \ref{assumption_consistency}-\ref{assumption_random_treatment_initiation} and (b) the same baseline assumptions and the information-borrowing Assumptions~\ref{assumption_limited_anticipation},\ref{assumption_invariance_time_shifts}-\ref{assumption_washout}. Solid and hollow circles represent treated and untreated observations, respectively. In panel (a), blue observations are Ideal Experiment observations, and white observations are not yet considered or restricted. In panel (b),  yellow, green, brown, and red observations are justified by Assumptions \ref{assumption_limited_anticipation},  \ref{assumption_invariance_time_shifts},  \ref{assumption_delay}, and \ref{assumption_washout}, respectively. The labels \textsf{T} and \textsf{C} indicate whether an observation contributes to the treatment or control component of the contrast.}
    \label{fig:decomp_usual}
    \end{figure}

    Figure~\ref{fig:decomp_usual} illustrates this comparison using a simple example with six independent units observed from 2000 to 2005, with unit 6 never receiving treatment. Assumptions~\ref{assumption_consistency}--\ref{assumption_random_treatment_initiation}, together with $\Fcal_i=\emptyset$, formalize the unconditional randomized setup described above. For $t_1=2002$ and $t_y=2003$, the estimator of $\ate^\Scal_{2003}(2002,\infty)$ uses only the outcomes observed in 2003 for units assigned to the $2002$ and $\infty$ treatment arms, shown in blue. The labels \textsf{T} and \textsf{C} denote whether the observation is part of the treatment component or the control component in the contrast, respectively.

    Under the parallel trends route, suppose that unconditional versions of Assumptions~\ref{assumption_limited_anticipation}--\ref{assumption_two_way_parallel} hold. Let $\Delta^{l} Y_{it_y}(t) = Y_{it_y}(t) - Y_{it_y - l}(t)$ denote the change in unit $i$'s potential outcome from the period $t_y-l$ to $t_y$, under treatment initiation at time $t$. We use the reference period $t_y-l$ as a baseline and subtract the potential outcome at that time from the potential outcome at $t_y$. The corresponding observed change is $\Delta^{l} Y^{\text{obs}}_{it_y} = \sum_{t \in \Tcal^{+}} \mathbbm{1}\{G_i = t\} \Delta^{l} Y_{it_y}(t)$. Assumption~\ref{assumption_two_way_parallel} gives mean exchangeability of these potential outcome differences across treatment arms, allowing us to formulate an analogous hypothetical experiment that compares outcome changes rather than levels. One structural interpretation is that unobserved confounding enters potential outcomes in an additive, time-invariant way, which is differenced out when constructing $\Delta^{l}Y_{it_y}(t)$. This interpretation is closely related to effect aliasing \citep{rosenbaum2026effect}, in which suitably constructed contrasts are insensitive to specific confounding components.

\begin{figure}[h!tp]
        \centering
        \begin{subfigure}[b]{0.47\textwidth}
        \centering
        \includegraphics[width=\textwidth]{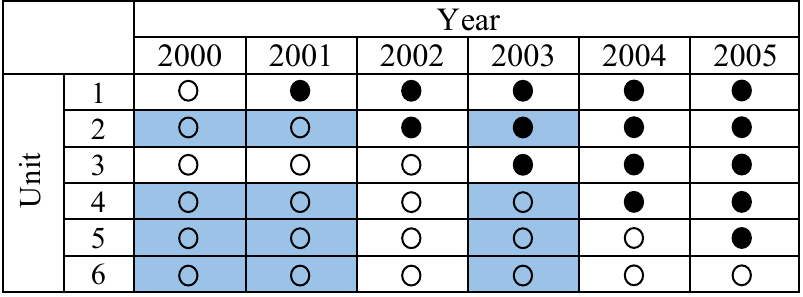}
        \caption{Assumptions~\ref{assumption_consistency}-\ref{assumption_positive_treatment_initiation}, \ref{assumption_limited_anticipation}-\ref{assumption_two_way_parallel}} 
        \label{fig:decomp_PT_a} 
        \end{subfigure}
        \hfill
        \begin{subfigure}[b]{0.47\textwidth}
        \centering
        \includegraphics[width=\textwidth]{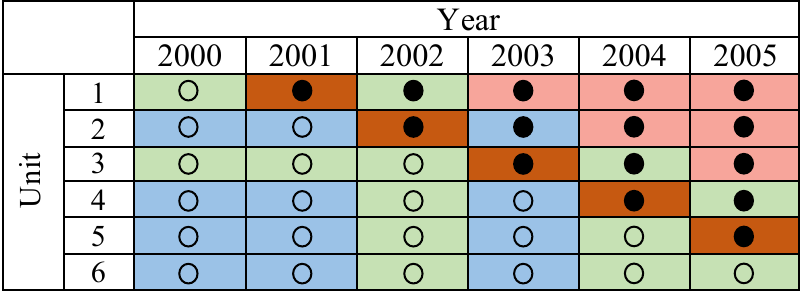}
        \caption{Assumptions \ref{assumption_consistency}-\ref{assumption_positive_treatment_initiation}, \ref{assumption_limited_anticipation}-\ref{assumption_washout}} 
        \label{fig:decomp_PT_b} 
        \end{subfigure}
        \caption{Observations used to estimate $\ate^\Scal_{2003}(2002, \infty)$ under (a) Assumptions \ref{assumption_consistency}--\ref{assumption_positive_treatment_initiation}, \ref{assumption_limited_anticipation}--\ref{assumption_two_way_parallel}, allowing general pre-treatment reference periods, and (b) the same baseline assumptions and the information-borrowing Assumptions~\ref{assumption_invariance_time_shifts}-\ref{assumption_washout}. Solid and hollow circles represent treated and untreated observations, respectively. In panel (a), blue observations are Ideal Experiment observations, and white observations are not yet considered or restricted. In panel (b), green, brown, and red observations are additionally justified by Assumptions \ref{assumption_invariance_time_shifts}, \ref{assumption_delay}, and \ref{assumption_washout}, respectively.}
        \label{fig:decomp_PT}
    \end{figure}

    The difference-in-means estimator constructed from $\Delta^{l^*+ \kappa + 1}Y^{\obs}_{it_y}$ is unbiased for $\ate^\Scal_{t_y}(t_1,\infty)$:
    \begin{equation}\label{equation_idealexperiment_PT_ideal_a}
    \widehat{\ate}^{\Scal}_{t_y}(t_1,\infty)
    = \frac{1}{n_{t_1}} \sum_{i = 1}^{n} \mathbbm{1}(G_i = t_1) \Delta^{l^* + \kappa + 1} Y^{\obs}_{it_y} - \frac{1}{n_{\infty}} \sum_{i = 1}^{n} \mathbbm{1}(G_i = \infty) \Delta^{l^* + \kappa + 1} Y^{\obs}_{it_y}.
    \end{equation}
    Here, $l^*=t_y-t_1$, and both components use the reference period $t_1-\kappa-1$. Assumption~\ref{assumption_limited_anticipation} implies a zero mean treatment effect at the reference period. Thus, $\ate^{\Scal}_{t_y}(t_1,\infty)$ can be expressed as the contrast between mean within-unit potential outcome changes from $t_1-\kappa-1$ to $t_y$ under initiation at $t_1$ and under never treatment. Equation~\eqref{equation_idealexperiment_PT_ideal_a} has the same treatment-minus-control structure as Equation~\eqref{equation_idealexperiment_standard}, with outcomes replaced by within-unit differences. Under Assumptions~\ref{assumption_limited_anticipation}--\ref{assumption_two_way_parallel}, the control component can also include not-yet-treated units satisfying $G_i>t_y+\kappa$. We can further replace the single reference period with a weighted average of outcomes over observed periods $s<t_1-\kappa$. For fixed weights satisfying $\sum_{s<t_1-\kappa}v_s=1$, the expanded estimator
    \begin{equation}\label{equation_idealexperiment_PT_ideal_c}  
        \begin{aligned}[b]
        \widehat{\ate}^{\Scal}_{t_y}(t_1,\infty)
        &=
        \frac{1}{n_{t_1}}\sum_{i=1}^n
        \mathbbm{1}\{G_i=t_1\}
        \left(
        Y^{\obs}_{it_y}
        -
        \sum_{s<t_1-\kappa}v_sY^{\obs}_{is}
        \right) \\
        &\quad -
        \frac{1}{n_{>t_y+\kappa}}\sum_{i=1}^n
        \mathbbm{1}\{G_i>t_y+\kappa\}
        \left(
        Y^{\obs}_{it_y}
        -
        \sum_{s<t_1-\kappa} v_sY^{\obs}_{is}
        \right),
        \end{aligned}  
    \end{equation}
    is also unbiased, where $n_{>t_y+\kappa}=\sum_{i=1}^n \mathbbm{1}\{G_i>t_y+\kappa\}$. Figure~\ref{fig:decomp_PT} illustrates the observations used by this estimator in the toy example introduced earlier. 

\subsection{Borrowing information}
\label{sec2:borrow_assumption}

The contrasts in Section~\ref{sec_hypothetical_experiment} use observations justified by the baseline identification assumptions in Section~\ref{sec_causal_identification}. In some settings, this limited information may make estimation infeasible or imprecise. We therefore introduce information-borrowing assumptions that allow the inclusion of additional observations by restricting treatment effect dynamics and potential outcome evolution. To state these assumptions once for the two frameworks, unconfoundedness and parallel trends, we let $Y^*_{it_y}(t)$ denote $Y_{it_y}(t)$ under unconfoundedness framework and $\Delta^lY_{it_y}(t)$ for a generic lag $l$ under parallel trends framework, and let $Y^{*\text{obs}}_{it_y}$ denote its observed counterpart. We explain the implications of these assumptions using parallel trends as the running framework, where we operate on within-unit outcome differences. Analogous arguments apply under unconfoundedness framework; the corresponding derivations are provided in Section~\ref{sec:app_borrowing_proof} of the Supplement.

For a fixed target estimand and identification route, let $\mathcal I_{\mathrm{ideal}}$ denote the observations used by the baseline contrast, with reference periods specified when outcomes are differenced. We use subscripts to record further expansions under the borrowing assumptions below.

\begin{assumption}[Invariance to time shifts]
\label{assumption_invariance_time_shifts}
For each time shift $h$ such that $t_1 + h$ and $t_y + h$ lie in $\Tcal$, $\E_{\S}\{Y^*_{it_y}(t_1) - Y^*_{it_y}(\infty) \mid \bX_i \} = \E_{\S}\{Y^*_{it_y+h}(t_1+h) - Y^*_{it_y+h}(\infty) \mid \bX_i \}$. 
\end{assumption}
Assumption~\ref{assumption_invariance_time_shifts} states that, conditional on baseline covariates, the average treatment effect depends only on how long a unit has been exposed to treatment, and not on when in calendar time the treatment occurred. Under parallel trends, this means we can use outcome differences from units treated earlier or later, as long as they are observed at the same time since treatment began. It is a conditional version of treatment-effect homogeneity across treatment cohorts \citep{sun2021estimating}. These additional observations are shown in green in Figure~\ref{fig:decomp_PT_b}.

\begin{assumption}[Delayed treatment onset]
\label{assumption_delay}
There is a known $\phi \geq 0$ such that $\E_{\S}\{Y^*_{it}(t - \phi^\prime) - Y^*_{it}(\infty) \mid \bX_i \} = 0$ for any $0 \leq \phi^\prime \leq \phi$ and any $t \in \Tcal$. 
\end{assumption}
Assumption~\ref{assumption_delay} states that, conditional on baseline covariates, the average treatment effect is zero for up to $\phi$ periods after treatment initiation. Under parallel trends, outcome differences from these recently treated units may enter the control component. A larger value of $\phi$ imposes a stronger restriction by requiring the treatment effect to remain absent for longer. These additional observations are shown in brown in Figure~\ref{fig:decomp_PT_b}.  

\begin{assumption}[Treatment effect dissipation]
\label{assumption_washout}
There is a known $\xi \geq 0$ such that $\E_{\S}\{Y^*_{it+\xi^\prime}(t) - Y^*_{it+\xi^\prime}(\infty)\mid \bX_i \} = 0$ for any $\xi^\prime \geq \xi$ and any $t \in \Tcal$. 
\end{assumption}
Assumption~\ref{assumption_washout} states that, conditional on baseline covariates, the average treatment effect has dissipated by $\xi$ periods after treatment initiation. Under parallel trends, this means that sufficiently long post-treatment outcomes can be ``recycled'' as if they were untreated, allowing outcome differences from units far enough from treatment onset to enter the control component. A smaller value of $\xi$ imposes a stronger restriction by requiring this return to baseline to occur more quickly. This transient-effect condition is also a central assumption in case-crossover designs \citep{shahn2023formal}, and it is a stronger version of the eventual-attenuation condition used in work that expands the comparison group \citep{cengiz2019, dube2025local}. In Figure~\ref{fig:decomp_PT_b}, these additional observations are shown in red.

Under unconfoundedness framework, when the adjustment set $\Fcal_i$ is observed, Assumptions~\ref{assumption_invariance_time_shifts}, \ref{assumption_limited_anticipation}, \ref{assumption_delay}, and \ref{assumption_washout} justify borrowing additional observations beyond those under the baseline Assumptions~\ref{assumption_consistency}--\ref{assumption_random_treatment_initiation}. In Figure~\ref{fig:decomp_usual_b}, the observations justified by these information-borrowing assumptions are shown in green, yellow, brown, and red, respectively. 



Including each additional group of observations requires a substantive justification for the corresponding assumption. The values of $\kappa$, $\phi$, and $\xi$ should be specified using subject-matter knowledge about the anticipated effect dynamics. Supplement Section~\ref{sec:app_simulation} examines the consequences of misspecifying delayed onset or effect dissipation. Section~\ref{sec:exact_decomp} takes the complementary approach: it recovers the observations and weights implied by an existing estimator so that its comparisons can be assessed against the intended design.


%% file: sec3.tex
\section{Building contrasts with weighting}
\label{sec3:weight_estimator}

For a specified target estimand, identification route, and eligible information set, we develop a robust weighting (RW) estimator that minimizes weight dispersion subject to approximate covariate balance and nonnegativity constraints. It estimates $\ate^{\Scal}_{t_y}(t_1,\infty)$ under either identification route. Under parallel trends, we use $Y^*_{it_y}(g)=\Delta^{l^*+\kappa+1}Y_{it_y}(g)$, where $l^*=t_y-t_1$. These changes are measured from the reference period $t_1-\kappa-1$, which must be observed. Section~\ref{sec:app_genestimand} of the Supplement extends the approach to the general estimand $\ate^\Pcal_{t_y}(t_1, t_\mathbf{p})$, allowing for an arbitrary target population $\Pcal$ and alternative treatment-initiation time $t_{\mathbf p}$.


\subsection{Ideal weighted contrasts}
\label{sec_buildingpure}
To begin, we posit only the baseline identification assumptions of Section \ref{sec_causal_identification}, under either framework. Then, we can write $\ate^\Scal_{t_y}(t_1, \infty) = \mu_1 -\mu_0$, where $\mu_1 = \E_\S \{Y^*_{it_y}(t_1)\}$ and $\mu_0 = \E_\S \{ Y^*_{it_y}(\infty)\}$. 
Further, $\mu_1$ and $\mu_0$ admit the following weighting representations
\begin{equation}
    \mu_1 = \frac{\E_\S \left\{\frac{\mathbbm{1}(G_i = t_1)}{\Pr(G_i = t_1|\bX_i)}Y^{*\obs}_{i{t_y}} \right\}}{\E_\S \left\{\frac{\mathbbm{1}(G_i = t_1)}{\Pr(G_i = t_1|\bX_i)}\right\}} \quad \textrm{and} \quad \mu_0 = \frac{\E_\S \left\{\frac{\mathbbm{1}(G_i = \infty)}{\Pr(G_i = \infty|\bX_i)}Y^{*\obs}_{i{t_y}} \right\}}{\E_\S \left\{\frac{\mathbbm{1}(G_i = \infty)}{\Pr(G_i = \infty|\bX_i)}\right\}}.
    \label{eq_ipw}
\end{equation}
From here, it is natural to estimate $\ate^\Scal_{t_y}(t_1, \infty)$ with a H\'{a}jek estimator
\begin{equation}
\widehat{\ate}^\Scal_{t_y}(t_1, \infty) = \frac{\sum_{i:G_i = t_1}\frac{Y^{*\obs}_{i{t_y}}}{\Pr(G_i = t_1|\bX_i)}}{\sum_{i:G_i = t_1}\frac{1}{\Pr(G_i = t_1|\bX_i)}} - \frac{\sum_{i:G_i = \infty}\frac{Y^{*\obs}_{i{t_y}}}{\Pr(G_i = \infty|\bX_i)}}{\sum_{i:G_i = \infty}\frac{1}{\Pr(G_i = \infty|\bX_i)}}.    
\end{equation}
More broadly, we consider a generic H\'{a}jek estimator of the form
\begin{equation}
    \widehat{\ate}^\Scal_{t_y}(t_1, \infty) =\sum_{i:G_i = t_1}w_{it_y}Y^{*\obs}_{it_y} - \sum_{i:G_i = \infty}w_{it_y}Y^{*\obs}_{it_y},
    \label{eq_contrast0}
\end{equation}
where $\sum_{i:G_i = t_1}w_{it_y} = \sum_{i:G_i = \infty}w_{it_y} = 1$. 
That is, $\widehat{\ate}^\Scal_{t_y}(t_1, \infty)$ is a weighted contrast between units that initiated treatment at $t_1$ and units that never initiated treatment, using observations measured at $t_y$. Under the parallel trends framework, the contrast also includes observations at the reference period $t_1 - \kappa - 1$. This weighted contrast is analogous to the unweighted contrast in Equation (\ref{equation_idealexperiment_standard}) for the ideal experiment in the sense that it uses minimal assumptions to yield a valid estimator of the average treatment effect. 
We refer to $\widehat{\ate}^\Scal_{t_y}(t_1, \infty)$ in Equation (\ref{eq_contrast0}) as an ideal weighted contrast.

Write $m_{t_1,t_y}(\bm{x}) = \E_\S\{Y^*_{i{t_y}}(t_1)|\bX_i = \bm{x}\}$ and $m_{\infty,t_y}(\bm{x}) = \E_\S\{Y^*_{i{t_y}}(\infty)|\bX_i = \bm{x}\}$. 
We can decompose the bias of the H\'{a}jek estimator (\ref{eq_contrast0}), $\E_\S\{\widehat{\ate}^\Scal_{t_y}(t_1, \infty)\} - \ate^\Scal_{t_y}(t_1, \infty)$, as 
\begin{align}
    \E_\S\left\{\sum_{i:G_i = t_1}w_{it_y}m_{t_1,t_y}(\bX_i) - \frac{1}{n}\sum_{i=1}^{n}m_{t_1,t_y}(\bX_i)\right\} - \E_\S\left\{\sum_{i:G_i = \infty}w_{it_y}m_{\infty,t_y}(\bX_i) - \frac{1}{n}\sum_{i=1}^{n}m_{\infty,t_y}(\bX_i)\right\}. \nonumber
\end{align}
Therefore, to remove the bias of $\widehat{\ate}^\Scal_{t_y}(t_1, \infty)$, the weights need to balance the function $m_{t_1,t_y}(\bm{x})$ in the treatment component, relative to the overall sample, and $m_{\infty,t_y}(\bm{x})$ in the control component, relative to the overall sample. We can thus construct these weights using standard minimum-dispersion balancing methods developed for observational studies with time-invariant treatments (see \citealt{ben2021balancing} for a review). We adapt these methods to event studies by linking the eligible observations to the identifying and borrowing assumptions and using within-unit outcome differences under parallel trends. Together with the implied-weight representations, this formulation clarifies how regression models borrow strength across units and time in panel data and the assumptions that justify this borrowing. In particular, the weights $\{w_{it_y}: G_i = t_1\}$ can be obtained by solving an optimization problem with the following general form
\begin{align}\label{minimal_1}
\min_{\bm{w}}\left\{\sum_{i: G_i=t_1}\psi(w_{it_y}) : \max\limits_{m(\cdot) \in \mathcal{M}}\left| \sum_{i:G_i = t_1}w_{it_y}m(\bX_i) - \frac{1}{n}\sum_{i=1}^{n}m(\bX_i)\right|\leq \bm{\delta} \right\} 
\end{align}
where $\mathcal{M}$ is a class of functions containing $m_{t_1,t_y}(\cdot)$, and $\psi(\cdot)$ is a convex function quantifying the dispersion of the weights \citep{wang2020minimal}. In other words, \eqref{minimal_1} finds the weights by minimizing their dispersion subject to a constraint on the worst-case imbalance in the weighted sample across the class $\mathcal{M}$ up to a parameter $\delta$. We can obtain the weights $\{w_{it_y}: G_i = \infty\}$ by solving an analogous optimization problem.

In practice, we assume that $\mathcal{M}$ can be spanned by a set of $K$ basis functions that grow with the sample size, i.e., $\mathcal{M} = \text{span}\left\{B_k(\cdot), k=1,2,...,K \right\}$. 
We set $\psi(\cdot)$ as the squared norm of the weights and constrain them to be non-negative and sum to one. 
Thus, the optimal weights are normalized, of minimum squared norm (or equivalently, variance), and they approximately balance the basis functions $\bm{B}(\cdot) = (B_1(\cdot),...,B_K(\cdot))^\top$ in $\mathcal{M}$ relative to the overall sample. Also, the weights are non-negative, which ensures that the resulting estimators are sample-bounded \citep{robins2007comment}. 

Finally, using the one-step weighting approach of \cite{chattopadhyay2024one}, our proposed weighting method can be directly extended to handle generalization and transportation problems. See Section \ref{sec:app_robust_generalization} of the Supplement for formal details. Roughly speaking, for an arbitrary target population $\mathcal{P}$, the weights are now constrained to approximately balance the set of basis functions relative to $\mathcal{P}$, instead of the study population.

\subsection{Expanded weighted contrasts}
\label{sec_buildinglarger}

The ideal weighted contrast compares one treatment cohort with never-treated units at a given time. We now expand this contrast to include observations at the same event time in other cohorts. Using them to estimate the same target effect requires invariance to time shifts (Assumption~\ref{assumption_invariance_time_shifts}). This assumption is not needed to define an aggregate estimand that averages potentially different cohort effects. All sums and unions over $r$ below are restricted to initiation times in $\Tcal$ for which $r+l^*\in\Tcal$. Under parallel trends, we also require $r-\kappa-1\in\Tcal$, so the reference period is observed. Under Assumption ~\ref{assumption_invariance_time_shifts}, we can write 
\begin{equation}
    \ate^\Scal_{t_y}(t_1,\infty)
    =
    \sum_r \lambda_r
    \E_\S\{Y^*_{i(r+l^*)}(r)-Y^*_{i(r+l^*)}(\infty)\}. 
    \label{eq_convex}
\end{equation}
Here, $l^* = t_y -  t_1$, and the nonnegative weights $\lambda_r$ sum to one over the included cohorts. A heuristic choice is each cohort's proportion of units among these cohorts. Equation (\ref{eq_convex}) represents $\ate^\Scal_{t_y}(t_1, \infty)$ as a convex combination of average treatment effects, with outcome measurement taking place $l^*$ periods after treatment initiation.

Denote $\Tilde{\mu}_1 = \sum_{r} \lambda_r \E_\S\{ Y^*_{i(r+l^*)}(r)\}$ and $\Tilde{\mu}_0 = \sum_{r}\lambda_r \E_\S \{ Y^*_{i(r+l^*)}(\infty) \}$, so $\ate^\Scal_{t_y}(t_1, \infty) = \Tilde{\mu}_1 - \Tilde{\mu}_0$. 
For any $r,s \in \mathcal{T}^+$, define $\mathcal{C}_{r,s} = \{(i,t): G_i = r, t = s\}$ as the set of observations where treatment is initiated at time $r$ and outcome is measured at time $s$. 

Finally, let $\mathcal{C} = \cup_{r} \mathcal{C}_{r,r+l^*}$ and $\mathcal{C}' = \cup_{r} \mathcal{C}_{\infty,r+l^*}$. 
As before, we consider the following generic weighted estimators for $\Tilde{\mu}_1$ and $\Tilde{\mu}_0$, 
\begin{equation}
    \hat{\Tilde{\mu}}_1 = \sum_{(i,t) \in \mathcal{C}}w_{it}Y^{*\obs}_{it} \quad \textrm{and} \quad \hat{\Tilde{\mu}}_0 = \sum_{(i,t) \in \mathcal{C}'}w_{it}Y^{*\obs}_{it}.
\end{equation}
Decomposing the bias of $\hat{\Tilde{\mu}}_1$, we can show that
\begin{align}
    \E_\S(\hat{\Tilde{\mu}}_1) - \Tilde{\mu}_1 = \sum_{r} \E_\S \left[\sum_{(i,t)\in \mathcal{C}_{r,r+l^*}} w_{it}m_{r,r+l^*}(\bX_i) - \lambda_r\left\{\frac{1}{n}\sum_{i=1}^{n}m_{r,r+l^*}(\bX_i)\right\} \right].
    \label{eq_bias_1}
\end{align}
The weights that satisfy $\sum_{(i,t)\in \mathcal{C}_{r,r+l^*}} w_{it}m_{r,r+l^*}(\bX_i) = \lambda_r\left\{\frac{1}{n}\sum_{i=1}^{n}m_{r,r+l^*}(\bX_i)\right\}$ for every included cohort $r$ completely remove the bias of $\hat{\Tilde{\mu}}_1$. 
Therefore, we can compute the weights $w_{it}$ by first approximating $m_{r,r+l^*}(\bm{x})$ with suitable basis functions and then including the above conditions as balancing constraints in an analogous optimization program to \eqref{minimal_1}. 
The weights for estimating $\Tilde{\mu}_0$ can be derived similarly. 
By construction, the resulting estimator of $\ate^\Scal_{t_y}(t_1, \infty)$ now incorporates observations measured at time points other than $t_y$.

This procedure requires sufficient observations in $\mathcal{C}_{r,r+l^*}$ to satisfy the $T - l^*$ balancing constraints for each included cohort $r$. By adopting stronger assumptions, we can relax these constraints, and thereby improve the efficiency of the resulting estimator.
In particular, we consider a stronger version of Assumption \ref{assumption_invariance_time_shifts} that requires $\E_\S\{Y^*_{it_y}(t_1)\mid \bX_i \} = \E_\S\{Y^*_{it_y+s}(t_1+s) \mid \bX_i \}$ and $\E_\S\{Y^*_{it_y}(\infty) \mid \bX_i \} = \E_\S\{Y^*_{it_y+s}(\infty) \mid \bX_i \}$ for all shifts $s$ for which the required outcome and reference periods are observed. Under this assumption, $m_{r,r+l^*}(\bm{x}) = m_{l^*}(\bm{x})$ for all $r$. 
Then, we can rewrite the bias of $\hat{\Tilde{\mu}}_1$ as
\begin{align}
    \E_\S(\hat{\Tilde{\mu}}_1) - \Tilde{\mu}_1 = \E_\S \left\{\sum_{(i,t)\in \mathcal{C}} w_{it}m_{l^*}(\bX_i) - \frac{1}{n}\sum_{i=1}^{n}m_{l^*}(\bX_i)\right\} .
    \label{eq_bias_2}
\end{align}
Equation (\ref{eq_bias_2}) says that the weights will ideally balance the mean $m_{l^*}(\bm{x})$ in $\mathcal{C}$ towards the overall sample. 
Therefore, under the stronger version of Assumption \ref{assumption_invariance_time_shifts}, the cohort-specific constraints reduce to a single balancing constraint, enhancing the feasibility and efficiency of the weighting method. 
In this case, we can obtain the weights by solving the following analog of the optimization problem in \eqref{minimal_1}
\begin{align}\label{minimal_2}
\min_{\bm{w}}\left\{\sum_{(i,t) \in \mathcal{C}} \psi(w_{it}) : \max\limits_{m(\cdot) \in \mathcal{M}}\left| \sum_{(i,t) \in \mathcal{C}}w_{it}m(\bX_i) - \frac{1}{n}\sum_{i=1}^{n}m(\bX_i)\right|\leq \bm{\delta} \right\}. 
\end{align}

Under the stronger version of Assumption \ref{assumption_invariance_time_shifts}, we can further expand the control component of the weighted contrast. Under parallel trends, we set $\gamma_t=0$ and $w_{it}=0$ whenever $t-(l^*+\kappa+1)\notin\Tcal$ and omit these periods from the sums below. Here, $\Tilde{\mu}_0 = \E_\S\{Y^*_{it}(\infty)\}$ for all $t$ and $\Tilde{\mu}_0 = \sum_{t=1}^{T} \gamma_t \E_\S\{Y^*_{it}(\infty)\}$, where $\gamma_1,...,\gamma_T$ are arbitrary non-negative scalars that sum to one. 
Thus, we can estimate $\Tilde{\mu}_0$ using the weighting estimator $\hat{\tilde{\mu}}_0 = \sum_{(i,t) :G_i  =\infty} w_{it} Y^{*\obs}_{it}$. 
This estimator uses information from the never-treated units across all time points, as opposed to the time points in the set $\mathcal{C}'$ only. 
The bias of this estimator is
\begin{align}
    \E_\S(\hat{\Tilde{\mu}}_0) - \Tilde{\mu}_0 = \E_\S \left\{\sum_{(i,t): G_i = \infty} w_{it}m_{\infty}(\bX_i) - \frac{1}{n}\sum_{i=1}^{n}m_{\infty}(\bX_i)\right\}
    \label{eq_bias_3}
\end{align}
where $m_{\infty}(\bm{x}) = m_{\infty,t}(\bm{x})$ is the common mean function over the included periods. 
Hence, similar to the estimation of $\tilde{\mu}_1$, we can find the weights in Equation (\ref{eq_bias_3}) by solving an analog of the optimization problem in \eqref{minimal_2}. 
Ultimately, the resulting expanded weighted contrast for $\ate^\Scal_{t_y}(t_1, \infty)$ is given by
\begin{equation}
    \widehat{\ate}^\Scal_{t_y}(t_1, \infty) = \sum_{(i,t) \in \mathcal{C}}w_{it}Y^{*\obs}_{it} -  \sum_{(i,t):G_i = \infty}w_{it}Y^{*\obs}_{it}, \label{eq_contrast1}
\end{equation}
where the weights within each component are non-negative and sum to one. 

\subsection{Asymptotic properties}
\label{sec_large}

Leveraging the asymptotic properties of one-step weighting estimators \citep{chattopadhyay2024one, wang2020minimal}, we can establish similar properties for the proposed estimators.
We first discuss the asymptotic properties of the ideal weighted contrasts, as defined in Equation (\ref{eq_contrast0}).
To this end, we make two assumptions, one on the assignment mechanism and the other on the potential outcomes.
In both cases, we denote $t$ as a generic time point in $\mathcal{T}^+$.
\begin{assumption}[Approximability of the treatment initiation model]
\label{assump_consistency1}
The conditional probability for treatment initiation at time $t$ satisfies $\Pr(G_i = t|\bX_i = \bm{x}) = [n \rho'\{g^{*}_t(\bm{x})\}]^{-1}$ for all $\bm{x} \in \text{Supp}(\bX_i)$, where $\rho(\cdot)$ is a smooth function obtained from the dual of \eqref{minimal_1}, and $g^*_t(\cdot)$ is a smooth function that satisfies $\sup_{\bm{x} \in \text{Supp}(\bX_i)}|g^*_t(\bm{x}) - \bm{B}(\bm{x})^\top \bm{\lambda}^*_{1t}| = O(K^{-r_t})$ for some $\bm{\lambda}^*_{1t} \in \mathbb{R}^K$ and $r_t>1$.
\end{assumption}

\begin{assumption}[Approximability of the potential outcome model]
\label{assump_consistency2}
The conditional mean function of the potential outcome $Y^*_{it_y}(t)$ satisfies $\sup_{\bm{x}\in \text{Supp}(\bX_i)}|m_{t,t_y}(\bm{x}) - \bm{B}(\bm{x})^\top \bm{\lambda}^*_{2t}| = O(K^{-s_t}) $ for some $\bm{\lambda}^*_{2t} \in \mathbb{R}^K$ and $s_t>3/4$, with $||\bm{\lambda}^*_{2t}||_2 ||\bm{\delta}||_2 = o(1)$.
\end{assumption}

Assumption \ref{assump_consistency1} states that the conditional probabilities of treatment initiation $\Pr(G_i = t|\bX_i = \bm{x})$ can be uniformly well-approximated by the basis functions in the functional class $\mathcal{M}$ with a known link function. 
We say that the treatment initiation model for time $t$ is correctly specified if Assumption \ref{assump_consistency1} holds for $t$.
Likewise, Assumption \ref{assump_consistency2} states that the conditional mean functions of the potential outcomes $m_{t,t_y}(\bm{x})$ can be uniformly well-approximated by basis functions in $\mathcal{M}$. 
We say that the potential outcome model for time $t$ is correctly specified if Assumption \ref{assump_consistency2} holds for $t$.
With these two assumptions, and some additional regularity conditions (see Assumptions \ref{assump_regularity1} and \ref{assump_regularity2} in the Supplement),  $\widehat{\ate}^\Scal_{t_y}(t_1, \infty)$ is consistent under multiple non-nested conditions, asymptotically Normal, and semiparametrically efficient for $\ate^\Scal_{t_y}(t_1, \infty)$. 
The formal results are as follows.

\begin{theorem}[Consistency] \normalfont
\label{thm_consistency}
Suppose that Assumptions \ref{assump_consistency1} and \ref{assump_regularity1} hold. Then
the weighting estimator $\widehat{\ate}^\Scal_{t_y}(t_1, \infty) =\sum_{i:G_i = t_1}w_{it_y}Y^{*\obs}_{it_y} - \sum_{i:G_i = \infty}w_{it_y}Y^{*\obs}_{it_y}$ is consistent for $\ate^\Scal_{t_y}(t_1, \infty)$ if any of the following conditions are satisfied:
(a) The treatment initiation models are correctly specified for both time $t_1$ and $\infty$, (b) the potential outcome models are correctly specified for both time $t_1$ and $\infty$, (c) the treatment initiation model for time $t_1$ and the potential outcome model for time $\infty$ are both correctly specified, (d) the treatment initiation model for time $\infty$ and the potential outcome model for time $t_1$ are both correctly specified.
\end{theorem}

\begin{theorem}[Asymptotic Normality and semiparametric efficiency]\normalfont
\label{thm_normality} 
Suppose that Assumptions \ref{assump_consistency1}, \ref{assump_consistency2}, \ref{assump_regularity1}, and \ref{assump_regularity2} hold. Then, $n^{1/2}\{\widehat{\ate}^\Scal_{t_y}(t_1, \infty) - \ate^\Scal_{t_y}(t_1, \infty)\} \xrightarrow{d} N(0, V_\tau)$, where $V_\tau$ is the semiparametric efficiency bound for $\ate^\Scal_{t_y}(t_1, \infty)$.
\end{theorem}

The asymptotic properties discussed so far carry over to the expanded weighted contrasts in Equation (\ref{eq_contrast1}) under the same model for the potential outcomes and a slightly modified model for the treatment initiation. 
More specifically, in the previous case of ideal contrasts, we require specific functional forms for $\Pr(G_i = \infty|\bX_i = \bm{x})$ and $\Pr(G_i = t_1|\bX_i = \bm{x})$, as given in Assumption \ref{assump_consistency1}.
In the case of extended contrasts, we require these functional forms for $\Pr(G_i = \infty|\bX_i = \bm{x})$ and $\Pr(G_i \in \{1,2,...,T - l^*\}|\bX_i = \bm{x})$, respectively, with the initiation-time range restricted to admissible values. Under the modified models and the additional regularity conditions as before, we can show analogously that the expanded weighting estimator $\widehat{\ate}^\Scal_{t_y}(t_1, \infty)$ is asymptotically Normal, multiply robust, and semiparametrically efficient for $\ate^\Scal_{t_y}(t_1, \infty)$. 
In practice, we conduct inference using a unit-level bootstrap, which is what we implement in the case study of Section \ref{sec:case_study}.


%% file: sec4.tex
\section{Decompositions and diagnostics}
\label{sec:diagnostics}

Section~\ref{sec3:weight_estimator} starts with a stated design and constructs an estimator by choosing weights for the eligible observations. We now start with an estimator and recover the observations it uses and the weights it assigns to them. These exact representations as weighted contrasts of within-unit outcome differences provide a common basis for comparing estimators and developing diagnostics for how they use information across the panel \citep{chattopadhyay2023implied}.

\subsection{Exact decompositions of event study estimators}
\label{sec:exact_decomp}

\subsubsection{Dynamic TWFE estimator}
Two-way fixed effects (TWFE) regression, one of the most widely used approaches to event study estimation, admits both static and dynamic specifications:
\begin{align}
    & \text{Static,} & Y_{i t} & =\alpha_i+\beta_t+ \tau \mathbbm{1}(t \geq G_i) +\epsilon_{it}, \quad \forall i, t; \\
    & \text{Dynamic,} & Y_{i t} & =\alpha_i+\beta_t+ \sum\limits_{\substack{l=a \\ l \neq -1}}^{b} \tau_{l} \mathbbm{1}(t - G_i = l) +\epsilon_{it}, \quad \forall i, t. \label{equation_TWFE_dynamic}
\end{align}
Here $\alpha_i$ and $\beta_t$ are unit and time fixed effects. The static specification includes a single coefficient $\tau$ on treatment status, whereas the dynamic specification includes a separate coefficient $\tau_l$ for each event time in the window $[a,b]$. The period $l = -1$ is conventionally omitted as the reference. Throughout, we focus on the dynamic specification in Equation~\eqref{equation_TWFE_dynamic} and omit covariates for clarity, although the decomposition extends directly to the static specification and to specifications that include covariates.

Suppose the dynamic TWFE regression is correctly specified. Then, for $l^* = t_y - t_1$, $\hat{\tau}_{l^*}$ is unbiased for $\ate^{\Scal_{t}}_{t_y}(t_1, \infty)$. The estimator can also be rewritten exactly as a weighted contrast of within-unit outcome differences:
\begin{align}
\label{equation_lmw_tau}
    \hat{\tau}_{l^*}
    = \sum_{i = 1}^n \sum_{t = 1}^T w_{it}\, \Delta^{l^*+1} Y_{it}^{\mathrm{obs}} \mathbbm{1}\{t - G_i = l^*\} - \sum_{i = 1}^n \sum_{t = 1}^T w_{it} \Delta^{t-G_i+1} Y_{it}^{\mathrm{obs}} \mathbbm{1}\{t - G_i \notin \{-1, l^*\} \}.
\end{align}
Here $\Delta^{t-G_i+1} Y_{it}^{\mathrm{obs}} = Y_{it}^{\mathrm{obs}} - Y_{i, G_i - 1}^{\mathrm{obs}}$ is the difference in unit $i$'s outcome from the period immediately before treatment initiation to period $t$, and $w_{it}$ are the implied regression weights of \citet{chattopadhyay2023implied}. In Equation~\eqref{equation_lmw_tau}, the first component uses observations at the target event time $l^*$, whereas the second draws on observations at all other event times except the reference period. For never-treated units, the outcome difference can be defined relative to any reference period. 

The weights are normalized so that $\sum_{i = 1}^n \sum_{t = 1}^T w_{it} \mathbbm{1}\{ t - G_i = l^* \} = 1$, while $\sum_{i = 1}^n \sum_{t = 1}^T w_{it} \mathbbm{1}\{ t - G_i = l\} = 0$ for every
$l \in [a, b] \setminus \{-1, l^*\}$. The weights also enforce finite-sample balance for unit fixed effects, time fixed effects, and non-target event-time indicators. Supplement Section~\ref{sec:app_twfe_decomp_proof} proves the representation and gives the explicit form of the weights.

Related decompositions are provided by \citet{goodman2021difference}, who writes the static
TWFE estimator as a weighted average of $2 \times 2$ difference-in-differences comparisons, by \citet{sun2021estimating}, who write $\hat{\tau}_{l^*}$ as a weighted average of cohort-specific effects, and by \citet{imai2021use}, who show that the multi-period difference-in-differences estimator equals a weighted TWFE estimator with some negative weights; see also \citet{de2023two}, \citet{li2024guide}, and \citet{caetano2024difference}. Whereas the work above rewrites the TWFE estimator in terms of more aggregated objects, our decomposition operates at the unit-period level. It shows which observations form the two components of the contrast, how they are weighted, and how TWFE borrows information across units, time periods, and treatment statuses.

\subsubsection{Interaction-weighted estimator}

Next, we consider the interaction-weighted estimator proposed by \cite{sun2021estimating}, which allows dynamic treatment effects to vary with the timing of treatment initiation. The estimator first obtains cohort-specific effects from a dynamic TWFE regression that interacts relative-time indicators with treatment-initiation indicators:
\begin{equation}
    \label{equation_SA_PT}
    Y_{it}=\alpha_i+\beta_t+\sum_{g \in \Tcal} \sum\limits_{\substack{l=a \\ l \neq -1}}^{b} \tau_{g, l}\mathbbm{1}(G_i=g, t - G_i = l) +\epsilon_{it}, \quad \forall i, t.
\end{equation}
Relative to the dynamic TWFE model in Equation~(\ref{equation_TWFE_dynamic}), this specification allows the effect at relative time $l$ to differ across treatment initiation times $g$. If the model is correctly specified, then $\hat{\tau}_{t_1, l^*}$ with $l^* = t_y - t_1$ is unbiased for $\ate^{\Scal_{t}}_{t_y}(t_1, \infty)$. Because the coefficient is indexed by cohort and event time, correct specification does not require effects at a given event time to be common across cohorts, so Assumption~\ref{assumption_invariance_time_shifts} is not invoked. Because $\hat{\tau}_{t_1, l^*}$ is again linear in the observed outcomes, the same argument characterizes it through its implied weights:
\begin{equation}
    \label{equation_SA_PT_decomp}
    \begin{aligned}[b]
    \hat{\tau}_{t_1, l^*}
    & = \sum_{i = 1}^n \sum_{t = 1}^T w_{it}\, \Delta^{l^*+1} Y_{it}^{\mathrm{obs}}\, \mathbbm{1}\{ G_i = t_1,\ t - G_i = l^* \} \\
    & \quad - \sum_{i = 1}^n \sum_{t = 1}^T w_{it}\, \Delta^{t-G_i + 1} Y^\obs_{it}\, \mathbbm{1}\{ (G_i,\, t - G_i) \neq (t_1,\, l^*),\ t - G_i \neq -1 \},
    \end{aligned}
\end{equation}
where $w_{it}$ denote the implied weights under this specification, which differ from those in Equation~\eqref{equation_lmw_tau} though we reuse the symbol. The weights on observations in the target cohort-by-event-time cell satisfy $\sum_{i=1}^n\sum_{t=1}^T w_{it}\mathbbm{1}\{G_i=t_1,\ t-G_i=l^* \}=1$, while for every other treated cohort-by-event-time cell $(g,l)\neq(t_1,l^*)$, $\sum_{i=1}^n \sum_{t=1}^T w_{it}\mathbbm{1}\{G_i=g,\ t-G_i=l\}=0$. As for the dynamic TWFE estimator, the weights also enforce exact balance between the two components on the unit fixed effects, the time fixed effects, and the remaining cohort-by-event-time indicators. Equation (\ref{equation_SA_PT_decomp}) expresses the coefficient as a weighted contrast of within-unit outcome differences. In a balanced panel without covariates, the saturated specification with never-treated units as controls yields the difference in mean changes between the target cohort and never-treated units. Each observation in a non-target treated cohort receives zero weight. Supplement Section~\ref{sec:app_sa_decomp_proof} derives this simplification and the exact weights.

\subsubsection{Doubly robust estimator}

Finally, we consider the doubly robust estimator of \citet{sant2020doubly}, extended by \citet{callaway2021difference} to settings with multiple periods and staggered treatment adoption. This estimator combines outcome regression with a propensity score model for treatment initiation. For the estimand $\ate^{\Scal_t}_{t_y}(t_1,\infty)$, it takes the form
\begin{equation}
    \label{equation_DR_callaway}
    \widehat{\ate}^{\Scal_t}_{t_y}(t_1,\infty) =\frac{1}{n} \sum_{i = 1}^n \left[\left(\widehat{q}_{it_1}^{\text {treat}}-\widehat{q}_{it_1}^{\text {never}}\right)\left(Y^\obs_{it_y}-Y^\obs_{it_1-\kappa-1}-\widehat{m}_{t_1, t_y, \kappa}\left(\bX_i \right)\right)\right], 
\end{equation}
where the weights are defined as:
\begin{align*}
\widehat{q}_{it_1}^{\text{treat}} = \frac{\mathbbm{1}\{G_i = t_1\}}{\frac{1}{n} \sum_{j = 1}^n \mathbbm{1}\{G_j = t_1\}}, \qquad
\widehat{q}_{it_1}^{\text{never}} = \frac{\widehat{e}_{t_1}(\bX_i)\, \mathbbm{1}\{G_i = \infty\} / \{1 - \widehat{e}_{t_1}(\bX_i)\}}{\frac{1}{n} \sum_{j = 1}^n \widehat{e}_{t_1}(\bX_j)\, \mathbbm{1}\{G_j = \infty\} / \{1 - \widehat{e}_{t_1}(\bX_j)\}},
\end{align*}
where $\widehat{e}_{t_1}(\bX_i) = \widehat{\Pr} \left(G_i = t_1 \mid \bX_i,\, G_i \in \{t_1, \infty\} \right)$ denotes the estimated probability of initiating treatment at $t_1$, rather than remaining never treated, conditional on $\bX_i$ among units in these two groups, and $\widehat{m}_{t_1, t_y, \kappa}(\bX_i) = \widehat{\E}_{\S} \{ Y^\obs_{it_y} - Y^\obs_{it_1 - \kappa - 1} \mid \bX_i,\, G_i = \infty \}$ denotes the estimated conditional expectation of the difference in outcomes for never-treated units, accounting for anticipation behavior with horizon $\kappa$, which may be nonzero. 

The estimator in Equation~(\ref{equation_DR_callaway}) has the usual doubly robust form, with the outcome model applied to the within-unit outcome difference, $\Delta^{t_y-t_1+\kappa+1}Y^\obs_{it_y}=Y^\obs_{it_y}-Y^\obs_{it_1-\kappa-1}$. When the outcome model is linear in $\bX_i$ and estimated by least squares, Theorem~3 of \citet{chattopadhyay2023implied} gives a weighting representation with the treatment-minus-control structure:
\begin{equation}
    \label{equation_DR_decomp}
    \widehat{\ate}^{\Scal_{t}}_{t_y}(t_1,\infty)
    =\sum_{i:\,G_i=t_1}w_i^{\text{treat}}\,\Delta^{t_y-t_1+\kappa+1}Y^\obs_{it_y}
    -\sum_{i:\,G_i=\infty}w_i^{\text{never}}\,\Delta^{t_y-t_1+\kappa+1}Y^\obs_{it_y}.
\end{equation}
The treated weights are simply the normalized weights for cohort $t_1$:
\[
  w_i^{\text{treat}}=\tfrac{1}{n}\,\widehat{q}_{it_1}^{\text{treat}}
  =\frac{\mathbbm 1\{G_i=t_1\}}{\sum_{j}\mathbbm 1\{G_j=t_1\}}. 
\]
For never-treated units, $w_i^{\text{never}}$ equals the propensity-score base weight, $\frac{1}{n}\widehat{q}_{it_1}^{\text{never}}$, plus the correction implied by the outcome regression. In the contrast in Equation~\eqref{equation_DR_decomp}, the treated and never-treated weights each sum to one and yield exact balance in $\bX_i$. Unit fixed effects are balanced by construction through within-unit differencing, while time fixed effects are balanced because both components use the same two time periods. 

Finally, the RW estimator's weights are the solution to Equation~\eqref{minimal_1} or~\eqref{minimal_2}. Because it constructs the treatment and control components explicitly, no decomposition is needed and the diagnostics of Section \ref{sec:diagnostic_metrics} apply directly.

\subsection{Diagnostics}
\label{sec:diagnostic_metrics}

The weights show which observation groups each estimator uses and how much weight each group receives. We first summarize the distribution of these weights, then examine how deleting an observation changes the estimate. We assess information borrowing using the effective sample size (ESS) of each observation group. This measure accounts for both the number of observations in the group and the dispersion of their weights. We define the group-wise ESS as
\begin{equation}
    {n}_{\text{eff}, g} = \frac{(\sum^{n}_{i=1} \sum^{T}_{t=1} |w_{it}| \mathbbm{1} \{ (i, t) \in g\} )^{2} }{\sum^{n}_{i=1} \sum^{T}_{t=1} w_{it}^{2} \mathbbm{1}\{ (i, t) \in g\} }, 
\end{equation}
where $g \in \mathcal{G}$ indexes a group in a chosen partition of the observations, such as by time period, unit, or treatment cohort. This metric measures the effective sample size that a group contributes to the final estimator. To quantify the information contributed by each group, we compute the ratio of the group-wise ESS to the sum of group-wise ESS as 
\begin{equation}
    {p}_{\text{info}, g} = \frac{{n}_{\text{eff}, g}}{\sum_{g\in \mathcal{G}} {n}_{\text{eff}, g}}.  
\end{equation}
We further define the squared-weight share of group $g$ as
\begin{equation}
\label{eq:D_g}
    D_g = \frac{\sum^{n}_{i=1} \sum^{T}_{t=1} w_{it}^{2} \mathbbm{1}\{ (i, t) \in g\}}{\sum^{n}_{i=1} \sum^{T}_{t=1} w_{it}^{2}}.
\end{equation}
The measure $D_g$ captures the share of the total squared weight attributed to the group $g$. Unlike the information share $p_{\text{info},g}$, which captures both the number of weighted observations and the within-group dispersion of the implied weights, $D_g$ decomposes the total squared weight such that $\sum_{g \in \mathcal{G}} D_g = 1$. A group with high $p_{\text{info},g}$ but low $D_g$ contains many observations with small weights, while a group with high $D_g$ but low $p_{\text{info},g}$ contains few observations with large weights that disproportionately influence the estimate.

To assess the influence of individual observations on the estimator, we compute the change in the point estimate when each observation is removed. The change in the point estimate for the observation indexed by $(i,t)$ is defined as $\operatorname{PE}_{it}= \hat{\tau}_{-(it)} - \hat{\tau}$, where $\hat{\tau}_{-(it)}$ is the estimate obtained after deleting observation $(i,t)$ and re-estimating the estimator. A larger magnitude of $\mathrm{PE}_{it}$ indicates that the corresponding observation has a greater influence on the estimator. A useful extension would delete an entire group of correlated observations rather than a single observation. We leave this direction for future work and refer interested readers to the related literature on group deletion and influence diagnostics \citep{banerjee1997influence, zhu2012perturbation}.

Additional diagnostic measures, such as covariate balance, sign reversal \citep{small2017instrumental}, and the target population implied by the weight adjustments, offer further insight into the estimators. We illustrate these diagnostic measures in Section \ref{sec_app_further} of the Supplement. The proposed diagnostics can also be applied to other methods that admit a linear, outcome-weighted representation \citep{knaus2024treatment}. We leave the derivation of these weighting representations to future work. 


%% file: sec5.tex
\section{Case study: Castle Doctrine and homicides}
\label{sec:case_study}

\subsection{Study overview}
\label{sec:case_overview}

To illustrate our approach, we revisit \citeauthor{cheng2013does}'s (\citeyear{cheng2013does}) study on the impact of Castle Doctrine laws on homicide rates in the United States from 2000 to 2010. Castle Doctrine laws, also known as ``stand your ground'' laws, expand the right to use lethal force in self-defense without a duty to retreat, potentially affecting homicide rates through both deterrence and escalation channels.

In this study, the outcome is the log annual state-level homicide rate, and the treatment is the enactment of a Castle Doctrine law in a state. Each observation is a specific state in a given year, resulting in a dataset of 550 observations from 50 states over 11 years. Among these states, 21 adopted Castle Doctrine laws between 2005 and 2009 across five treatment cohorts: one state in 2005, thirteen in 2006, four in 2007, two in 2008, and one in 2009. The remaining 29 states did not enact such laws during this period. We analyze the study under the parallel trends framework and compare estimators that differ in the observations they use and whether they incorporate covariate adjustment. We first present their event study estimates and then use their weights to identify the observations and comparisons that determine those estimates.

\subsection{Estimation of dynamic treatment effects}
\label{sec:case_estimation}

We first turn to the estimation of dynamic treatment effects. The previous sections analyzed identification and estimation for the cohort-specific estimand $\ate_{t_y}^{\Scal_t}(t_1,\infty)$, which measures the effect in year $t_y$ among states that first enacted a Castle Doctrine law in year $t_1$, relative to the counterfactual in which those states never enacted the law. This estimand often serves as a building block for an aggregate event study parameter. Examining each building block separately helps clarify which comparisons and observations are justified for each enactment cohort and outcome year before the effects are combined across cohorts. We examine both levels in this case study. This section focuses on the aggregate estimand while Supplement Section~\ref{sec_app_further} reports the cohort-specific estimates and diagnostics.

At each event time $l$, we define the aggregate estimand $\ate^{\Scal_t}_{\mathrm{agg}}(l)$ as the cohort-size-weighted average of $\ate^{\Scal_g}_{g+l}(g,\infty)$ over enactment cohorts observed at $l$. We compare four estimators. The first is a dynamic TWFE regression with event-time indicators and state and year fixed-effects, using the year immediately before enactment as the omitted reference period. The other three are the interaction-weighted (IW) estimator of \cite{sun2021estimating}, the doubly robust (DR) estimator of \citeauthor{sant2020doubly} (\citeyear{sant2020doubly}; \citealt{callaway2021difference}), and our proposed robust weighting (RW) estimator. IW and DR first estimate the effect for each enactment cohort and then aggregate these estimates. RW can be implemented in the same two-step manner, but it also admits a direct one-step implementation that pools the treated and control components across enactment cohorts before solving a single optimization problem.

\begin{figure}[htbp]
  \centering
  \begin{subfigure}{0.48\textwidth}
    \centering
    \includegraphics[width=\textwidth]{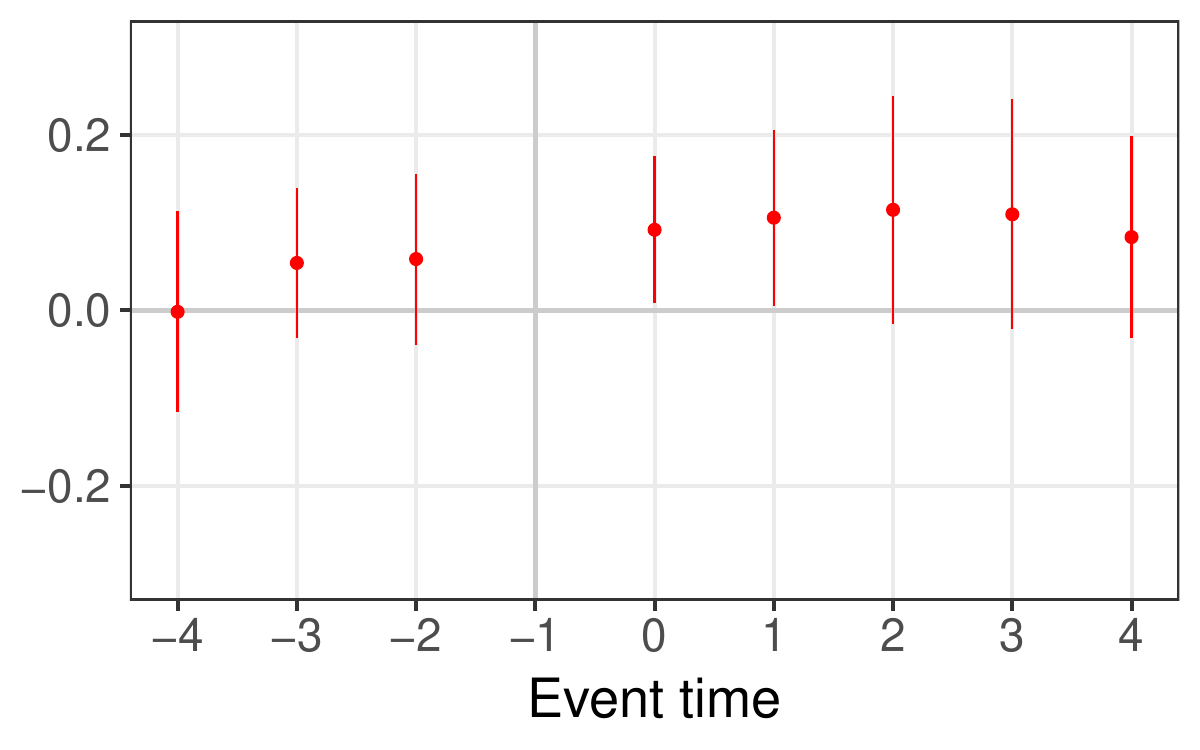}
    \caption{Dynamic TWFE}
    \label{fig:event_study_plot_twfe}
  \end{subfigure}%
  \hfill
  \begin{subfigure}{0.48\textwidth}
    \centering
    \includegraphics[width=\textwidth]{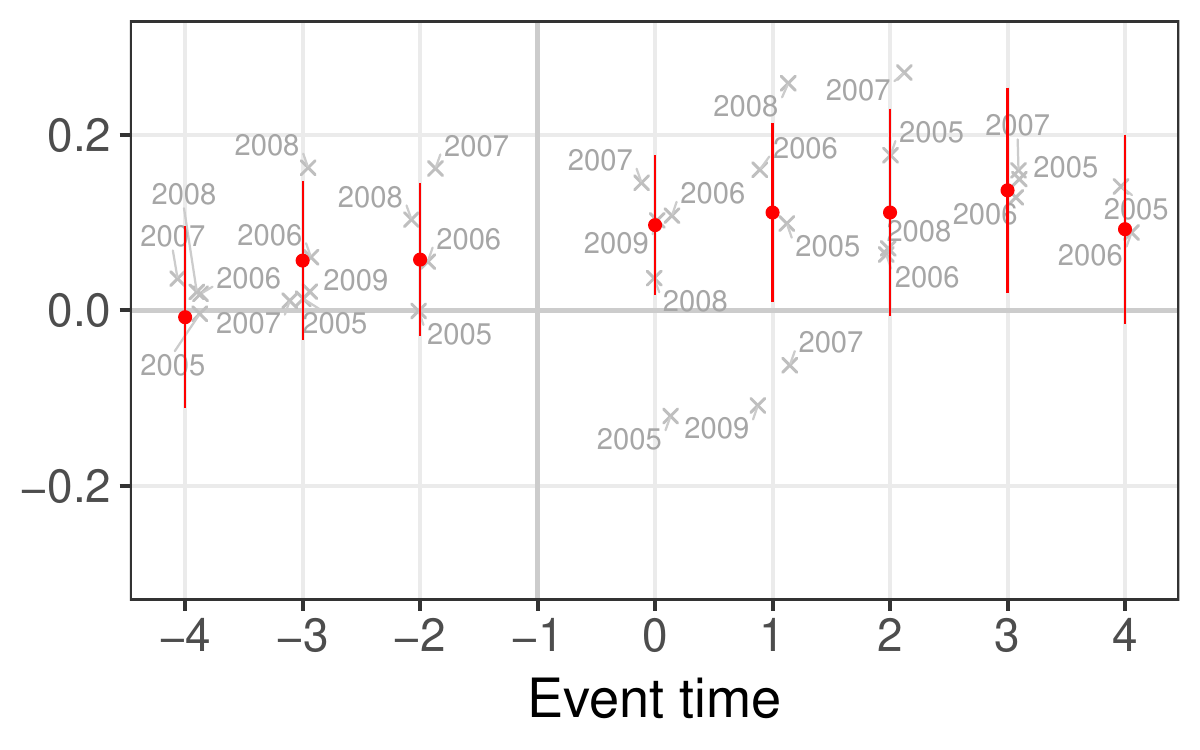}
    \caption{IW}
    \label{fig:event_study_plot_sa}
  \end{subfigure}
  \vspace{0.5em}
  \begin{subfigure}{0.48\textwidth}
   \centering
   \includegraphics[width=\textwidth]{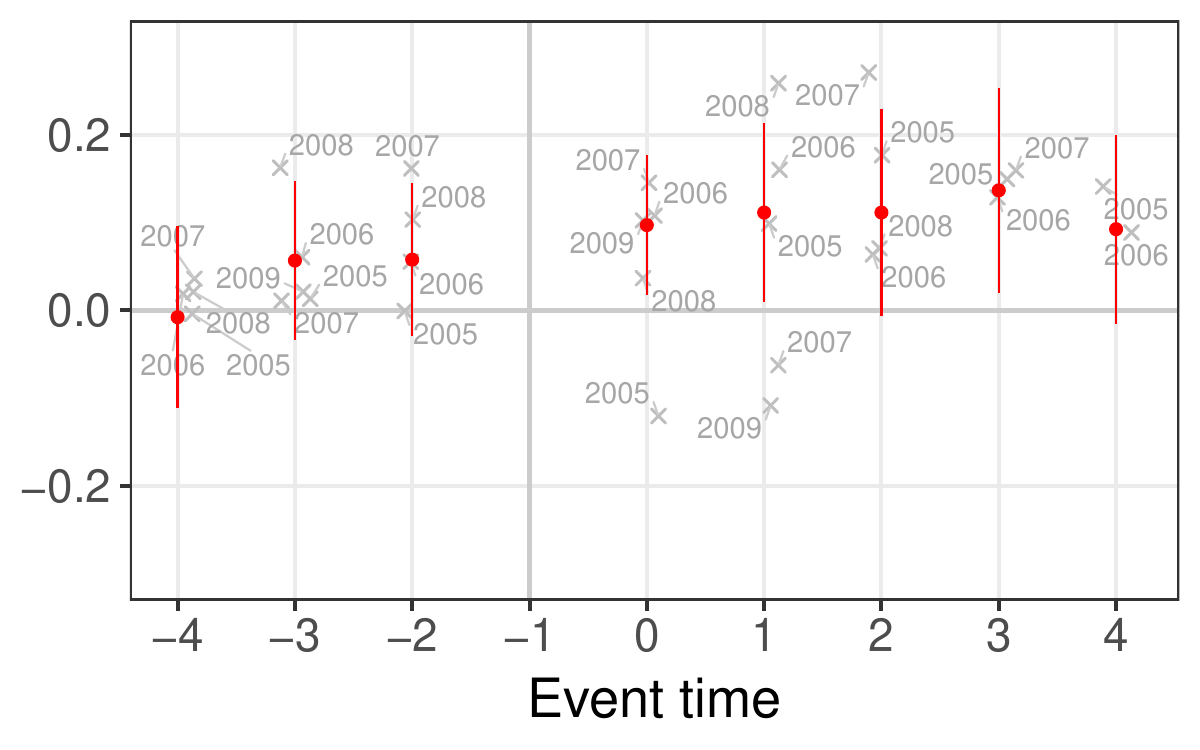}
   \caption{DR, no covariate adjustment}
   \label{fig:event_study_plot_dr_null}
  \end{subfigure}%
  \hfill
  \begin{subfigure}{0.48\textwidth}
    \centering
    \includegraphics[width=\textwidth]{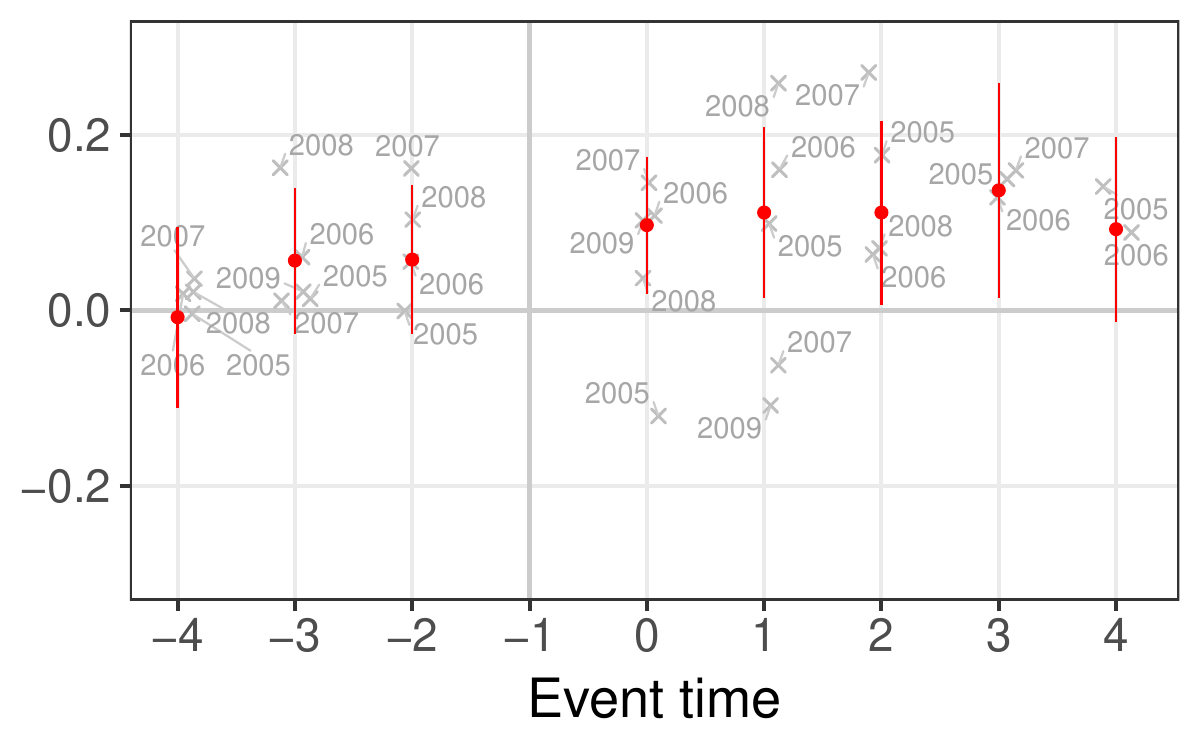}
    \caption{RW, no covariate adjustment}
    \label{fig:event_study_plot_rw_null}
  \end{subfigure}
  \vspace{0.5em}
  \begin{subfigure}{0.48\textwidth}
    \centering
    \includegraphics[width=\textwidth]{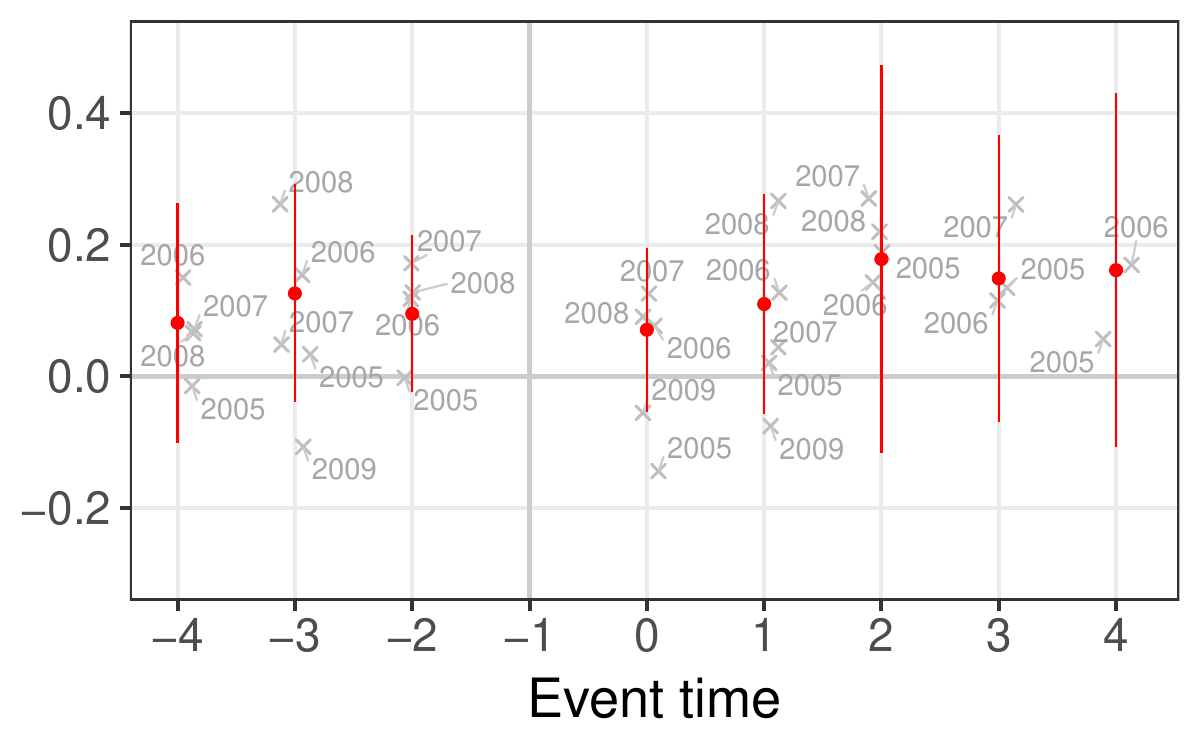}
    \caption{DR, baseline covariate adjustment}
    \label{fig:event_study_plot_dr_cov}
  \end{subfigure}%
  \hfill
  \begin{subfigure}{0.48\textwidth}
    \centering
    \includegraphics[width=\textwidth]{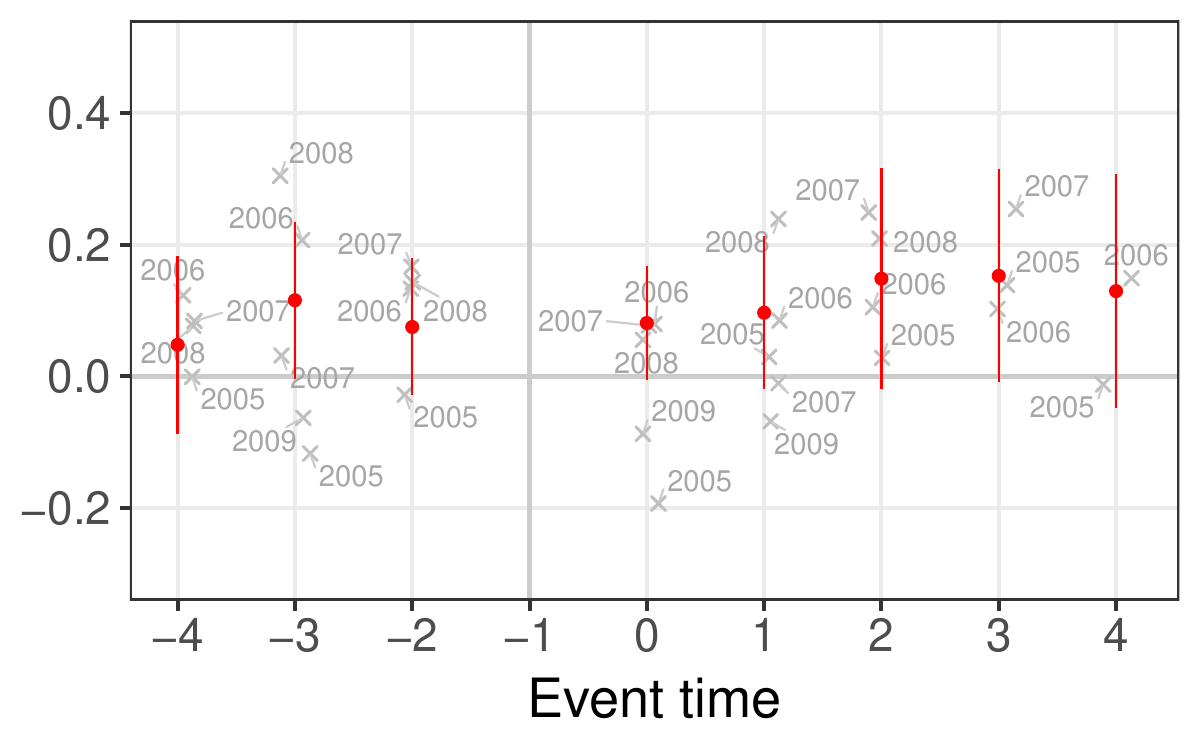}
    \caption{RW, baseline covariate adjustment}
    \label{fig:event_study_plot_rw_cov}
  \end{subfigure}
  \caption{
  Event study estimates of the effects of Castle Doctrine laws on log homicide rates. Points with error bars report aggregate event study parameter $\ate_{\mathrm{agg}}(l)$ and 95\% confidence intervals for IW, DR, and RW; the dynamic TWFE panel reports the conventional event-time coefficient $\hat{\tau}_l$. Crosses report cohort-specific building-block estimates where applicable. DR and RW are shown both with and without adjustment for baseline covariates. Standard errors are obtained from 500 state-level bootstrap replications.
  }
  \label{fig:event_study_plot_estimators}
\end{figure}

Figure~\ref{fig:event_study_plot_estimators} compares the resulting event study estimates. The points and confidence intervals show the aggregate estimates, while the crosses show the cohort-specific estimates where applicable. We report point-wise confidence intervals and leave the development of uniform confidence bands to future work. For DR and RW, we show results with and without adjustment for baseline log counts of police officers and prisoners and baseline log income. Before enactment, none of the four estimators shows a clear trend in homicide rates. After enactment, the estimates without covariate adjustment (panels~\ref{fig:event_study_plot_twfe}--\ref{fig:event_study_plot_rw_null}) are positive and roughly stable across event times, suggesting a persistent increase in homicide rates following the laws. The IW, DR, and RW estimates coincide exactly without covariate adjustment. These results broadly align with \cite{cheng2013does}, who report an approximately $8\%$ increase in homicides using a static TWFE specification and an approximately $10\%$ increase using a dynamic specification. After covariate adjustment, the aggregate post-enactment estimates remain positive and follow a similar pattern, but are generally larger and less precise (panels~\ref{fig:event_study_plot_dr_cov} and \ref{fig:event_study_plot_rw_cov}). The one-step RW estimator also produces smaller standard errors than DR in this application, consistent with imposing balance once across the pooled components rather than separately within each enactment cohort.

\subsection{Information and influence diagnostics}
\label{sec:case_diagnostics}

We now compare how the four estimators use individual observations to estimate the aggregate effect two years after enactment, $\ate^{\Scal_t}_{\mathrm{agg}}(l = 2)$. The diagnostics distinguish differences in the observations included from differences in their weights. Supplement Section~\ref{sec_app_further} reports the corresponding results for the 2006 cohort.

\subsubsection{The implied experiment}
To show which observations contribute to the dynamic TWFE estimate of the effect two years after enactment, we decompose $\widehat{\tau}_2$ as,
\begin{equation}
\label{case_study_twfe_lmw}
\widehat{\tau}_{2}=\sum_{i=1}^{50} \sum_{t=2000}^{2010} w_{i t} \Delta^{3} Y_{i t}^{\mathrm{obs}} \mathbbm{1}\left\{t-G_i=2\right\} - \sum_{i=1}^{50} \sum_{t=2000}^{2010} w_{i t} \Delta^{t-G_i+1} Y_{i t}^{\mathrm{obs}} \mathbbm{1}\left\{t-G_i \notin\left\{-1, 2\right\}\right\}.
\end{equation}
The treated component includes observations from all enactment cohorts observed two years after enactment. For each cohort, the ideal contrast compares the three-year outcome change from the year before enactment to two years after with the corresponding change among never-treated states. We use this ideal contrast as the benchmark in the case study, although eligible not-yet-treated observations, labeled no anticipation, may also be included under the baseline identification assumptions. For the 2006 cohort, this means comparing changes from 2005 to 2008 for states that enacted the law in 2006, $\Delta^3 Y_{i,2008}^{\mathrm{obs}}\mathbbm{1}\{G_i=2006\}$, with those for never-treated states, $\Delta^3 Y_{i,2008}^{\mathrm{obs}}\mathbbm{1}\{G_i=\infty\}$. The TWFE control component also draws on already treated states at other exposure durations. A causal interpretation of the resulting contrast depends on additional assumptions about treatment effect dynamics that the investigator may not intend to make.

\begin{sidewaysfigure}
    \centering
    \includegraphics[width=0.93\textheight]{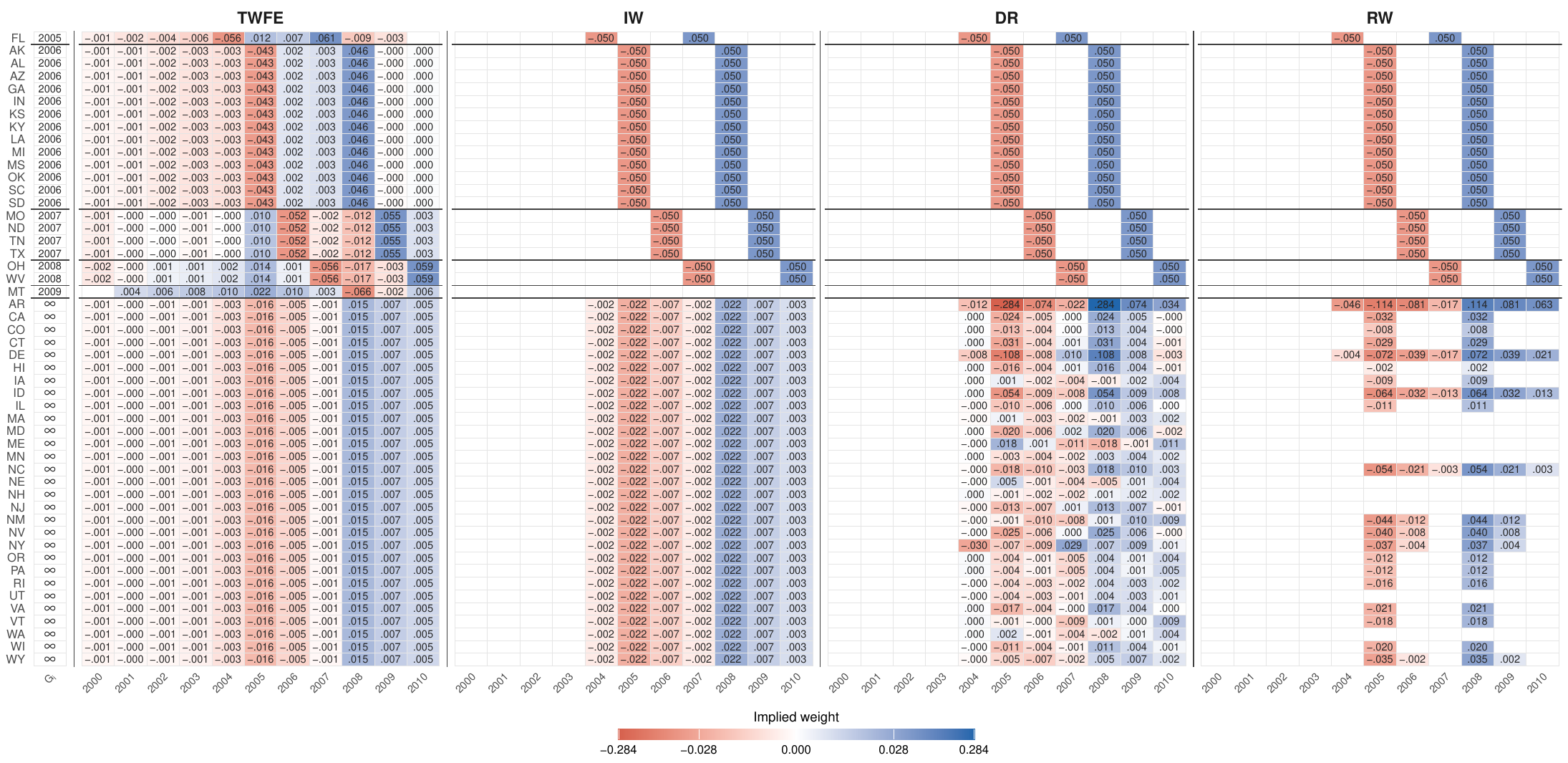}
    \caption{Implied observation-level weights for the aggregate lag-2 estimand $\ate^{\Scal_t}_{\mathrm{agg}}(l=2)$. The column $G_i$ reports each state's enactment year, and horizontal lines separate enactment cohorts. Blank cells have exactly zero weight; displayed values of $0.000$ are nonzero weights smaller than $0.0005$ in absolute value. TWFE draws on nearly the entire panel, whereas IW, DR, and RW use only observations from the cohort-specific ideal experiments. IW weights eligible controls uniformly, while DR and RW reweight never-treated states to balance baseline covariates. IW and RW are sample bounded, whereas DR permits extrapolation through negative control weights.}\label{fig:map_agg}
\end{sidewaysfigure}

The corresponding decompositions for the IW, DR, and RW estimators recover their implied weighted contrasts, shown in Figure~\ref{fig:map_agg}. TWFE assigns nonzero weight to nearly the entire panel, whereas IW, DR, and RW use only observations implied by the ideal experiment per enactment cohort. In the adjusted specifications shown here, DR and RW reweight never-treated states to balance baseline covariates, while IW gives them equal weights within each cohort comparison.

\subsubsection{Information}

We next examine how concentrated the implied weights are across observations using the effective sample size $n_{\text{eff},g}$, information share $p_{\text{info},g}$, and squared-weight share $D_g$ reported in Table~\ref{tab:ess_aggregate}. For TWFE, the target treated group, with 20 observations, accounts for only $8.5\%$ of the information share. In comparison, the delay and dissipation groups, whose use requires the strongest additional assumptions, together account for $12.9\%$, more than one-third larger than the target treated group. Turning to the squared-weight shares, the target treated group contributes $70.8\%$ of $D_g$; its observations receive larger individual weights. The delay and dissipation groups together account for only $2.4\%$ of $D_g$, indicating that their contribution is spread across many observations carrying small individual weights. Together, these measures show whether a group's weights are spread across many observations or concentrated in a few. They allow the investigator to assess reliance on each group alongside the assumptions that justify its inclusion.

The IW, DR, and RW estimators restrict attention to the 136 observations in the ideal contrasts: 20 ideal treated observations at the target event time and 116 never-treated observations. IW assigns uniform control weights within each cohort comparison, yielding an effective control sample size of $n_{\text{eff}} = 61.05$ and a squared-weight share of $D_g = 0.247$. Covariate adjustment makes the control weights more concentrated: DR's combined propensity-score and outcome-regression weights yield $n_{\mathrm{eff}}=10.63$ and $D_g=0.686$, while RW's nonnegative minimum-dispersion weights yield $n_{\mathrm{eff}}=19.40$ and $D_g=0.508$. If the treated and control states differ in baseline covariates and the investigator chooses to adjust for those differences, the smaller effective sample size is the cost of constructing a more comparable control group. The diagnostics make this tradeoff explicit, allowing the investigator to weigh the improvement in covariate balance against the loss in effective sample size.

\begin{table}[!t]
\centering
\caption{\label{tab:ess_aggregate}
Group-wise effective sample sizes (ESS), information shares, and squared-weight shares for estimating the aggregate lag-2 estimand $\ate^{\Scal_t}_{\mathrm{agg}}(l=2)$.}
\renewcommand{\arraystretch}{1.2}
\resizebox{\textwidth}{!}{%
\begin{tabular}{@{}rcccc@{\qquad}rcccc@{}}
\toprule
\multicolumn{5}{c}{\textbf{TWFE}}
& \multicolumn{5}{c}{\textbf{IW}} \\
\cmidrule(r){1-5} \cmidrule(l){6-10}
Observation group ($g$) & $N_g$ & $n_{\text{eff},g}$
& $p_{\text{info},g}$ & $D_g$
& Observation group ($g$) & $N_g$ & $n_{\text{eff},g}$
& $p_{\text{info},g}$ & $D_g$ \\
\midrule
$l = 2$ (ideal treated) & 20 & 19.78 & 0.085 & 0.708
& $l = 2$ (ideal treated) & 20 & 20.00 & 0.247 & 0.753 \\
$l < -1$ (no anticipation) & 115 & 41.97 & 0.180 & 0.027
& $l < -1$ (no anticipation) & 0 & \multicolumn{3}{c}{---} \\
$0 \leq l < 2$ (delay) & 42 & 23.34 & 0.100 & 0.022
& $0 \leq l < 2$ (delay) & 0 & \multicolumn{3}{c}{---} \\
$l > 2$ (dissipate) & 33 & 6.71 & 0.029 & 0.002
& $l > 2$ (dissipate) & 0 & \multicolumn{3}{c}{---} \\
$G_i = \infty$ (never treated) & 290 & 141.25 & 0.606 & 0.241
& $G_i = \infty$ (never treated) & 116 & 61.05 & 0.753 & 0.247 \\
Total & 500 & 233.04 & 1.000 & 1.000
& Total & 136 & 81.05 & 1.000 & 1.000 \\
\midrule
\multicolumn{5}{c}{\textbf{DR}}
& \multicolumn{5}{c}{\textbf{RW}} \\
\cmidrule(r){1-5} \cmidrule(l){6-10}
Observation group ($g$) & $N_g$ & $n_{\text{eff},g}$
& $p_{\text{info},g}$ & $D_g$
& Observation group ($g$) & $N_g$ & $n_{\text{eff},g}$
& $p_{\text{info},g}$ & $D_g$ \\
\midrule
$l = 2$ (ideal treated) & 20 & 20.00 & 0.653 & 0.314
& $l = 2$ (ideal treated) & 20 & 20.00 & 0.508 & 0.492 \\
$l < -1$ (no anticipation) & 0 & \multicolumn{3}{c}{---}
& $l < -1$ (no anticipation) & 0 & \multicolumn{3}{c}{---} \\
$0 \leq l < 2$ (delay) & 0 & \multicolumn{3}{c}{---}
& $0 \leq l < 2$ (delay) & 0 & \multicolumn{3}{c}{---} \\
$l > 2$ (dissipate) & 0 & \multicolumn{3}{c}{---}
& $l > 2$ (dissipate) & 0 & \multicolumn{3}{c}{---} \\
$G_i = \infty$ (never treated) & 116 & 10.63 & 0.347 & 0.686
& $G_i = \infty$ (never treated) & 116 & 19.40 & 0.492 & 0.508 \\
Total & 136 & 30.63 & 1.000 & 1.000
& Total & 136 & 39.40 & 1.000 & 1.000 \\
\bottomrule
\end{tabular}%
}
\end{table}

\subsubsection{Influence}
\begin{figure}[h!tbp]
    \centering
    \includegraphics[scale = 0.5]{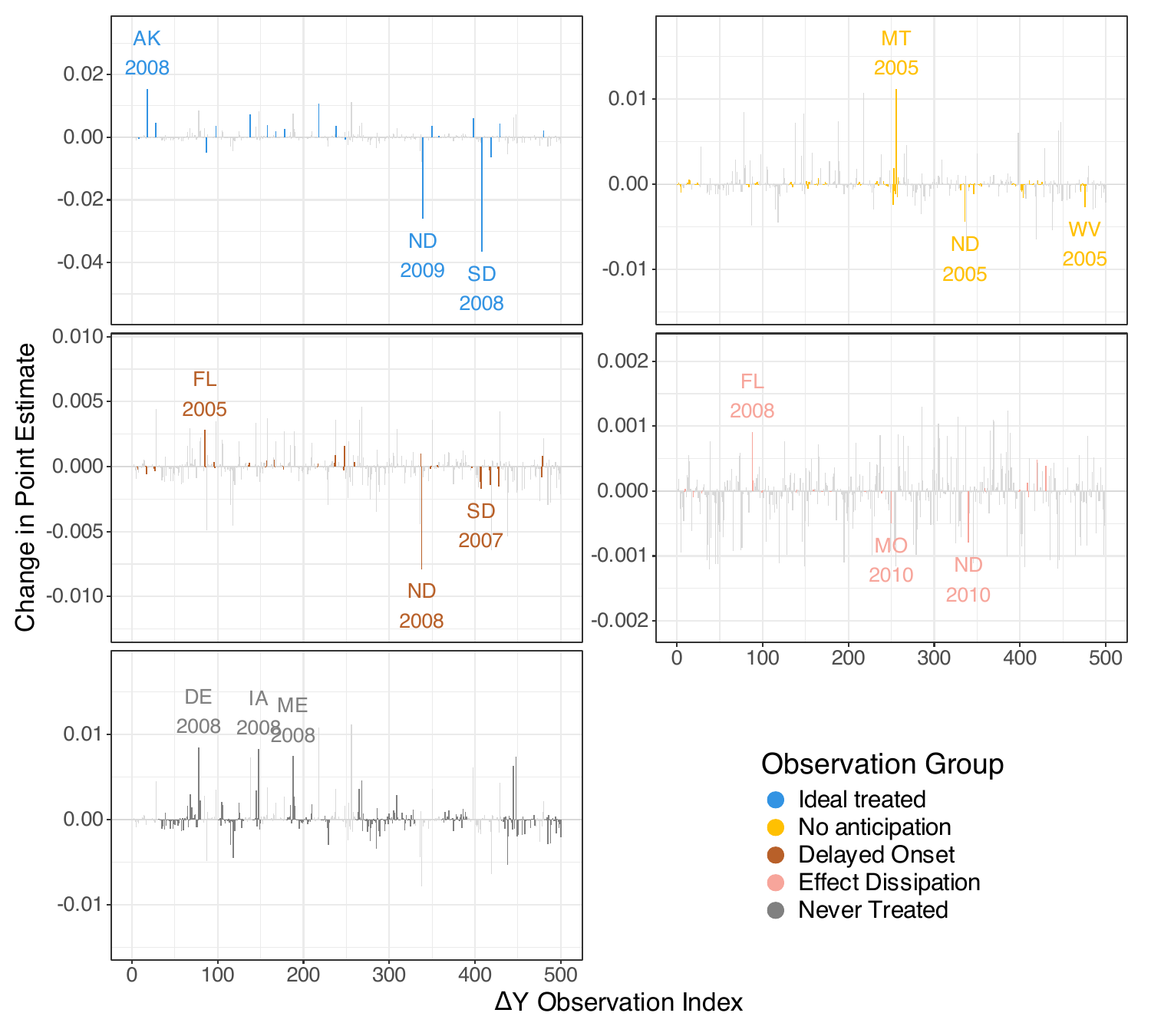}
    \caption{Leave-one-observation-out influence for the TWFE estimator estimating $\ate^{\Scal_t}_{\mathrm{agg}}(l=2)$. Each segment represents the change in $\widehat{\tau}_2$ from removing one observation. Observations are colored by group membership, and the three observations with the largest absolute changes per group are labeled.}
    \label{fig:twfe_influence_agg}
\end{figure}

The weight summaries show how information is distributed across groups. We now examine sensitivity to individual observations by deleting each outcome difference and recomputing the estimate. Figure~\ref{fig:twfe_influence_agg} presents the TWFE influence diagnostics. The most influential observations belong to the target treated group, with individual removals shifting $\widehat{\tau}_2$ by up to $-0.037$ and $-0.026$, or $31.8\%$ and $22.7\%$ of the point estimate at $0.115$. The most influential observation in the delay group shifts the estimate by $-0.008$, or $6.9\%$. Because these observations enter the TWFE contrast under the delay assumption, their influence shows the sensitivity of the estimator to an assumption that the investigator may not intend to invoke. 

The leave-one-observation-out diagnostic measures sensitivity to individual observations rather than the collective contribution of an entire group. Because the delayed-onset and effect-dissipation groups contain 42 and 33 observations, respectively, their influence may be spread across many observations, with no single removal moving $\widehat{\tau}_2$ very far. Supplement Table~\ref{tab:supp_twfe_rw} provides a complementary illustration for the 2006-cohort target by changing the information set and its balancing constraints. The ideal contrast yields $0.064$. Adding observations under invariance to time shifts gives $0.119$; adding the not-yet-treated observations under no anticipation gives $0.106$; adding the delayed-onset group gives $0.115$; and the final expansion reproduces the TWFE estimate of $0.115$. These changes reflect both the observations included and the constraints used to weight them. A separate subset-deletion diagnostic is beyond the scope of this paper. For IW, DR, and RW, which only use the target treated and never-treated groups, the influence diagnostics results are reported in Supplement Section~\ref{sec_app_case_influence}.

The case study illustrates why both the information set and the weights matter. TWFE draws on observations outside the ideal cohort-specific comparisons, and the diagnostics quantify their role in the estimate. In the specifications used here, IW, DR, and RW use treated observations at the target event time and never-treated controls. Covariate adjustment under DR and RW then concentrates the control weights and reduces the effective sample size. These diagnostics reveal the anatomy of an event study estimate by showing which observations enter the contrast, how they are weighted, and how much they influence the result.

%% file: sec7.tex
\section{Summary and remarks}
\label{sec:conclusion}

We develop a framework grounded in experimental design principles for observational studies that encompasses several widely used event study estimators. It characterizes each estimator by the observations it uses and the weights it assigns to them, distinguishing the assumptions that identify the target effect from those that justify borrowing additional information. This common representation makes explicit how estimators draw on information across units and time periods.

The exact finite-sample weighting representations provide a common basis for comparing estimators and assessing whether their implied comparisons match the intended design. For TWFE, this includes examining the role of treated observations in the control component. The resulting diagnostics quantify reliance on different observation groups, assess covariate balance and study representativeness, and reveal sensitivity to individual observations. Together, they allow investigators to assess the empirical consequences of information borrowing and covariate adjustment.

The framework also supports constructing estimators for a stated design. Our proposed weighting estimator minimizes weight dispersion subject to approximate covariate balance over a specified set of eligible observations. It gives investigators control over which observations enter the contrast and which covariates are balanced, uses nonnegative weights to produce sample-bounded estimates, and accommodates generalization to a target population.

Future work could extend the weighting decomposition and diagnostics to other panel data methods, including marginal structural models \citep{robins2000marginal}, interactive fixed effects \citep{bai2009panel}, matrix completion \citep{athey2021matrix}, and synthetic control methods \citep{abadie2010synthetic, arkhangelsky2021synthetic, ben2022synthetic}. A second direction is to broaden the identification foundations of event study analyses through partial identification \citep{ye2024negative} and the integration of multiple identification assumptions \citep{ding2019bracketing}.


%% file: sec_supp.tex

\ifsuppstandalone\else
  \clearpage
  \phantomsection
  \addcontentsline{toc}{section}{Supplementary Materials}
\fi

\setcounter{section}{0}    \renewcommand{\thesection}{S\arabic{section}}
\setcounter{table}{0}      \renewcommand{\thetable}{S\arabic{table}}
\setcounter{figure}{0}     \renewcommand{\thefigure}{S\arabic{figure}}
\setcounter{equation}{0}   \renewcommand{\theequation}{S\arabic{equation}}
\setcounter{theorem}{0}    \renewcommand{\thetheorem}{S\arabic{theorem}}
\setcounter{assumption}{0} \renewcommand{\theassumption}{S\arabic{assumption}}

\ifsuppstandalone\else
  \renewcommand{\theHsection}{S\arabic{section}}
  \renewcommand{\theHtable}{S\arabic{table}}
  \renewcommand{\theHfigure}{S\arabic{figure}}
  \renewcommand{\theHequation}{S\arabic{equation}}
  \renewcommand{\theHtheorem}{S\arabic{theorem}}
  \renewcommand{\theHassumption}{S\arabic{assumption}}
\fi

\setcounter{page}{1}

\ifsuppstandalone\else
  \section*{Supplementary Materials for An Anatomy of Event Studies:
  Hypothetical Experiments, Exact Decomposition, and Weighting Diagnostics}
\fi

\input{supp0_overview}
\input{supp1_identification}

\input{supp2_weighting}
\input{supp3_decomposition}
\input{supp4_diagnostics}
\input{supp5_case_study}
\input{supp6_simulation}

%% file: supp0_overview.tex

\section{Overview and notation}
\label{sec:app_overview}

This supplement follows the order of the main text. Section~\ref{sec:app_identification} proves the identification results of Section~\ref{sec_causal_identification} and shows, one assumption at a time, which observations each information borrowing assumption adds to the contrast. Section~\ref{sec:app_rw} develops the robust weighting estimator: the regularity conditions
and proofs behind Theorems~\ref{thm_consistency} and~\ref{thm_normality}, the extension to a
target population other than the study population, the extension to a general reference
regime, and implementation details. Section~\ref{sec:app_decomp} derives the exact weighting
representations of the dynamic TWFE, interaction-weighted, and doubly robust estimators.
Section~\ref{sec:app_diagnostics} defines the diagnostic quantities of
Section~\ref{sec:diagnostic_metrics} and records their properties.
Section~\ref{sec_app_further} reports additional results for the Castle Doctrine case study.

Notation follows the main text throughout. Table~\ref{tab:app_notation} collects the symbols
that recur most often.

\begin{table}[h!tbp]
\caption{\label{tab:app_notation} Notation used in the main text and this supplement.}
\centering
\small
\begin{tabular}{ll}
\toprule
Symbol & Meaning \\
\midrule
$i$, $t$, $(i,t)$ & unit, time period, and observation (a unit--period pair) \\
$n$, $T$, $N$ & number of units, number of periods, number of observed cells \\
$\Tcal$, $\Tcal^{+}$ & $\{1,\dots,T\}$, and $\Tcal \cup \{\infty\}$ \\
$G_i \in \Tcal^{+}$ & treatment initiation time; $G_i = \infty$ if never treated \\
$Z_{it}$ & treatment indicator, $Z_{it} = \mathbbm{1}\{G_i \leq t\}$ \\
$Y_{it}(g)$, $Y^{\obs}_{it}$ & potential outcome under initiation at $g$; observed outcome \\
$\Delta^{l} Y_{it_y}(t)$ & $Y_{it_y}(t) - Y_{i,t_y-l}(t)$, the $l$-period within-unit change in potential outcomes \\
$Y^{*}_{it_y}(t)$, $Y^{*\obs}_{it_y}$ & $Y_{it_y}(t)$ under unconfoundedness; $\Delta^{l} Y_{it_y}(t)$ under parallel trends \\
$\bX_i$, $\bU_i$, $\Fcal_i$ & baseline covariates, unit-level unobservables, adjustment set \\
$\Scal$, $\Pcal$ & study and target populations, with measures $\S$ and $\P$ \\
$\Scal_t$, $\Pcal_t$ & subpopulations initiating treatment at time $t$ \\
$\ate^{\Pcal}_{t_y}(t_1,t_0)$ & effect at $t_y$ of initiation at $t_1$ versus $t_0$, in population $\Pcal$ \\
$t_{\mathbf{p}}$, $p_t$ & stochastic reference regime, and its initiation probabilities \\
$l^{*}$ & target event time, $l^{*} = t_y - t_1$ \\
$\kappa$, $\phi$, $\xi$ & anticipation window, delayed-onset window, dissipation horizon \\
$m_{t,t_y}(\bm{x})$ & $\E\{Y^{*}_{it_y}(t) \mid \bX_i = \bm{x}\}$, the outcome mean function \\
$e_t(\bm{x})$ & $\Pr_{\mathbb{S}}(G_i = t \mid \bX_i = \bm{x})$, the treatment initiation probability in the study population \\
$w_{it}$ & weight attached to observation $(i,t)$ \\
$\Ccal_{r,s}$, $\Ccal$, $\Ccal'$ & observation sets used by the expanded contrast \\
$\Mcal$, $B_k(\cdot)$, $K$, $\bm{\delta}$ & function class, basis functions, their number, and balance tolerances \\
$n_{\text{eff},g}$, $p_{\text{info},g}$, $D_g$ & group-wise effective sample size, information share, squared-weight share \\
$\bm A$, $\bm B$, $\bm X$, $\bm F$ & unit, time, and covariate design blocks, with $\bm F = [\,\bm A\ \bm B\ \bm X\,]$ \\
$\bm D$, $\tilde{\bm D}$, $M$ & event-time design, its residualization $M \bm D$, and $M = I - P_{\bm F}$ \\
\bottomrule
\end{tabular}
\end{table}

%% file: supp1_identification.tex

\section{Identification and observation groups}
\label{sec:app_identification}

This section proves the identification results of Section~\ref{sec_causal_identification} and then works through the information borrowing assumptions of Section~\ref{sec2:borrow_assumption} one at a time, showing which observations each borrowing assumption brings into the contrast. Section~\ref{sec:app_identification_proof} proves the identification.  Sections~\ref{sec:app_borrowing_proof} and~\ref{sec:app_borrowing_proof_pt} then show which observations each borrowing assumption justifies to include in the contrast building under the unconfoundedness framework and then under parallel trends framework. Sections~\ref{sec:app_twfe_decomposition_proof_Y} and ~\ref{sec:app_twfe_decomposition_proof_Ydiff} apply these groups to the dynamic TWFE estimator.

\subsection{Identification of the target estimands}
\label{sec:app_identification_proof}

We prove identification of $\ate^\Pcal_{t_y}(t_1, t_0)$ for any fixed $t_0 \in \Tcal^+$ and of $\ate^\Pcal_{t_y}(t_1, t_{\mathbf{p}})$, for a general target population $\Pcal$. Recall that
\begin{equation*}
    \begin{aligned}
        \ate^\Pcal_{t_y}(t_1, t_0) &= \E_{\P}\{ Y_{it_y}(t_1) \} - \E_{\P}\{ Y_{it_y}(t_0) \}, \\
        \ate^\Pcal_{t_y}(t_1, t_\mathbf{p}) &= \E_{\P}\{ Y_{it_y}(t_1) \} - \sum_{t \in \Tcal^+} p_t \E_{\P}\{ Y_{it_y}(t) \}.
    \end{aligned}
\end{equation*}
Both reduce to identifying $\E_{\P}\{ Y_{it_y}(t) \}$ for an arbitrary $t \in \Tcal^+$. We consider the two settings of Section~\ref{sec_causal_identification}, $\Fcal_i = \bX_i$ and $\Fcal_i = \bU_i$. First, we state the two conditions that establish the relationship between the study and target populations that allows for the transportation of effects estimated in a study population to a target population. 

\begin{assumption}[Common support of covariates across populations]
    \label{assumption_common_support}
    $\mathrm{supp}(\S_{\bX}) = \mathrm{supp}(\P_{\bX})$.
\end{assumption}

\begin{assumption}[Conditional exchangeability across populations]
    \label{assumption_exchangable_populations}
    $\S_{Y_{t_y}(t) \mid \bX } = \P_{Y_{t_y}(t) \mid \bX}$.
\end{assumption}
Assumption~\ref{assumption_common_support} states that the covariate distributions in the study and target populations share the common support. When Assumption~\ref{assumption_exchangable_populations} holds, all differences in the distribution of potential outcomes between the study and the target populations are removed by adjusting for the differences in the observed covariate distributions between the two populations.

\paragraph{Unconfoundedness framework.}
Under $\Fcal_i = \bX_i$, Assumptions~\ref{assumption_consistency}--\ref{assumption_random_treatment_initiation} together with Assumptions~\ref{assumption_common_support} and~\ref{assumption_exchangable_populations} give
\begin{equation}
\label{equation_identification_proof}
   \begin{aligned}[b]
       \E_{\P}\{ Y_{it_y}(t) \}
       &= \E_{\P_{\bX}} [ \E_{\P_{Y_{t_y}(t) \mid \bX}} \{ Y_{it_y}(t) \mid \bX_i \} ] \\
       &= \E_{\P_{\bX}} [ \E_{\S_{Y_{t_y}(t) \mid \bX}} \{ Y_{it_y}(t) \mid \bX_i \} ] \\
       &= \E_{\P_{\bX}} [ \E_{\S_{Y_{t_y}(t) \mid \bX}} \{ Y_{it_y}(t) \mid \bX_i, G_i = t \} ] \\
       &= \E_{\P_{\bX}} [ \E_{\S_{Y_{t_y}(t) \mid \bX}} \{ Y_{it_y}^{\obs} \mid \bX_i, G_i = t \} ],
   \end{aligned}
\end{equation}
where the second equality uses Assumption~\ref{assumption_exchangable_populations}, the
third uses Assumption~\ref{assumption_random_treatment_initiation}, and the last uses
Assumption~\ref{assumption_consistency}. Since this holds for every $t \in \Tcal^+$, both
$\ate^\Pcal_{t_y}(t_1, t_0)$ and $\ate^\Pcal_{t_y}(t_1, t_{\mathbf{p}})$ are identified.

\paragraph{Parallel trends framework.}
Under the parallel trends framework, we work with within-unit differences instead. Let $l^*=t_y-t_1$ denote the target event time and suppose that $t_1-\kappa-1\in\Tcal$. Define
\begin{equation}
    \Delta^{l^*+\kappa+1}Y_{it_y}(t_1)
    =
    Y_{it_y}(t_1)-Y_{i,t_1-\kappa-1}(t_1),
\end{equation}
\begin{equation}
    \Delta^{l^*+\kappa+1}Y_{it_y}(\infty)
    =
    Y_{it_y}(\infty)-Y_{i,t_1-\kappa-1}(\infty),
\end{equation}
and let
\begin{equation}
    \Delta^{l^*+\kappa+1}Y_{it_y}^{\obs}
    =
    Y_{it_y}^{\obs}-Y_{i,t_1-\kappa-1}^{\obs}
\end{equation}
denote the corresponding observed outcome difference. Under Assumptions~\ref{assumption_consistency}--\ref{assumption_positive_treatment_initiation} and \ref{assumption_limited_anticipation}--\ref{assumption_two_way_parallel}, together with Assumptions~\ref{assumption_common_support} and~\ref{assumption_exchangable_populations}, we have
\allowdisplaybreaks
\begin{align}
\ate^\Pcal_{t_y}(t_1,\infty) 
&= \E_{\P} \left\{ Y_{it_y}(t_1)-Y_{it_y}(\infty) \right\} \nonumber \\
&= \E_{\P} \left\{ \Delta^{l^*+\kappa+1}Y_{it_y}(t_1) - \Delta^{l^*+\kappa+1}Y_{it_y}(\infty) \right\} \nonumber \\
&= \E_{\P_{\bX}} \left[ \E_{\P} \left\{ \Delta^{l^*+\kappa+1}Y_{it_y}(t_1) \mid \bX_i \right\} - \E_{\P} \left\{ \Delta^{l^*+\kappa+1}Y_{it_y}(\infty) \mid \bX_i \right\} \right] \nonumber \\
&= \E_{\P_{\bX}} \left[ \E_{\S} \left\{ \Delta^{l^*+\kappa+1}Y_{it_y}(t_1) \mid \bX_i \right\} - \E_{\S} \left\{ \Delta^{l^*+\kappa+1}Y_{it_y}(\infty) \mid \bX_i \right\} \right] \nonumber \\
&= \E_{\P_{\bX}} \left[ \E_{\S} \left\{ \Delta^{l^*+\kappa+1}Y_{it_y}(t_1) \mid \bX_i,G_i=t_1 \right\} - \E_{\S} \left\{ \Delta^{l^*+\kappa+1}Y_{it_y}(\infty) \mid \bX_i,G_i=\infty \right\} \right] \nonumber \\
&= \E_{\P_{\bX}} \left[ \E_{\S} \left\{ \Delta^{l^*+\kappa+1}Y_{it_y}^{\obs} \mid \bX_i,G_i=t_1 \right\} - \E_{\S} \left\{ \Delta^{l^*+\kappa+1}Y_{it_y}^{\obs} \mid \bX_i,G_i=\infty \right\} \right]. \label{equation_identification_proof_PT_4}
\end{align}
The first equality is the definition of the target estimand. The second uses Assumption~\ref{assumption_limited_anticipation}, together with Assumption~\ref{assumption_exchangable_populations}, to equate the expected potential outcomes under initiation at $t_1$ and never initiation at the reference period $t_1-\kappa-1$. The third equality follows from the law of iterated expectations. The fourth uses Assumption~\ref{assumption_exchangable_populations}, the fifth uses Assumption~\ref{assumption_two_way_parallel}, and the last uses Assumption~\ref{assumption_consistency}. Assumption \ref{assumption_common_support} ensures that the relevant conditional expectations are defined over the covariate support of the target population, Assumption~\ref{assumption_positive_treatment_initiation} ensures that the two treatment groups have positive probability, and Assumption~\ref{assumption_random_sample} connects the study population expectations to the observed study sample.

Finally, for pre-specified weights $p_t$ with $t \in \Tcal^+$,
\begin{equation}
\label{equation_identification_proof_PT_5}
   \begin{aligned}[b]
       \ate^\Pcal_{t_y}(t_1,t_{\mathbf p})
       &= \left[ \E_{\P}\{ Y_{it_y}(t_1) \} - \E_{\P}\{ Y_{it_y}(\infty) \} \right]
       - \sum_{t \in \Tcal^+} p_t \left[ \E_{\P}\{ Y_{it_y}(t) \} - \E_{\P}\{ Y_{it_y}(\infty) \} \right] \\
       &= \ate^\Pcal_{t_y}(t_1,\infty) - \sum_{t \in \Tcal^+} p_t\, \ate^\Pcal_{t_y}(t,\infty).
   \end{aligned}
\end{equation}
Each term on the right is identified by Equation~\eqref{equation_identification_proof_PT_4}, so $\ate^\Pcal_{t_y}(t_1,t_{\mathbf p})$ is also identified.

\subsection{Observation groups under the unconfoundedness framework}
\label{sec:app_borrowing_proof}

We now focus on the estimand at $\ate^\Pcal_{t_y}(t_1, \infty)$ and show how each borrowing assumption enlarges the set of observations that can enter a valid contrast. Throughout this subsection we take $\Pcal = \Scal$ and $\Fcal_i = \emptyset$, while the conditional versions follow similarly by applying  a further layer of iterated expectation over $\bX_i$.

Under Assumptions~\ref{assumption_consistency}--\ref{assumption_random_treatment_initiation},
\begin{equation}
     \ate^\Scal_{t_y}(t_1, \infty)  = \E_{\S} ( Y_{it_y} \mid G_i = t_1 ) - \E_{\S} ( Y_{it_y}\mid G_i = \infty ),
\end{equation}
and, following Section~\ref{sec_hypothetical_experiment}, an unbiased estimator is
\begin{equation}
    \widehat{\ate}^\Scal_{t_y}(t_1, \infty) = \frac{1}{n_{t_1}} \sum_{i = 1}^n \mathbbm{1}(G_i = t_1) Y^{\obs}_{it_y} - \frac{1}{n_{\infty}} \sum_{i = 1}^n \mathbbm{1}(G_i = \infty) Y^{\obs}_{it_y}.
\end{equation}
This uses only outcomes measured at $t_y$ in the $t_1$ and $\infty$ arms. These observations form the Ideal Experiment group, and the estimator is the ideal contrast.

\paragraph{Invariance to time shifts.}
Under Assumption~\ref{assumption_invariance_time_shifts}, for each shift $h$ such that $t_1+h$ and $t_y+h$ lie in $\Tcal$
\begin{equation}
    \label{supp_eq:invariance_eq_1}
    \begin{aligned}[b]
    \ate^\Scal_{t_y}(t_1, \infty)
    &= \E_{\S}\{Y_{it_y}(t_1) - Y_{it_y}(\infty) \} \\
    &= \E_{\S}\{Y_{i,t_y+h}(t_1+h) - Y_{i,t_y+h}(\infty)\} \\
    &= \ate^\Scal_{t_y+h}(t_1+h, \infty).
    \end{aligned}
\end{equation}
Any unbiased estimator of the shifted estimand is also unbiased for the original estimand, and in particular
\begin{equation}
    \label{supp_eq:invariance_eq_2_estimator}
    \widehat{\ate}^\Scal_{t_y+h}(t_1+h, \infty) = \frac{1}{n_{t_1+h}} \sum_{i = 1}^n \mathbbm{1}(G_i = t_1+h) Y^{\obs}_{i,t_y+h} - \frac{1}{n_{\infty}} \sum_{i = 1}^n \mathbbm{1}(G_i = \infty) Y^{\obs}_{i,t_y+h}.
\end{equation}
The observations newly included by Equation~\eqref{supp_eq:invariance_eq_2_estimator} form the Time Invariance group; they sit at the target event time $l^* = t_y - t_1$ but belong to treatment cohorts other than $t_1$.

\paragraph{Limited anticipation.}
Assumption~\ref{assumption_limited_anticipation} says
$\ate^\Scal_{t}(t + \kappa^\prime, \infty) = 0$ for any $\kappa^\prime > \kappa$. Taking $\kappa = 0$ for concreteness, for all $\kappa^\prime > 0$,
\begin{equation}
    \E_{\S}\left\{Y_{i t}\left(t+\kappa^{\prime}\right)\right\} 
    = \E_{\S}\left\{Y_{i t}\left(\infty\right)\right\} 
    = \E_{\S}\left(Y_{it}^{\obs}\mid G_i=t+\kappa^\prime\right)
    = \E_{\S}\left(Y_{it}^{\obs}\mid G_i=\infty\right).
\end{equation}
Hence $n_{t_y+\kappa^{\prime}}^{-1} \sum_{i} \mathbbm{1}(G_i = t_y+\kappa^{\prime}) Y^{\obs}_{it_y}$ is unbiased for $\E_{\S}\{Y_{i t_y}(\infty)\}$. Outcomes of not-yet-treated units may serve as controls. Combined with
Assumption~\ref{assumption_invariance_time_shifts}, analogous
not-yet-treated observations may be used in the corresponding shifted contrasts at other calendar periods. These observations form the Limited Anticipation group.

\paragraph{Delayed treatment onset.}
Assumption~\ref{assumption_delay} plays the same role for already-treated units. Setting $\phi = t_y - t_1 - 1$, for any $0 \leq \phi^\prime \leq \phi$,
\begin{equation}
    \E_{\S}\{Y_{it}(t - \phi^\prime) \} 
    = \E_{\S}\{Y_{it}(\infty) \}
    = \E_{\S} \left( Y^\obs_{i t} \mid G_i = t-\phi^\prime \right) 
    = \E_{\S} \left( Y^\obs_{i t}\mid G_i = \infty \right),
\end{equation}
so $n_{t_y-\phi^{\prime}}^{-1} \sum_{i} \mathbbm{1}(G_i = t_y-\phi^\prime) Y^{\obs}_{it_y}$ is also unbiased for $\E_{\S}\{Y_{it_y}(\infty)\}$. Therefore, units treated too recently for the effect to have appeared may enter the control component. Combined with
Assumption~\ref{assumption_invariance_time_shifts}, analogous
recently treated observations may enter shifted contrasts at other
calendar periods. These observations form the Delayed Onset group.

\paragraph{Treatment effect dissipation.}
Assumption~\ref{assumption_washout} makes the symmetric statement at longer time horizons. Setting $\xi = t_y - t_1 + 1$, for any $\xi^\prime \geq \xi$,
\begin{equation}
    \E_{\S}\{Y_{it}(t - \xi^\prime) \} 
    = \E_{\S}\{Y_{it}(\infty) \} 
    = \E_{\S} \left( Y^\obs_{i t} \mid G_i = t-\xi^\prime \right) 
    = \E_{\S} \left( Y^\obs_{i t}\mid G_i = \infty \right),
\end{equation}
so $n_{t_y-\xi^{\prime}}^{-1} \sum_{i} \mathbbm{1}(G_i = t_y-\xi^\prime) Y^{\obs}_{it_y}$ is also unbiased for $\E_{\S}\{Y_{it_y}(\infty)\}$. Therefore, units whose treatment began long enough ago that the effect has dissipated may also enter the control component. Combined with
Assumption~\ref{assumption_invariance_time_shifts}, analogous
long-treated observations may enter shifted contrasts at other
calendar periods. These observations form the Effect Dissipation group.

\subsection{Observation groups under the parallel trends framework}
\label{sec:app_borrowing_proof_pt}

We repeat the exercise under the parallel trends framework, where the objects being compared are within-unit outcome changes rather than levels. Unlike in Section~\ref{sec:app_borrowing_proof}, limited/no anticipation is part of the baseline identification framework rather than an additional information-borrowing assumption. We therefore first characterize the observations available under the baseline parallel-trends conditions and then consider the additional information-borrowing Assumptions \ref{assumption_invariance_time_shifts}--\ref{assumption_washout}. We again take $\Pcal = \Scal$ and use unconditional versions of the assumptions for notational simplicity, while the conditional versions follow by iterated expectation over $\bX_i$. 

Let $l^* = t_y - t_1$, and under Assumptions~\ref{assumption_consistency}--\ref{assumption_positive_treatment_initiation} and \ref{assumption_limited_anticipation}--\ref{assumption_two_way_parallel}, the identification result in Section \ref{sec:app_identification_proof} gives
\begin{equation} 
\label{eq_supp:PT_ideal_contrast} 
\begin{aligned}[b] 
\ate^\Scal_{t_y}(t_1,\infty) 
&= \E_{\S}\left\{ \Delta^{l^*+\kappa+1}Y_{it_y}^{\obs} \mid G_i=t_1 \right\} - \E_{\S}\left\{ \Delta^{l^*+\kappa+1}Y_{it_y}^{\obs} \mid G_i=\infty \right\}, 
\end{aligned} 
\end{equation} 
where $\Delta^{l^*+\kappa+1}Y_{it_y}^{\obs} = Y_{it_y}^{\obs}-Y_{i,t_1-\kappa-1}^{\obs}$. An unbiased estimator is
\begin{align}
    \hat{\ate}^{\Scal}_{t_y}(t_1, \infty)
    & = \frac{1}{n_{t_1}} \sum_{i = 1}^{n} \mathbbm{1}(G_i = t_1) \Delta^{l^*+\kappa+1} Y^{\obs}_{it_y} - \frac{1}{n_{\infty}} \sum_{i = 1}^{n} \mathbbm{1}(G_i = \infty) \Delta^{l^*+\kappa+1} Y^{\obs}_{it_y} \notag \\
    & = \frac{1}{n_{t_1}} \sum_{i = 1}^{n} \mathbbm{1}(G_i = t_1) Y^{\obs}_{it_y} - \frac{1}{n_{\infty}} \sum_{i = 1}^{n} \mathbbm{1}(G_i = \infty) Y^{\obs}_{it_y} \nonumber \\
    & \quad - \left\{ \frac{1}{n_{t_1}} \sum_{i = 1}^{n} \mathbbm{1}(G_i = t_1) Y^{\obs}_{i t_1 - \kappa - 1} - \frac{1}{n_{\infty}} \sum_{i = 1}^{n} \mathbbm{1}(G_i = \infty) Y^{\obs}_{i t_1-\kappa - 1} \right\}. \label{equation_idealexperiment_PT_ideal}    
    \end{align}
This defines a treatment component and a control component in terms of the outcome differences. In Equation~\eqref{equation_idealexperiment_PT_ideal}, the control component includes only the never-treated observations, but it does not include all the valid control observations. 

Under Assumptions~\ref{assumption_limited_anticipation} and \ref{assumption_two_way_parallel}, units with $G_i>t_y+\kappa$ may also enter the control component. Limited anticipation implies that their outcomes through $t_y$ are unaffected in expectation by their future treatment, while parallel trends makes the corresponding within-unit outcome differences comparable across treatment cohorts. The control component in Equation~\eqref{equation_idealexperiment_PT_ideal} can be expanded by including both never-treated units and not-yet-treated units as controls:
    \begin{equation*}
        \begin{aligned}
        \widehat{\ate}^{\Scal}_{t_y}(t_1,\infty)
        &=
        \frac{1}{n_{t_1}}\sum_{i=1}^n
        \mathbbm{1}\{G_i=t_1\}Y^{\obs}_{it_y}
        -
        \frac{1}{n_{>t_y+\kappa}}\sum_{i=1}^n
        \mathbbm{1}\{G_i>t_y+\kappa\}Y^{\obs}_{it_y}  \\
        &\quad -
        \left\{
        \frac{1}{n_{t_1}}\sum_{i=1}^n
        \mathbbm{1}\{G_i=t_1\}Y^{\obs}_{it_1-\kappa-1}
        -
        \frac{1}{n_{>t_y+\kappa}}\sum_{i=1}^n
        \mathbbm{1}\{G_i>t_y+\kappa\}Y^{\obs}_{it_1-\kappa-1}
        \right\},
        \end{aligned}
    \end{equation*}
    where $n_{>t_y+\kappa}=\sum_{i=1}^n \mathbbm{1}\{G_i>t_y+\kappa\}$. More generally, the reference period need not be restricted to a single pre-treatment time point. Under the baseline conditions, any period $s<t_1-\kappa$ can serve as a valid reference period. Thus, we may define a weighted combination of pre-treatment periods, $\tilde{Y}_{i}^{\text{ref}}(t) = \sum_{s < t_1 - \kappa} v_s Y_{is}(t)$, where ${v_s}$ are weights that sum to one. Using the expanded control component, the corresponding estimator is
    \begin{equation}
        \begin{aligned}[b]
        \widehat{\ate}^{\Scal}_{t_y}(t_1,\infty)
        &=
        \frac{1}{n_{t_1}}\sum_{i=1}^n
        \mathbbm{1}\{G_i=t_1\}
        \left(
        Y^{\obs}_{it_y}
        -
        \sum_{s<t_1-\kappa}v_sY^{\obs}_{is}
        \right) \\
        &\quad -
        \frac{1}{n_{>t_y+\kappa}}\sum_{i=1}^n
        \mathbbm{1}\{G_i>t_y+\kappa\}
        \left(
        Y^{\obs}_{it_y}
        -
        \sum_{s<t_1-\kappa}v_sY^{\obs}_{is}
        \right).
        \end{aligned}
    \end{equation}
    This estimator is also unbiased and aggregates information across all valid pre-treatment reference periods. 
    
\paragraph{Invariance to time shifts.}
Assumption~\ref{assumption_invariance_time_shifts} states that for each shift $h$ such that $t_1+h$, $t_y+h$, and $t_1+h-\kappa-1$ lie in $\Tcal$,
\begin{equation}
    \label{eq_supp:PT_invariance}
    \begin{aligned}[b]
    \ate^\Scal_{t_y}(t_1,\infty) 
    &= \E_{\S}\left\{ \Delta^{l^*+\kappa+1}Y_{it_y}(t_1) - \Delta^{l^*+\kappa+1}Y_{it_y}(\infty) \right\} \\
    &= \E_{\S}\left\{ \Delta^{l^*+\kappa+1}Y_{i,t_y+h}(t_1+h) - \Delta^{l^*+\kappa+1}Y_{i,t_y+h}(\infty) \right\} \\
    &= \E_{\S}\left\{ \Delta^{l^*+\kappa+1}Y_{i,t_y+h}^{\obs} \mid G_i=t_1+h \right\} \\
    &\quad - \E_{\S}\left\{ \Delta^{l^*+\kappa+1}Y_{i,t_y+h}^{\obs} \mid G_i=\infty \right\} \\
    &= \ate^\Scal_{t_y+h}(t_1+h,\infty).
    \end{aligned}
\end{equation}
Any unbiased estimator of the shifted estimand is also unbiased for the original estimand, and in particular
\begin{equation}
    \begin{aligned}[b]
    \widehat{\ate}^\Scal_{t_y+h}(t_1+h, \infty) 
        & = \frac{1}{n_{t_1 + h}}\sum_{i=1}^n
        \mathbbm{1}\{G_i=t_1+h\}
        \left(
        Y^{\obs}_{i,t_y+h} - \sum_{s<t_1 + h -\kappa}v_sY^{\obs}_{is} \right) \\
        &\quad -
        \frac{1}{n_{>t_y+ h +\kappa}}\sum_{i=1}^n
        \mathbbm{1}\{G_i>t_y+ h +\kappa\}
        \left(
        Y^{\obs}_{i,t_y+h}
        -
        \sum_{s<t_1+ h -\kappa}v_sY^{\obs}_{is}
        \right), 
    \end{aligned}
\end{equation}
where the reference period weights satisfy $\sum_{s<t_1+h-\kappa}v_s=1$. Thus, Assumption~\ref{assumption_invariance_time_shifts} justifies treated outcome differences from shifted cohorts at event time $l^*$, together with the corresponding shifted control differences.

For the remaining control groups, let $t\in\Tcal$ denote a generic treatment-initiation time and let $l=t_y-t$ denote its event time at the target outcome period. The corresponding within-unit potential outcome difference is $\Delta^{l+1}Y_{it_y}(g) = Y_{it_y}(g)-Y_{i,t-1}(g), g\in\Tcal^+$. For observations from cohort $t$ to enter the control component, it is sufficient that the treatment effect on this outcome difference is zero:
\begin{equation}
\label{eq_supp:PT_zero_effect}
\E_{\S}\left\{ \Delta^{l+1}Y_{it_y}(t) - \Delta^{l+1}Y_{it_y}(\infty) \right\} =0.
\end{equation}
Whenever Equation~\eqref{eq_supp:PT_zero_effect} holds,
\begin{equation}
\label{eq_supp:PT_control_chain}
\begin{aligned}[b]
\E_{\S}\left\{ \Delta^{l+1}Y_{it_y}^{\obs}\mid G_i=t \right\} &= \E_{\S}\left\{ \Delta^{l+1}Y_{it_y}(t)\mid G_i=t
\right\} \\
&= \E_{\S}\left\{ \Delta^{l+1}Y_{it_y}(t) \right\} \\
&= \E_{\S}\left\{ \Delta^{l+1}Y_{it_y}(\infty) \right\} \\
&= \E_{\S}\left\{ \Delta^{l+1}Y_{it_y}(\infty)\mid G_i=\infty \right\} \\
&= \E_{\S}\left\{ \Delta^{l+1}Y_{it_y}^{\obs}\mid G_i=\infty
\right\}.
\end{aligned}
\end{equation}
The first and last equalities use Assumption~\ref{assumption_consistency}, the second and fourth use Assumption~\ref{assumption_two_way_parallel}, and the middle equality is provided by the relevant information-borrowing assumption.

\paragraph{Delayed treatment onset.}
For a cohort $t$ satisfying $0\leq l=t_y-t\leq\phi$,
Assumption~\ref{assumption_delay}, applied with $\phi^\prime=l$
and with $Y^*_{it_y}(g)=\Delta^{l+1}Y_{it_y}(g)$, gives
\begin{equation}
\label{eq_supp:PT_delay}
\E_{\S}\left\{
\Delta^{l+1}Y_{it_y}(t)
-
\Delta^{l+1}Y_{it_y}(\infty)
\right\}=0.
\end{equation}
Equation~\eqref{eq_supp:PT_control_chain} permits outcome differences from units whose treatment began too recently for its effect to have appeared to enter the control component. These observations form the Delayed Onset group.

\paragraph{Treatment effect dissipation.}
For a cohort $t$ satisfying $l=t_y-t \geq\xi$, Assumption~\ref{assumption_washout}, applied with $\xi^\prime=l$ and with $Y^*_{it_y}(g)=\Delta^{l+1}Y_{it_y}(g)$, gives
\begin{equation}
\label{eq_supp:PT_dissipation}
\E_{\S}\left\{ \Delta^{l+1}Y_{it_y}(t) - \Delta^{l+1}Y_{it_y}(\infty) \right\} =0.
\end{equation}
Equation~\eqref{eq_supp:PT_dissipation} permits outcome differences from units whose treatment effect has dissipated to enter the control component. These observations form the Effect Dissipation group.

\subsection{Observation groups of the dynamic TWFE estimator}
\label{sec:app_twfe_decomposition_proof}

The preceding subsections characterize which observations may enter a valid contrast under the unconfoundedness and parallel trends frameworks. We now map those results to the dynamic TWFE coefficient. Algebraically, the coefficient admits two equivalent representations: one in terms of weighted outcome levels and another in terms of weighted within-unit outcome differences. The first representation is useful for connecting the TWFE to the unconfoundedness framework. The second is the representation used in the main text under parallel trends framework. The two representations are algebraically equivalent, but their causal interpretations follow the two different identification routes.

\subsubsection{Outcome-level representation}
\label{sec:app_twfe_decomposition_proof_Y}

We begin with the outcome-level representation. We set $t-G_i=-\infty$ when $G_i=\infty$. The dynamic TWFE regression estimator in Equation~\eqref{equation_TWFE_dynamic} can be written as
\begin{align}
\label{equation_implied_tau}
    \widehat{\tau}_{l^*} 
    & = \sum_{i = 1}^n \sum_{t = 1}^T w_{it} Y_{it}^{\mathrm{obs}} \mathbbm{1}(t - G_i = l^*) -\sum_{i = 1}^n \sum_{t = 1}^T w_{it} Y_{it}^{\mathrm{obs}} \mathbbm{1}(t - G_i \neq l^*), 
\end{align}
where $w_{it}$ are the regression  weights in \cite{chattopadhyay2023implied}. These weights are constructed such that $\sum_{i = 1}^n \sum_{t = 1}^T w_{it} \mathbbm{1}(t - G_i = l^*) = \sum_{i = 1}^n \sum_{t = 1}^T w_{it} \mathbbm{1}(t - G_i \neq l^*) = 1$. In Equation (\ref{equation_implied_tau}), the parameter $\widehat{\tau}_{l^*}$ is expressed as a contrast between a treatment component and a control component of weighted outcome levels. 

We further break down the treatment and control components of the dynamic TWFE estimator, clarifying the assumptions required for including each observation group. The treatment component can be expressed as:
\begin{align}
        \sum_{i,t} w_{it} Y_{it}^{\mathrm{obs}} \mathbbm{1}(t - G_i = l^*)
        & = \sum_{i} w_{it_y} Y_{it_y}^{\mathrm{obs}} \mathbbm{1}(t_y - G_i = l^*) \label{equation_usual_tau_ideal_treat_Y}
        \\
        & + \sum_{i} \sum_{t \neq t_y} w_{it} Y_{it}^{\mathrm{obs}} \mathbbm{1}(t - G_i = l^*), \label{equation_usual_tau_inv_treat_Y}
\end{align}
and the control component can be decomposed as: 
\allowdisplaybreaks
\begin{align}
        \sum_{i,t} w_{it} Y_{it}^{\mathrm{obs}} \mathbbm{1}(t - G_i \neq l^*)
        & = \sum_{i} w_{it_y} Y_{it_y}^{\mathrm{obs}} \mathbbm{1}(t_y - G_i = -\infty) \label{equation_usual_tau_ideal_control_Y} \\
        & + \sum_{i} w_{it_y} Y_{it_y}^{\mathrm{obs}} \mathbbm{1}( -\infty < t_y - G_i < 0) \label{equation_usual_tau_forbid_1_Y} \\
        & + \sum_{i} w_{it_y} Y_{it_y}^{\mathrm{obs}} \mathbbm{1}(  0 \leq t_y - G_i < l^*) \label{equation_usual_tau_forbid_2_Y} \\
        & + \sum_{i} w_{it_y} Y_{it_y}^{\mathrm{obs}} \mathbbm{1}(t_y - G_i > l^*) \label{equation_usual_tau_forbid_3_Y} \\
        & + \sum_{i} \sum_{t \neq t_y} w_{it} Y_{it}^{\mathrm{obs}} \mathbbm{1}(t - G_i = -\infty) \label{equation_usual_tau_inv_control_Y} \\
        & + \sum_{i} \sum_{t \neq t_y} w_{it} Y_{it}^{\mathrm{obs}} \mathbbm{1}( -\infty < t - G_i < 0) \label{equation_usual_tau_inv_forbid_1_Y} \\
        & + \sum_{i} \sum_{t \neq t_y} w_{it} Y_{it}^{\mathrm{obs}} \mathbbm{1}( 0 \leq t - G_i < l^*) \label{equation_usual_tau_inv_forbid_2_Y} \\
        & + \sum_{i} \sum_{t \neq t_y} w_{it} Y_{it}^{\mathrm{obs}} \mathbbm{1}(t - G_i > l^*). \label{equation_usual_tau_inv_forbid_3_Y} 
\end{align}
Under Assumptions \ref{assumption_consistency}-\ref{assumption_random_treatment_initiation}, for the estimand $\ate^{\Scal_t}_{t_y}(t_1, \infty)$, the observations in \eqref{equation_usual_tau_ideal_treat_Y} serve as valid treated observations and observations in \eqref{equation_usual_tau_ideal_control_Y} serve as valid control observations. Assumption~\ref{assumption_invariance_time_shifts} makes the shifted treatment and never-treated terms \eqref{equation_usual_tau_inv_treat_Y} and \eqref{equation_usual_tau_inv_control_Y} jointly form valid shifted contrasts for the same target effect. Furthermore, Assumptions \ref{assumption_limited_anticipation} ($\kappa = 0$), Assumptions \ref{assumption_delay} ($\phi = l^* - 1$)
 and Assumptions \ref{assumption_washout} ($\xi = l^* + 1$) bring in observations in \eqref{equation_usual_tau_forbid_1_Y}, \eqref{equation_usual_tau_forbid_2_Y} and \eqref{equation_usual_tau_forbid_3_Y} as the valid control observations. Combined with Assumption \ref{assumption_invariance_time_shifts}, they also justify the inclusion of observations in \eqref{equation_usual_tau_inv_forbid_1_Y}, \eqref{equation_usual_tau_inv_forbid_2_Y} and \eqref{equation_usual_tau_inv_forbid_3_Y} as the valid control observations, respectively.  

\subsubsection{Outcome-difference representation}
\label{sec:app_twfe_decomposition_proof_Ydiff}

We focus on a post-treatment target event time $l^*=t_y-t_1$ and use the conventional omitted period $l=-1$, corresponding to $\kappa=0$. For ever-treated units, write $l=t-G_i$ for event time. For never-treated units, fix any observed reference period and, with a slight abuse of notation, write $\Delta^{t-G_i+1}Y_{it}^{\obs}$ as the difference between $Y_{it}^{\obs}$ and the outcome at that reference period. For ever-treated units at $l=-1$, $\Delta^{t-G_i+1}Y_{it}^{\obs}=\Delta^0Y_{it}^{\obs}=0$. Splitting Equation~\eqref{equation_lmw_tau} accordingly, the treatment component is
\begin{align}
        \sum_{i,t} w_{it}\, \Delta^{l^*+1} Y_{it}^{\obs}\, \mathbbm{1}\{t - G_i = l^*\}
        & = \sum_{i} w_{it_y}\, \Delta^{l^*+1} Y_{it_y}^{\obs}\, \mathbbm{1}\{G_i = t_1\} \label{equation_usual_tau_ideal_treat} \\
        & \quad + \sum_{i} \sum_{t \neq t_y} w_{it}\, \Delta^{l^*+1} Y_{it}^{\obs}\, \mathbbm{1}\{t - G_i = l^*\}, \label{equation_usual_tau_inv_treat}
\end{align}
and the control component is
\allowdisplaybreaks
\begin{align}
        \sum_{i,t} w_{it}\, \Delta^{t-G_i+1} Y_{it}^{\obs}\, \mathbbm{1}\{t - G_i \notin \{-1, l^*\}\}
        & = \sum_{i} w_{it_y}\, \Delta^{t_y-G_i+1} Y_{it_y}^{\obs}\, \mathbbm{1}( G_i = \infty) \label{equation_usual_tau_ideal_control} \\
        & \quad + \sum_{i} w_{it_y}\, \Delta^{t_y-G_i+1} Y_{it_y}^{\obs}\, \mathbbm{1}( -\infty < t_y - G_i < -1 ) \label{equation_usual_tau_forbid_1} \\
        & \quad + \sum_{i} w_{it_y}\, \Delta^{t_y-G_i+1} Y_{it_y}^{\obs}\, \mathbbm{1}( 0 \leq t_y - G_i < l^*) \label{equation_usual_tau_forbid_2} \\
        & \quad + \sum_{i} w_{it_y}\, \Delta^{t_y-G_i+1} Y_{it_y}^{\obs}\, \mathbbm{1}(t_y - G_i > l^*) \label{equation_usual_tau_forbid_3} \\
        & \quad + \sum_{i} \sum_{t \neq t_y} w_{it}\, \Delta^{t-G_i+1} Y_{it}^{\obs}\, \mathbbm{1}( G_i = \infty) \label{equation_usual_tau_inv_control} \\
        & \quad + \sum_{i} \sum_{t \neq t_y} w_{it}\, \Delta^{t-G_i+1} Y_{it}^{\obs}\, \mathbbm{1}(-\infty < t - G_i < -1) \label{equation_usual_tau_inv_forbid_1} \\
        & \quad + \sum_{i} \sum_{t \neq t_y} w_{it}\, \Delta^{t-G_i+1} Y_{it}^{\obs}\, \mathbbm{1}( 0 \leq t - G_i < l^*) \label{equation_usual_tau_inv_forbid_2} \\
        & \quad + \sum_{i} \sum_{t \neq t_y} w_{it}\, \Delta^{t-G_i+1} Y_{it}^{\obs}\, \mathbbm{1}(t - G_i > l^*). \label{equation_usual_tau_inv_forbid_3}
\end{align}
Terms~\eqref{equation_usual_tau_ideal_treat} and
\eqref{equation_usual_tau_ideal_control} give the target-cohort and never-treated components of the baseline contrast. Assumption~\ref{assumption_limited_anticipation} with $\kappa=0$ also justifies \eqref{equation_usual_tau_forbid_1}, corresponding to sufficiently not-yet-treated observations at $t_y$.

Maintaining these baseline conditions, Assumption~\ref{assumption_invariance_time_shifts} justifies \eqref{equation_usual_tau_inv_treat} and \eqref{equation_usual_tau_inv_control}, and, together with
Assumption~\ref{assumption_limited_anticipation}, also justifies \eqref{equation_usual_tau_inv_forbid_1}. Assumption~\ref{assumption_delay} with $\phi=t_y-t_1-1$ justifies \eqref{equation_usual_tau_forbid_2}, while Assumption~\ref{assumption_washout} with $\xi=t_y-t_1+1$ justifies \eqref{equation_usual_tau_forbid_3}. Combined with
Assumption~\ref{assumption_invariance_time_shifts}, they also admit \eqref{equation_usual_tau_inv_forbid_2} and \eqref{equation_usual_tau_inv_forbid_3}, respectively.

Under this interpretation, treating every non-target treated difference as a valid control is justified by the baseline parallel trends conditions together with the corresponding additional information borrowing assumptions. Table~\ref{tab:app_groups} summarizes the six observation groups and the assumptions that justify their inclusion. For the control groups, using observations from periods other than $t_y$ additionally requires Assumption~\ref{assumption_invariance_time_shifts}.
\begin{table}[h!tbp]
\caption{\label{tab:app_groups}
Observation groups in the dynamic TWFE contrast for
$\ate^{\Scal_t}_{t_y}(t_1,\infty)$, with $l=t-G_i$ and
$l^*=t_y-t_1$.}
\centering
\footnotesize
\begin{tabular}{lll}
\toprule
Group & Observations & Conditions \\
\midrule
\multicolumn{3}{l}{Baseline parallel-trends comparison} \\[2pt]
Ideal Experiment & $l=l^*$, $G_i=t_1$ & Baseline conditions \\
Never Treated & $G_i=\infty$ & Baseline conditions \\
No Anticipation & $-\infty <l<-1$ & Baseline conditions ($\kappa=0$) \\
\midrule
\multicolumn{3}{l}{Additional information borrowing} \\[2pt]
Time Invariance & $l=l^*$, $G_i\neq t_1$ & Assumption~\ref{assumption_invariance_time_shifts} \\
Delayed Onset & $0\leq l<l^*$ & Assumption~\ref{assumption_delay}, $\phi=l^*-1$ \\
Effect Dissipation & $l>l^*$ & Assumption~\ref{assumption_washout}, $\xi=l^*+1$ \\
\bottomrule
\end{tabular}
\end{table}

%% file: supp2_weighting.tex

\section{Robust weighting: asymptotics, extensions, and implementation}
\label{sec:app_rw}

This section completes the details on robust weighting (RW) estimator of Section~\ref{sec3:weight_estimator}. Section~\ref{sec:app_rw_asymptotics} gives the dual of the balancing problem, the regularity conditions behind Theorems~\ref{thm_consistency} and~\ref{thm_normality}, and their proofs. Section~\ref{sec:app_robust_generalization} extends the estimator to a target population distinct from the study population, and Section~\ref{sec:app_genestimand} extends it to a general reference regime $t_{\mathbf{p}}$. Section~\ref{sec:app_rw_onestep} defines the aggregate event study estimand and the one-step RW estimator used for it in the case study, and Section~\ref{sec:app_rw_implementation} documents implementation choices.

\subsection{Dual problem, regularity conditions, and proofs}
\label{sec:app_rw_asymptotics}

Throughout this subsection $\Pcal = \Scal$ and the weights solve \eqref{minimal_1}. We first show the dual of that problem, since Assumption~\ref{assump_consistency1} is stated in terms of the function $\rho(\cdot)$ it produces.

\begin{theorem}
\label{thm_dual_study}
The dual problem of \eqref{minimal_1} with $\Mcal = \text{span}\left\{B_k(\cdot), k=1,2,...,K \right\}$ is equivalent to the empirical loss minimization problem with $L_1$ regularization:
\begin{align} \label{dual_study}
& \underset{\bm{\lambda}}{\text{minimize}}
& & n^{-1}\sum_{i=1}^{n}\Big[-n\,\mathbbm{1}(G_i=t)\rho\{\bm{B}(\bX_i)^\top \bm{\lambda} \} + \bm{B}(\bX_i)^\top \bm{\lambda} \Big] + |\bm{\lambda}|^\top\bm{\delta},
\end{align}
where $\bm{\lambda}$ is a $K \times 1$ vector of dual variables corresponding to the $K$ balancing constraints, $|\bm{\lambda}|$ is the vector of component-wise absolute values of $\bm{\lambda}$, and $\rho(v) = v/n - v(h')^{-1}(v) - h\{(h')^{-1}(v)\}$, with $h(v) = \psi(1/n-v)$.
\end{theorem}

Problem \eqref{dual_study} is the study-population case of Theorem~\ref{thm_dual1} below, with $n^*$ replaced by $n$ and the study-inclusion indicator removed.

We now state the regularity conditions required by Theorems~\ref{thm_consistency} and~\ref{thm_normality}. These conditions are exact analogs of those in Assumption 4 of \cite{chattopadhyay2024one}.

\begin{assumption}
\label{assump_regularity1}
For $t \in \{t_1,\infty\}$,
\begin{enumerate}[label=(\alph*)]
\item There exist constants $c_0$, $c_1$, $c_2$ with $0<c_0<1/2$ and $c_1<c_2<0$, such that $c_1 \leq n\rho''(v) \leq c_2$ for all $v$ in a neighborhood of $\bm{B}(\bm{x})^\top \bm{\lambda}^*_{1t}$. Also, $c_0 \leq 1/\{n\rho'(v)\} \leq 1-c_0$ for all $v = \bm{B}(\bm{x})^\top \bm{\lambda}$, $\bm{x} \in \text{Supp}(\bX_i)$, $\bm{\lambda}$.

\item $\sup\limits_{\bm{x} \in \text{Supp}(\bX_i)}\|\bm{B}(\bm{x})\|_2 \leq CK^{1/2}$ and $\|\E_{\S}\{\bm{B}(\bX_i) \bm{B}(\bX_i)^\top\}\|_F \leq C$, for some constant $C>0$, where $\|\cdot\|_F$ denotes the Frobenius norm.

\item $K = O(n^\alpha)$ for some $0<\alpha<2/3$.

\item For some constant $C>0$,
$\lambda_{\text{min}}\Big[ \E_{\S}\{\mathbbm{1}(G_i = t)\bm{B}(\bX_i)\bm{B}(\bX_i)^\top\}\Big]>C$,
where $\lambda_{\min}(\bm{A})$ denotes the smallest eigenvalue of $\bm{A}$.

\item $\|\bm{\delta}\|_2 = O_P\left[K^{1/4}\left\{(\log K) / n\right\}^{1/2} + K^{- r_t + 1/2}\right]$.
\end{enumerate}
\end{assumption}

Here, Assumption \ref{assump_regularity1}(a) controls the slope and curvature of the function $\rho(\cdot)$ in \eqref{dual_study}, which enables us to translate the convergence of the dual solution to the convergence of the weights. Assumption \ref{assump_regularity1}(b) and (c) restrict the rate of growth of the magnitude and the
number of the basis functions. Assumption \ref{assump_regularity1}(d) ensures non-singularity of the covariance matrices of the basis functions within each treatment group. Finally, \ref{assump_regularity1}(e) controls the magnitude of the degree of approximate balance.

To show asymptotic Normality we require a few additional regularity conditions. These conditions are exact analogs of those in Assumption 5 in \cite{chattopadhyay2024one}.

\begin{assumption}
\label{assump_regularity2}
For $t \in \{t_1,\infty\}$,
\begin{enumerate}[label=(\alph*)]
\item $\E_{\S}[\{Y^{*}_{it_y}(t)\}^2]<\infty$.

\item Let $g^*_t(\cdot) \in \Gcal_t$. $\Gcal_t$ satisfies
$\log N_{[]}\{\epsilon, \Gcal_t, L_2(\S)\} \leq C_1(1/\epsilon)^{1/k_1}$ for some constants $C_1>0$ and $k_1>1/2$, where $N_{[]}\{\epsilon, \Gcal_t, L_2(\S)\}$ is the covering number of $\Gcal_t$ by epsilon brackets.

\item Let $m_{t,t_y}(\cdot) \in \Mcal_t$. $\Mcal_t$ satisfies
$\log N_{[]}\{\epsilon, \Mcal_t, L_2(\S)\} \leq C_2(1/\epsilon)^{1/k_2}$ for some constants $C_2>0$ and $k_2>1/2$, where $N_{[]}\{\epsilon, \Mcal_t, L_2(\S)\}$ is the covering number of $\Mcal_t$ by epsilon brackets.

\item $n^{\{2(r_t + s_t - 0.5)\}^{-1}} = o(K)$, where $r_t, s_t$ are the constants in Assumptions~\ref{assump_consistency1} and~\ref{assump_consistency2}, respectively.
\end{enumerate}
\end{assumption}

Here, Assumption \ref{assump_regularity2}(a) requires finite second moment of $Y^{*}_{it_y}(t)$. Assumptions \ref{assump_regularity2}(b) and (c) control the complexity of
the function classes $\Gcal_t$ and $\Mcal_t$. Assumption \ref{assump_regularity2}(d) restricts the growth rate of the number of basis functions $K$ as a function of $n$.

\begin{proof}[Proof of Theorem~\ref{thm_consistency}]
Under Assumption~\ref{assump_regularity1}, problem \eqref{minimal_1} is the one-step weighting problem of \cite{chattopadhyay2024one} with target sample equal to the study sample, so their Theorem 4.1 applies with $n^*=n$ and $\pi(\bm{x}) \equiv 1$. Their result gives consistency of each arm-specific estimator whenever, for that arm, either the
treatment initiation model or the potential outcome model is correctly specified in the sense of Assumptions~\ref{assump_consistency1} and~\ref{assump_consistency2}. Since $\widehat{\ate}^\Scal_{t_y}(t_1, \infty)$ is the difference of the two arm-specific estimators, it is consistent when this holds in each arm separately, which gives the four
non-nested conditions (a)--(d).
\end{proof}

\begin{proof}[Proof of Theorem~\ref{thm_normality}]
Under Assumptions~\ref{assump_consistency1}, \ref{assump_consistency2},
\ref{assump_regularity1}, and~\ref{assump_regularity2}, Theorem 4.2 of
\cite{chattopadhyay2024one} applies with $n^*=n$ and $\pi(\bm{x}) \equiv 1$, giving
asymptotic Normality of each arm-specific estimator with influence function equal to the
efficient influence function for that arm. Differencing the two and setting
$\pi(\bm{x}) \equiv 1$ in the variance expression of Theorem~\ref{thm_normality_gen} gives
\begin{equation*}
V_\tau = \E_{\S}\left(\frac{\sigma^2_1(\bX)}{e_{t_1}(\bX)}
+ \frac{\sigma^2_0(\bX)}{e_\infty(\bX)} + \{m_{t_1,t_y}(\bX)-m_{\infty,t_y}(\bX) -\ate^\Scal_{t_y}(t_1,\infty)\}^2 \right),
\end{equation*}
where $\sigma^2_t(\bm{x}) = \Var_{\S}\{Y^{*}_{it_y}(t)|\bX=\bm{x}\}$, which is the
semiparametric efficiency bound for $\ate^\Scal_{t_y}(t_1, \infty)$.
\end{proof}

These results carry over to the expanded weighted contrast in
Equation~\eqref{eq_contrast1} under the same potential outcome model and a modified
treatment initiation model, as described in Section~\ref{sec_large}. The proofs are
unchanged, since they use the treatment initiation model only through membership in the two
components, which for \eqref{eq_contrast1} are $\Ccal$ and $\{(i,t): G_i = \infty\}$ rather
than $\{i: G_i = t_1\}$ and $\{i: G_i = \infty\}$.

\subsection{Robust weighting for generalization and transportation}
\label{sec:app_robust_generalization}

In this section, we present the robust weighting approach for generalization and
transportation problems and discuss its properties. Throughout we maintain
Assumptions~\ref{assumption_common_support} and~\ref{assumption_exchangable_populations} of
Section~\ref{sec:app_identification_proof}. For brevity, we focus on the method for
obtaining ideal weighted contrasts as estimators for the average treatment effect (akin to
Section \ref{sec_buildingpure}). Using similar approaches as in Section
\ref{sec_buildinglarger}, the contrast can be enlarged to accommodate additional
observations. Moreover, without loss of generality, here we focus on the unconfoundedness
framework and write outcomes in levels; the parallel trends case follows by reading
$Y^{*}_{it_y}(t)$ for $Y_{it_y}(t)$ throughout, as in Section~\ref{sec3:weight_estimator}.

We consider an arbitrary target population $\Pcal$ with corresponding measure $\P$.
Moreover, we assume that we have covariate data on a random sample of size $n^*$ from
$\Pcal$. Using the same notations as those in Section \ref{sec_buildinglarger}, we estimate
$\ate^\Pcal_{t_y}(t_1,\infty)$ by the weighted contrast
\begin{equation}
    \widehat{\ate}^\Pcal_{t_y}(t_1,\infty) =\sum_{i:G_i = t_1}w_{it_y}Y^{\obs}_{it_y} - \sum_{i:G_i = \infty}w_{it_y}Y^{\obs}_{it_y}.
    \label{eq_contrast_gen0}
\end{equation}

Denote $m_{t_1,t_y}(\bm{x}) = \E_{\P}\{Y_{{it_y}}(t_1)|\bX_i = \bm{x}\}$ and
$m_{\infty,t_y}(\bm{x}) = \E_{\P}\{Y_{{it_y}}(\infty)|\bX_i = \bm{x}\}$. We can decompose
the bias of $\widehat{\ate}^\Pcal_{t_y}(t_1,\infty)$ as follows,
\begin{align}
    &\E_{\P}\{ \widehat{\ate}^\Pcal_{t_y}(t_1,\infty)\} - \ate^\Pcal_{t_y}(t_1,\infty) = \nonumber \\
    & \E_{\P}\left\{\sum_{i:G_i = t_1}w_{it_y}m_{t_1,t_y}(\bX_i) - \frac{1}{n^*}\sum_{j=1}^{n^*}m_{t_1,t_y}(\bX_j)\right\} - \E_{\P}\left\{\sum_{i:G_i = \infty}w_{it_y}m_{\infty,t_y}(\bX_i) - \frac{1}{n^*}\sum_{j=1}^{n^*}m_{\infty,t_y}(\bX_j)\right\}.
    \label{eq_genbias}
\end{align}
Equation (\ref{eq_genbias}) shows that, to remove the bias of
$\widehat{\ate}^\Pcal_{t_y}(t_1,\infty)$, the weights need to balance the function
$m_{t_1,t_y}(\bm{x})$ in the treatment component, relative to the target sample, and
$m_{\infty,t_y}(\bm{x})$ in the control component, relative to the target sample. Thus, to
compute the weights in the group $\{i:G_i = t\}$ ($t \in \{t_1,\infty\}$), we modify the
optimization problem in \eqref{minimal_1} as follows.
\begin{align}\label{minimal_gen1}
\argmin_{\bm{w}}\sum_{i: G_i=t}\psi(w_{it_y}), \quad \text{subject to} \quad \max\limits_{m(\cdot) \in \Mcal}\left| \sum_{i:G_i = t}w_{it_y}m(\bX_i) - \frac{1}{n^*}\sum_{j=1}^{n^*}m(\bX_j)\right|\leq \bm{\delta}.
\end{align}
In other words, \eqref{minimal_gen1} ensures that the weights are of minimal dispersion,
subject to approximately balancing all the functions in $\Mcal$ in the weighted sample
$\{i:G_i = t\}$, relative to the target sample. In practice, we assume that $\Mcal$ can be
spanned by a set of $K$ basis functions, i.e.,
$\Mcal = \text{span}\left\{B_k(\cdot), k=1,2,...,K \right\}$. Also, like before, we set
$\psi(\cdot)$ as the squared norm of the weights and constrain the weights to be
non-negative and sum to one.

We now discuss the asymptotic properties of the proposed weighting estimator. We focus on a nested study setting, where the study sample is nested within a random sample of size $n^*$ from the target population. For a unit $i$ in the target sample, let $D_i$ be an indicator variable that equals one if the unit is included in the study and zero otherwise. Let $\pi(\bm{x})=\Pr(D_i=1\mid\bX_i=\bm{x})$ be the conditional probability of inclusion in the study, given the covariates. Following \cite{chattopadhyay2024one}, we first obtain the dual problem of \eqref{minimal_gen1} as follows.
\begin{theorem}
\label{thm_dual1}
The dual problem of \eqref{minimal_gen1} with $\Mcal = \text{span}\left\{B_k(\cdot), k=1,2,...,K \right\}$
is equivalent to the empirical loss minimization problem with $L_1$ regularization:
\begin{align} \label{dual1}
& \underset{\bm{\lambda}}{\text{minimize}}
& & (n^*)^{-1}\sum_{i=1}^{n^*}\Big[-n^*\mathbbm{1}(G_i=t)D_i\rho\{\bm{B}(\bX_i)^\top \bm{\lambda} \} + \bm{B}(\bX_i)^\top \bm{\lambda} \Big] + |\bm{\lambda}|^\top\bm{\delta},
\end{align}
where $\bm{\lambda}$ is a $K \times 1$ vector of dual variables corresponding to the $K$
balancing constraints, $|\bm{\lambda}|$ is the vector of component-wise absolute values of
$\bm{\lambda}$, and $\rho(v) = v/n^* - v(h')^{-1}(v) - h\{(h')^{-1}(v)\}$, with
$h(v) = \psi(1/n^*-v)$.
\end{theorem}

Akin to Section \ref{sec_buildingpure}, below we provide the modified versions of
Assumptions \ref{assump_consistency1} and \ref{assump_consistency2} for generalization and
transportation problems.
\begin{assumption}[Approximability of the treatment initiation model]
\label{assump_consistency_gen1}
The conditional probability for being in study and initiating treatment at time $t$ satisfies $\Pr(G_i = t, D_i = 1 \mid \bX_i = \bm{x}) = [n^* \rho'\{g^{*}_t(\bm{x})\}]^{-1}$ for all $\bm{x} \in \text{Supp}(\bX_i)$, where $\rho(\cdot)$ is same as that in \eqref{dual1}, and $g^*_t(\cdot)$ is a smooth function that satisfies $\sup_{\bm{x} \in \text{Supp}(\bX_i)}|g^*_t(\bm{x}) - \bm{B}(\bm{x})^\top \bm{\lambda}^*_{1t}| = O(K^{-r_t})$ for some $\bm{\lambda}^*_{1t} \in \R^K$ and $r_t>1$.
\end{assumption}
\begin{assumption}[Approximability of the potential outcome model]
\label{assump_consistency_gen2}
The conditional mean function of the potential outcome at $t_y$ under treatment initiation
at $t$ satisfies
$\sup_{\bm{x}\in \text{Supp}(\bX_i)}|m_{t,t_y}(\bm{x}) - \bm{B}(\bm{x})^\top \bm{\lambda}^*_{2t}| = O(K^{-s_t})$
for some $\bm{\lambda}^*_{2t} \in \R^K$ and $s_t>3/4$, with
$\|\bm{\lambda}^*_{2t}\|_2 \|\bm{\delta}\|_2 = o(1)$.
\end{assumption}
In addition, we require the regularity conditions of
Assumptions~\ref{assump_regularity1} and~\ref{assump_regularity2}, restated for the target
population.
\begin{assumption}
\label{assump_regularity1_gen}
Assumption~\ref{assump_regularity1} holds with $n$ replaced by $n^*$ throughout, with
$r_t$ the constant in Assumption~\ref{assump_consistency_gen1}, and with the expectation in
part (b) taken under $\P$.
\end{assumption}
\begin{assumption}
\label{assump_regularity2_gen}
Assumption~\ref{assump_regularity2} holds with $n$ replaced by $n^*$, with the outcome in
part (a) taken as $Y_{jt_y}(t)$ under $\P$, with the covering numbers in parts (b) and (c)
taken in $L_2(\P)$, and with $r_t, s_t$ the constants in
Assumptions~\ref{assump_consistency_gen1} and~\ref{assump_consistency_gen2}.
\end{assumption}

With Assumptions \ref{assump_consistency_gen1} and \ref{assump_regularity1_gen}, we can show that the proposed weighting estimator is multiply robust for $\ate^\Pcal_{t_y}(t_1,\infty)$. The proof follows directly from the proof of Theorem 4.1 in \cite{chattopadhyay2024one}.

\begin{theorem}[Consistency]
\label{thm_consistency_gen}
Suppose that Assumptions \ref{assump_consistency_gen1} and \ref{assump_regularity1_gen} hold. Then the proposed weighting estimator $\widehat{\ate}^\Pcal_{t_y}(t_1,\infty) =\sum_{i:G_i = t_1}w_{it_y}Y^{\obs}_{it_y} - \sum_{i:G_i = \infty}w_{it_y}Y^{\obs}_{it_y}$ is consistent for $\ate^\Pcal_{t_y}(t_1,\infty)$ if any of the following conditions are satisfied: (a) The treatment initiation models are correctly specified for both time $t_1$ and $\infty$, (b) the potential outcome models are correctly specified for both time $t_1$ and $\infty$, (c) the treatment initiation model for time $t_1$ and the potential outcome model for time $\infty$ are both correctly specified, (d) the treatment initiation model for time $\infty$ and the potential outcome model for time $t_1$ are both correctly specified.
\end{theorem}

With Assumptions \ref{assump_consistency_gen1}-\ref{assump_regularity2_gen}, we can derive
the asymptotic distribution of our proposed estimator. Let $e_t(\bm{x}) = \Pr(G_i = t|\bX_i = \bm{x})$ be the conditional probability of treatment initiation at time $t$ among study participants. Theorem \ref{thm_normality_gen} shows that $\widehat{\ate}^\Pcal_{t_y}(t_1,\infty)$ is asymptotically Normal and semiparametrically efficient for ${\ate}^\Pcal_{t_y}(t_1,\infty)$. The proof follows directly from the proof of Theorem 4.2 in \cite{chattopadhyay2024one}.
\begin{theorem}[Asymptotic Normality and semiparametric efficiency]
\label{thm_normality_gen}
Suppose that the study sample is nested within a random sample of size $n^*$ from the target
population and that Assumptions \ref{assump_consistency_gen1}-\ref{assump_regularity2_gen}
hold. Then, as $n^*$ tends to infinity,
$(n^*)^{1/2}\{\widehat{\ate}^\Pcal_{t_y}(t_1,\infty) - {\ate}^\Pcal_{t_y}(t_1,\infty)\}$
converges in distribution to a Normal random variable with mean 0 and variance $V^*_\tau$,
where
$V^*_\tau = \E_{\P}(\sigma^2_1(\bX)/ \{\pi(\bX)e_{t_1}(\bX)\} + {\sigma^2_0(\bX) / \{ \pi(\bX)e_{\infty}(\bX) \} } + \{m_{t_1,t_y}(\bX)-m_{\infty,t_y}(\bX) -\ate^\Pcal_{t_y}(t_1,\infty)\}^2 )$,
where $\sigma^2_t(\bm{x}) = \Var_{\P}\{Y_{jt_y}(t)|\bX=\bm{x}\}$ for
$t \in \{t_1,\infty\}$. Here, $V^*_\tau$ equals the semiparametric efficiency bound for
$\ate^\Pcal_{t_y}(t_1,\infty)$.
\end{theorem}

\subsection{Robust weighting for $\ate^\Pcal_{t_y}(t_1, t_\mathbf{p})$}
\label{sec:app_genestimand}

In this section, we focus on settings with $\Pcal = \Scal$. Our proposed methodology can be
adapted to a general target population $\Pcal$ by following the discussions in Section
\ref{sec:app_robust_generalization}. Moreover, without loss of generality, we focus on the
unconfoundedness framework.

Recall that,
\begin{equation}
    \begin{aligned}[b]
    \ate^\Pcal_{t_y}(t_1, t_\mathbf{p})
    &= \E_{\P}\{Y_{i{t_y}}(t_1)\} - \sum_{t \in \Tcal^{+}}p_{t}  \E_{\P}\{Y_{i{t_y}}(t)\} = \mu_1 - \mu_0,
    \end{aligned}
\end{equation}
where $\mu_1= \E_{\P}\{Y_{i{t_y}}(t_1)\}$ and
$\mu_0 = \sum_{t \in \Tcal^{+}}p_{t}  \E_{\P}\{Y_{i{t_y}}(t)\}$. The estimation of $\mu_1$
follows exactly the same steps as described in Section \ref{sec3:weight_estimator}. For
estimating $\mu_0$, we first consider the inverse probability weighting representation
\begin{equation*}
     \mu_0 = \sum_{t \in \Tcal^{+}}p_{t}\frac{\E_{\P} \left\{\frac{\mathbbm{1}(G_i = t)}{\Pr(G_i = t|\bX_i)}Y^{\obs}_{i{t_y}} \right\}}{\E_{\P} \left\{\frac{\mathbbm{1}(G_i = t)}{\Pr(G_i = t|\bX_i)}\right\}},
\end{equation*}
which prompts us to use a H\'{a}jek estimator of the form
\begin{equation*}
    \hat{\mu}_0 = \sum_{t \in \Tcal^{+}}p_{t}\sum_{i:G_i = t}w_{it_y,t}Y^{\obs}_{it_y},
\end{equation*}
where the weights satisfy $\sum_{i:G_i = t}w_{it_y,t} = 1$.

Denoting $m_{t,t_y}(\bm{x}) = \E_{\P}\{Y_{i{t_y}}(t)|\bX_i = \bm{x}\}$, we can decompose the
bias of this H\'{a}jek estimator as
\begin{align}
  \E_{\P}(\hat{\mu}_0) - \mu_0 =  \sum_{t \in \Tcal^{+}}p_{t} \E_{\P}\left\{ \sum_{i:G_i = t}w_{it_y,t}m_{t,t_y}(\bX_i) - \frac{1}{n}\sum_{i=1}^{n}m_{t,t_y}(\bX_i)\right\}.
  \label{eq_A_bias1}
\end{align}
Therefore, to remove the bias of $\hat{\mu}_0$, the weights need to balance the function
$m_{t,t_y}(\bm{x})$ in the group $\{i:G_i = t\}$ relative to the overall sample, for all
$t \in \Tcal^+$. This amounts to running a separate optimization problem of the following
form for each $t$.
\begin{align}\label{minimal_app_1}
\min_{\bm{w}_t}\left\{\sum_{i: G_i=t}\psi(w_{it_y,t}) : \max\limits_{m(\cdot) \in \Mcal}\left| \sum_{i:G_i = t}w_{it_y,t}m(\bX_i) - \frac{1}{n}\sum_{i=1}^{n}m(\bX_i)\right|\leq \bm{\delta} \right\}
\end{align}
With additional assumptions, we can condense the collection of optimization problems in
\eqref{minimal_app_1} to a single optimization problem. For instance, assume that
$m_{t,t_y}(\cdot)$ has the following additive structure
\begin{align}
    m_{t,t_y}(\bm{x}) = \alpha + \beta_{t_y} + \tau_{t_y -t} + g_{t_y}(\bm{x}).
    \label{eq_app_additive}
\end{align}
In this case, the bias in Equation (\ref{eq_A_bias1}) boils down to
\begin{align}
\E_{\P}(\hat{\mu}_0) - \mu_0 &=  \E_{\P}\left\{\sum_{t \in \Tcal^{+}}\sum_{i:G_i = t}p_{t}w_{it_y,t}g_{t_y}(\bX_i) - \frac{1}{n}\sum_{i=1}^{n}g_{t_y}(\bX_i)   \right\} \nonumber\\
& = \E_{\P}\left\{\sum_{i:G_i \in \Tcal^{+}}w_{it_y}g_{t_y}(\bX_i) - \frac{1}{n}\sum_{i=1}^{n}g_{t_y}(\bX_i)   \right\},
\end{align}
where $w_{it_y} = p_tw_{it_y,t}$ if $G_i = t$. Therefore, in this case, the weights
$w_{it_y}$ only needs to balance the function $g_{t_y}(\cdot)$ in the group
$\{i:G_i \in \Tcal^+\}$ relative to the overall sample. The additive structure in
\eqref{eq_app_additive} is the restriction that the dynamic TWFE specification in
Equation~\eqref{equation_TWFE_dynamic} imposes.

Finally, following the discussions in Section \ref{sec_buildinglarger}, we can expand the
weighted contrast to incorporate additional observations. In particular, under the stronger
version of Assumption \ref{assumption_invariance_time_shifts}, the resulting expanded
weighted contrast for $\ate^\Pcal_{t_y}(t_1, t_{\mathbf{p}})$ has the form
\begin{equation}
    \widehat{\ate}^\Pcal_{t_y}(t_1, t_{\mathbf{p}}) = \hat{\mu}_1 - \hat{\mu}_0 =  \sum_{(i,t) \in \Ccal}w_{it}Y^{\obs}_{it} -  \sum_{t' \in \Tcal^{+}}p_{t'}\sum_{(i,t) \in \Ccal_{t'}}w_{it,t'}Y^{\obs}_{it}, \label{eq_app_contrast2}
\end{equation}
where $\Ccal = \cup_{r=1}^{T - l^*}\Ccal_{r,r+l^*}$ and
$\Ccal_{t'} = \cup_{r=1}^{T - l_{t'}}\Ccal_{r,r+l_{t'}}$, with
$\Ccal_{r,s} = \{(i,t): G_i = r, t = s\}$, $l^* = t_y - t_1$, and $l_{t'} = t_y - t'$. For units that start treatment during follow-up, the definitions above include only observations whose corresponding outcome period is also observed. For the never-treated group, $t'=\infty$, let $\Ccal_\infty=\{(i,t): G_i=\infty,\ t\in\Tcal\}$, which includes all observed periods for never-treated units. In Equation~\eqref{eq_A_bias2}, use $m_\infty(x)$ as the mean outcome function for this group. For $(i,t) \in \Ccal$, the weights are computed in the same manner as described in Section
\ref{sec_buildinglarger}. For $(i,t) \in \Ccal_{t'}$, the weights $w_{it,t'}$ satisfy $\sum_{(i,t) \in \Ccal_{t'}}w_{it,t'} = 1$. Now,
\begin{align}
  \E_{\P}(\hat{\mu}_0) - \mu_0 =  \sum_{t' \in \Tcal^{+}}p_{t'} \E_{\P}\left(\sum_{(i,t) \in \Ccal_{t'}}w_{it,t'}m_{l_{t'}}(\bX_i) - \frac{1}{n}\sum_{i=1}^{n}m_{l_{t'}}(\bX_i)\right),
  \label{eq_A_bias2}
\end{align}
where, under the stronger version of Assumption~\ref{assumption_invariance_time_shifts}, $m_{l_{t'}}(\cdot)=m_{r,r+l_{t'}}(\cdot)$ for finite $t'$ and all admissible $r$. Therefore, to remove the bias of $\hat{\mu}_0$, the weights in the group $\{(i,t) \in \Ccal_{t'}\}$ should balance
$m_{l_{t'}}(\cdot)$ relative to the overall sample, which can be achieved by solving an
analog of the optimization problem in \eqref{minimal_app_1}.

\subsection{The aggregate event study estimand and the one-step estimator}
\label{sec:app_rw_onestep}

Event study plots report a single effect at each event time rather than one effect per
cohort. We define the aggregate estimand at event time $l$ as the cohort-size weighted
average of the cohort-specific estimands,
\begin{equation}
\label{eq:app_agg_estimand}
    \ate^{\Scal_t}_{\text{agg}}(l) = \sum_{g \in \Tcal} \omega_g\, \ate^{\Scal_g}_{g+l}(g, \infty),
    \qquad
    \omega_g = \frac{n_g\,\mathbbm{1}\{g + l \in \Tcal\}}{\sum_{g' \in \Tcal} n_{g'}\mathbbm{1}\{g' + l \in \Tcal\}},
\end{equation}
where $n_g$ is the number of units initiating treatment at $g$, and the indicator restricts the average to cohorts observed at event time $l$ within the panel. For estimation under parallel trends, we require $g-\kappa-1\in\Tcal$ for every cohort with $\omega_g>0$. Terms with zero weight are omitted.

Estimand \eqref{eq:app_agg_estimand} can be estimated in two ways, which differ in whether balance is imposed cohort by cohort or once across all cohorts. In the two-step approach, a separate contrast is formed for each cohort $g$ and the resulting estimates are averaged with weights $\omega_g$. This is how the IW and DR estimators produce their aggregate estimates, and the RW estimator can be implemented this way as well.

In the one-step approach, the treated and control components are pooled before the weights are computed. Let
\begin{equation*}
\begin{aligned}
\mathcal C(l) &= \{(i,t): t-G_i=l,\ t\in\Tcal,\ G_i-\kappa-1\in\Tcal\},\\
\mathcal C'(l) &= \{(i,t): G_i=\infty,\ t\in\Tcal,\
t-(l+\kappa+1)\in\Tcal\}.
\end{aligned}
\end{equation*}
These sets collect observations at event time $l$ and never-treated observations, respectively, for which both the outcome and reference periods are observed. The one-step estimator is
\begin{equation}
\label{eq:app_onestep}
    \widehat{\ate}^{\Scal_t}_{\text{agg}}(l) = \sum_{(i,t) \in \Ccal(l)}w_{it}\,\Delta^{l+\kappa+1} Y^{\obs}_{it} - \sum_{(i,t) \in \Ccal'(l)}w_{it}\,\Delta^{l+\kappa+1} Y^{\obs}_{it}.
\end{equation}
For the one-step implementation used in the case study, the treatment component uses uniform weights, preserving the cohort-size weights in~\eqref{eq:app_agg_estimand}. The control weights are obtained from an analog of \eqref{minimal_2}, using the pooled treatment component as the target and balancing the baseline covariates and outcome-year indicators as described in Section \ref{sec:app_rw_implementation}.

The two approaches impose different balance requirements. The two-step approach balances covariates within each cohort separately, so a cohort whose treated units are far from the never-treated units in covariate space may require control weights concentrated on a small number of units. The one-step approach instead balances covariates between the pooled treatment and control components, while also balancing outcome-year indicators. Thus, an imbalance in one cohort can be offset by the others. This greater flexibility may contribute to the difference in precision between the RW and DR aggregate estimates reported in Section~\ref{sec:case_estimation}. The corresponding weight dispersion diagnostics are reported in Section~\ref{sec_app_further}.

\subsection{Implementation}
\label{sec:app_rw_implementation}

\paragraph{Basis functions and balance tolerances.}
In the case study we take $\Mcal$ to be spanned by the baseline covariates themselves, so $\bm{B}(\bX_i) = \bX_i$ and $K = 3$. This corresponds to a mean balance on the three baseline covariates. The tolerances $\bm{\delta}$ are set to $0.01$ on each covariate, measured in units of the standard deviation of the group being reweighted. In the one-step weighting implementation, the outcome-year indicators are balanced exactly, with $\delta = 0$. Larger classes $\Mcal$ can be accommodated by appending transformations of $\bX_i$ to $\bm{B}(\cdot)$, at the cost of tighter computational feasibility.

\paragraph{Solving the weighting problem.}
With $\psi(\cdot)$ being the squared norm, Equation \eqref{minimal_1} is a quadratic program with linear constraints, solved by the \texttt{quadprog} solver and called through the \texttt{sbw} package. The weights are constrained to sum to one within each component. Non-negativity can be imposed, in which case the resulting estimates are sample-bounded \citep{robins2007comment}. In the case study, we impose the non-negativity constraint. When the optimization is infeasible at the specified $\bm{\delta}$, the balance tolerances can be relaxed until a solution exists and the achieved imbalance is reported alongside the estimate, as in the balance diagnostics of Section \ref{sec_app_further}.

\paragraph{Inference.}
Standard errors are obtained from a state-level bootstrap with $B = 500$ replications. Each replication resamples states with replacement, recomputes the weights on the resampled panel, and re-evaluates the contrast. The reported standard error is the standard deviation of the resulting estimates. Resampling states rather than observations preserves the within-state dependence in the panel, and recomputing the weights within each replication accounts for the fact that the weights are estimated rather than fixed. The same procedure is applied to the TWFE, IW, and DR estimators so that the uncertainty estimates in Table \ref{tab:granular_estimation} and Figure \ref{fig:event_study_plot_estimators} are comparable across estimators.

%% file: supp3_decomposition.tex

\section{Exact decompositions of event study estimators}
\label{sec:app_decomp}

This section derives the weighting representations stated in
Section~\ref{sec:exact_decomp}. Section~\ref{sec:app_decomp_setup} introduces the notation
common to all of them. Sections~\ref{sec:app_twfe_decomp_proof},
\ref{sec:app_sa_decomp_proof}, and~\ref{sec:app_dr_decomp_proof} derive the representations
for the dynamic TWFE, interaction-weighted, and doubly robust estimators. Section~\ref{sec:app_twfe_general_correction} shows that the dynamic TWFE coefficient does not correspond to the general estimand $\ate^\Pcal_{t_y}(t_1, t_{\mathbf{p}})$, and Section~\ref{sec:app_twfe_rw_equivalence} shows that the dynamic TWFE estimator is a special case of the robust weighting estimator of Section~\ref{sec3:weight_estimator}.

\subsection{Setup and implied weights}
\label{sec:app_decomp_setup}

Let $\Ical$ denote the index set of observed cells $(i,t)$, with $N = |\Ical|$. We stack the cells into vectors and matrices, keeping the $(i,t)$ label on each row. Let $\bm Y \in \R^{N}$ collect the observed outcomes $Y^{\obs}_{it}$, and let
\begin{equation}
    \bm A\in\R^{N\times n},\quad \bm A_{it,j}=\mathbbm{1}\{i=j\},\qquad
  \bm B\in\R^{N\times T},\quad \bm B_{it,s}=\mathbbm{1}\{t=s\},
\end{equation}
be the unit- and time-fixed effect design matrices. Let $\bm X\in\R^{N\times p}$ collect any additional regressors, with rows $\bX_{it}^{\top}$, and set $\bm F=[\,\bm A\ \ \bm B\ \ \bm X\,]$. Covariates that are constant within a unit lie in $\operatorname{col}(\bm A)$ and are absorbed by the unit fixed effects, so $\bm X$ is needed only when the specification includes time-varying regressors or interactions of baseline covariates with time. In the specifications of Section~\ref{sec:case_study}, $\bm F=[\,\bm A\ \ \bm B\,]$.

Let $\bm D\in\R^{N\times K}$ be the event-time dummies of
Equation~\eqref{equation_TWFE_dynamic}, with column $l$ given by
$\bm D_{it,l}=\mathbbm{1}\{t-G_i=l\}$. Let $\bm D$ contain the included event-time indicators. The reference period $l=-1$ is omitted, and never-treated observations have zeros in all columns of $\bm D$. After removing any linearly dependent columns, we assume that $\tilde{\bm D}^{\top}\tilde{\bm D}$ is nonsingular (defined below). The dynamic TWFE fit of Equation~\eqref{equation_TWFE_dynamic} is the ordinary least squares problem
\begin{equation}
    (\hat\gamma,\hat\beta) =\arg\min_{\gamma,\beta}\big\lVert \bm Y-\bm F\gamma-\bm D\beta\big\rVert^{2},
\label{eq:app_decomp_ols}
\end{equation}
and $\hat\tau_{l^*}=\hat\beta_{l^*}$ is the coefficient on the event-time-$l^*$ dummy.

Let $P_{\bm F}=\bm F(\bm F^{\top}\bm F)^{-1}\bm F^{\top}$ be the orthogonal projection onto $\operatorname{col}(\bm F)$, and let $M = I - P_{\bm F}$ be the orthogonal projection onto $\operatorname{col}(\bm F)^\perp$ (we remove redundant columns from $\bm F$ so that it has full column rank). Thus, $M=M^{\top}=M^2$ and $M\bm F=\bm 0$. Define the residualized event-time design matrix $\tilde{\bm D}=M\bm D$. By the Frisch--Waugh--Lovell theorem, after partialling out the controls $\bm F$, the normal equations for $\beta$ become $\bm D^{\top}M\bm D \hat\beta=\bm D^{\top}M \bm Y$. Equivalently, since $M=M^{\top}=M^2$,
\begin{equation}
    \tilde{\bm D}^{\top}\tilde{\bm D} \hat\beta =\tilde{\bm D}^{\top}\bm Y, \qquad\text{so}\qquad \hat\beta=(\tilde{\bm D}^{\top}\tilde{\bm D})^{-1}\tilde{\bm D}^{\top}\bm Y .
\label{eq:app_decomp_fwl}
\end{equation}
Hence $\hat\tau_{l^*}=\hat\beta_{l^*} = w^{\top}\bm Y$ with the implied-weight vector defined as
\begin{equation}
  w = \tilde{\bm D}\,(\tilde{\bm D}^{\top}\tilde{\bm D})^{-1}\bm{u}_{l^*}\in\R^{N},
  \label{eq:app_decomp_w}
\end{equation}
where $\bm{u}_{l^*}\in\R^{K}$ is the coordinate vector with a single one in the entry indexing the event-time-$l^*$ column of $\tilde{\bm D}$ and zeros elsewhere, so that $w^{\top}\bm Y$ gives the $l^*$ coefficient $\hat\beta_{l^*}$. We denote the entries of $w$ by $w_{it}$, which are the implied regression weights \citep{chattopadhyay2023implied} specific to the event-time coefficient $\hat\tau_{l^*}$. We thus obtain the weighting representation $\hat\tau_{l^*} = \sum_{i,t} w_{it}\, Y^{\obs}_{it}$, in which the implied weights are attached to the original outcomes.

\subsection{Exact decomposition of the dynamic TWFE regression estimator}
\label{sec:app_twfe_decomp_proof}

We now introduce the following lemma to establish the within-unit outcome difference representation of $\hat\tau_{l^*}$ and the exact balance properties achieved by the weights.

\begin{lemma}\label{lem:app_twfe_orth}
Every column of $\tilde{\bm D}$ sums to zero within each unit: $\bm A^{\top}\tilde{\bm D}=\bm 0$. Equivalently,
$\sum_{t:(j,t)\in\Ical}\tilde{\bm D}_{jt,l}=0$ for every unit $j$ and every event-time column $l$. Consequently, for any vector $c\in\R^{N}$ that is constant within each unit,
$\tilde{\bm D}^{\top}c=\bm 0$.
\end{lemma}

\begin{proof}
Each column of $\bm A$ is included in $\bm F$, so
$\operatorname{col}(\bm A)\subseteq \operatorname{col}(\bm F)$ and $M\bm A=\bm 0$. Hence
\[
\bm A^{\top}\tilde{\bm D}
=\bm A^{\top}M\bm D
=(M\bm A)^{\top}\bm D
=\bm 0.
\]
The $(j,l)$ entry of $\bm A^{\top}\tilde{\bm D}$ is $\sum_t \tilde{\bm D}_{jt,l}$, which
gives the within-unit zero-sum property. If $c$ is constant within each unit, then
$c=\bm A a$ for some $a\in\R^{n}$, and therefore
\begin{equation*}
    \tilde{\bm D}^{\top}c
    =\tilde{\bm D}^{\top}\bm A a
    =(\bm A^{\top}\tilde{\bm D})^{\top}a
    =\bm 0.
\end{equation*}
\end{proof}
For each unit, choose an observed base period $b_i$, taking
$b_i=G_i-1$ for ever-treated units and any observed period for
never-treated units. Define
\begin{equation}
    \Delta_{b_i}Y_{it}^{\obs} = Y_{it}^{\obs}-Y_{ib_i}^{\obs},
    \qquad \Delta\bm Y = \bm Y-\bm Y^{\mathrm{base}}, \qquad \bm Y^{\mathrm{base}}_{it} = Y_{ib_i}^{\obs}.    
\end{equation}
For an ever-treated unit observed at event time $l=t-G_i$,
\begin{equation}
    \Delta_{b_i}Y_{it}^{\obs} = \Delta^{l+1}Y_{it}^{\obs} = Y_{it}^{\obs}-Y_{i,G_i-1}^{\obs}.
\end{equation}
Since $\bm Y^{\mathrm{base}}$ is constant within each unit,
Lemma~\ref{lem:app_twfe_orth} implies $\tilde{\bm D}^{\top}\bm Y^{\mathrm{base}}=\bm 0$. Therefore, $\tilde{\bm D}^{\top}\Delta\bm Y = \tilde{\bm D}^{\top}\bm Y$, and equation~\eqref{eq:app_decomp_fwl} can be written as
\begin{equation}
\hat\beta = (\tilde{\bm D}^{\top}\tilde{\bm D})^{-1} \tilde{\bm D}^{\top}\Delta\bm Y, \qquad\text{or equivalently}\qquad \hat\tau_{l^*} = w^{\top}\Delta\bm Y = \sum_{(i,t)\in\Ical} w_{it}\Delta_{b_i}Y_{it}^{\obs}.
\label{eq:app_twfe_main}
\end{equation}
Thus, the TWFE coefficient can be written as a weighted linear
combination of within-unit outcome differences relative to
unit-specific base periods.

Splitting Equation~\eqref{eq:app_twfe_main} by event time and reversing the signs of the comparison-component weights gives the treatment-minus-control contrast we show in the main paper. Observations with $t-G_i=-1$ drop out, since $\Delta^{t-G_i+1}Y^{\obs}_{it}=0$ by convention. Observations with $t-G_i=l^*$ form the treatment component, and the rest form the comparison component, reported in Equation~\eqref{equation_lmw_tau}. The following proposition demonstrates the exact, finite-sample balance properties of the
implied weights.
\begin{proposition}
\label{prop:app_twfe_weights}
The implied weights $w$ in \eqref{eq:app_decomp_w} satisfy:
\begin{enumerate}
\item[(i)] \emph{(Normalization and balance of the treatment indicators.)} For every
event-time column $l$ of $\bm D$,
\[
\sum_{i=1}^{n}\sum_{t=1}^{T} w_{it}\,\mathbbm{1}\{t-G_i=l\}=\mathbbm{1}\{l=l^*\}.
\]
\item[(ii)] \emph{(Exact covariate balance.)} $\bm F^{\top}w=\bm 0$; that is,
\[
\sum_{t} w_{it}=0\ \ \text{for every unit }i,\qquad
\sum_{i} w_{it}=0\ \ \text{for every period }t,\qquad
\sum_{i=1}^{n}\sum_{t=1}^{T} w_{it}\,\bX_{it}=\bm 0 .
\]
\end{enumerate}
\end{proposition}

\begin{proof}
First we show that $w$ lies in $\operatorname{col}(M)$, equivalently $Mw=w$. Indeed, since $M^{2}=M$ we have $\tilde{\bm D}=M\bm D=M\tilde{\bm D}$, so every column of $\tilde{\bm D}$ lies in $\operatorname{col}(M)$, and hence so does their linear combination $w=\tilde{\bm D}(\tilde{\bm D}^{\top}\tilde{\bm D})^{-1}\bm{u}_{l^*}$.

For (ii), recall that $M$ is the orthogonal projection onto
$\operatorname{col}(\bm F)^{\perp}$, so $\operatorname{col}(M)=\operatorname{col}(\bm F)^\perp$. The membership $w\in\operatorname{col}(M)=\operatorname{col}(\bm F)^{\perp}$ is the same statement as orthogonality of $w$ to every column of $\bm F$, i.e.\ $\bm F^{\top}w=\bm 0$. Explicitly, using $Mw=w$, $M=M^{\top}$, and $M\bm F=\bm 0$,
\[
\bm F^{\top}w=\bm F^{\top}(Mw)=(M\bm F)^{\top}w=\bm 0 .
\]
Because
\[
\operatorname{col}(\bm A),
\operatorname{col}(\bm B),
\operatorname{col}(\bm X)
\subseteq
\operatorname{col}(\bm F),
\]
the identity $\bm F^{\top}w=\bm 0$ implies $\bm A^{\top}w=\bm 0$, $\bm B^{\top}w=\bm 0$, and $\bm X^{\top}w=\bm 0$. The $j$-th entry of $\bm A^{\top}w$ is $\sum_{t} w_{jt}$, the $s$-th entry of $\bm B^{\top}w$ is $\sum_{i} w_{is}$, and $\bm X^{\top}w=\sum_{i,t}w_{it}\bX_{it}$, which are the three displayed identities. Since $\bm 1=\bm A\bm 1_n$ is a column of $\operatorname{col}(\bm A)$, the total weight $\sum_{i,t}w_{it}=0$ as well. For (i), the same $Mw=w$ gives
\[
\bm D^{\top}w
=\bm D^{\top}(Mw)
=(M\bm D)^{\top}w
=\tilde{\bm D}^{\top}w
=\tilde{\bm D}^{\top}\tilde{\bm D}\,(\tilde{\bm D}^{\top}\tilde{\bm D})^{-1}\bm{u}_{l^*}
=\bm{u}_{l^*}.
\]
The $l$-th entry of $\bm D^{\top}w$ is $\sum_{i,t}w_{it}\mathbbm{1}\{t-G_i=l\}$ and the $l$-th entry of $\bm{u}_{l^*}$ is $\mathbbm{1}\{l=l^*\}$, giving (i).
\end{proof}

Property~(i) shows the weights place total mass $1$ on cells observed at event time $l^*$ and zero net mass on every other event-time bin, which represent the auxiliary treatment indicators. Property~(ii) shows the weights are exactly balanced, in finite samples, on the unit fixed effects, the time fixed effects, and the covariates. Equivalently, the treatment component and the control component of the contrast carry identical weighted means of every TWFE variable.

\subsection{Exact decomposition of the interaction-weighted estimator by \citet{sun2021estimating}}
\label{sec:app_sa_decomp_proof}

The same argument applies to the interaction-weighted estimator of
Equation~\eqref{equation_SA_PT}, with the event-time design matrix $\bm D$ replaced by the
cohort-by-event-time interaction design $\bm D^{\mathrm{IW}}\in\R^{N\times K}$, whose
columns are indexed by pairs $(g,l)$ with $g\neq\infty$, $l\neq-1$, and
\[
  \bm D^{\mathrm{IW}}_{it,(g,l)}=\mathbbm{1}\{G_i=g,\ t-G_i=l\}.
\]
The controls $\bm F=[\,\bm A\ \ \bm B\,]$ (with $\bm X$ appended if the specification includes covariates), the projection $M=I-P_{\bm F}$, and the residualization $\tilde{\bm D}^{\mathrm{IW}}=M\bm D^{\mathrm{IW}}$ are as above. The reference period $l=-1$ is omitted, and the never-treated cohort serves as the comparison group so that $(\tilde{\bm D}^{\mathrm{IW}})^{\top}\tilde{\bm D}^{\mathrm{IW}}$ is invertible. By Frisch--Waugh--Lovell, $\hat\tau_{t_1,l^*}=w^{\top}\bm Y$ with
\begin{equation*}
    w=\tilde{\bm D}^{\mathrm{IW}}\big((\tilde{\bm D}^{\mathrm{IW}})^{\top}
  \tilde{\bm D}^{\mathrm{IW}}\big)^{-1}\bm{u}_{(t_1,l^*)}.
\end{equation*}
Lemma~\ref{lem:app_twfe_orth} applies directly, and the proof of
Proposition~\ref{prop:app_twfe_weights} carries over with $\bm D^{\mathrm{IW}}$ replacing $\bm D$. Hence, by the same base-period invariance argument, $w^{\top}\bm Y=w^{\top}\Delta \bm Y$. Moreover,
\begin{equation*}
  (\bm D^{\mathrm{IW}})^{\top}w=\bm{u}_{(t_1,l^*)},\qquad \bm F^{\top}w=\bm 0,
\end{equation*}
so $\sum_{i,t}w_{it}\,\mathbbm 1\{G_i=g,\ t-G_i=l\}=\mathbbm 1\{(g,l)=(t_1,l^*)\}$ for every interaction cell, and the unit, time, and covariate balance of Proposition~\ref{prop:app_twfe_weights}(ii).

Splitting $\hat{\tau}_{t_1,l^*} =\sum_{i,t}w_{it}\Delta Y_{it}$
by cohort-by-event-time gives Equation~\eqref{equation_SA_PT_decomp}. Reference-period differences are zero by definition and drop out. The target cell receives total weight one, whereas every non-target treated cohort-by-event-time cell receives zero weight. The omitted comparison cohort and the reference-period observations supply the remaining outcome weights. 

\paragraph{Closed form under the saturated IW specification.}
The cell-level result simplifies further in a balanced panel when the never-treated cohort is the comparison group and the regression contains a separate indicator for every observed treated cohort-by-time cell except the reference period $l=-1$. Let
\begin{equation*}
    t_b=t_1-1,  \qquad  t_y=t_1+l^*,    \qquad  n_{t_1}=\sum_i\mathbbm{1}\{G_i=t_1\},   \qquad
    n_\infty=\sum_i\mathbbm{1}\{G_i=\infty\}. 
\end{equation*}
Define $\bm q\in\R^N$ by
\begin{equation}
\label{eq:app_iw_q}
    q_{it} = \left[ \mathbbm{1}\{G_i=t_1\} - \frac{n_{t_1}}{n_\infty}\mathbbm{1}\{G_i=\infty\} \right] \left[ \mathbbm{1}\{t=t_y\} - \mathbbm{1}\{t=t_b\} \right].
\end{equation}
Let $\bm d\in\R^N$ denote the target-cell indicator vector, with
$(i,t)$ entry $ d_{it} = \mathbbm{1}\{G_i=t_1,\ t=t_y\}$. Let $\bm H$ collect the unit and time fixed effects and all columns of
$\bm D^{\mathrm{IW}}$ other than the target column $\bm d$. By construction, $\bm H^\top \bm q =\bm 0$. Under the saturated specification, $\bm q \in\operatorname{col}([\bm H\ \bm d])$, and $\bm d^\top \bm q =n_{t_1}$. The Frisch--Waugh--Lovell weights for the target coefficient are therefore
\begin{equation}
\label{eq:app_iw_explicit_weights}
    w^{\mathrm{raw}}_{it} 
    = \frac{q_{it}}{n_{t_1}}
    = \left[ \frac{\mathbbm{1}\{G_i=t_1\}}{n_{t_1}} - \frac{\mathbbm{1}\{G_i=\infty\}}{n_\infty} \right] \left[ \mathbbm{1}\{t=t_y\} - \mathbbm{1}\{t=t_b\} \right].
\end{equation}
Thus, only observations from the target and never-treated cohorts at $t_y$ and $t_b$ receive nonzero raw-outcome weights. Substitution into $\widehat{\tau}_{t_1,l^*} =(w^{\mathrm{raw}})^\top\bm Y$ gives 
\begin{align}
    \widehat{\tau}_{t_1,l^*}
    &= \frac{1}{n_{t_1}} \sum_{i:G_i=t_1} \left( Y_{it_y}^{\obs}-Y_{i,t_1-1}^{\obs} \right) - \frac{1}{n_\infty} \sum_{i:G_i=\infty} \left( Y_{it_y}^{\obs}-Y_{i,t_1-1}^{\obs}
    \right) \notag\\
    &= \frac{1}{n_{t_1}} \sum_{i:G_i=t_1} \Delta^{l^*+1}Y_{it_y}^{\obs} - \frac{1}{n_\infty} \sum_{i:G_i=\infty} \Delta^{l^*+1}Y_{it_y}^{\obs}.
\label{eq:app_iw_did}
\end{align}
Hence, under the saturated specification, the cohort-specific IW
coefficient is exactly the difference in mean outcome changes between the target cohort and the never-treated cohort. In particular, every individual observation in a non-target treated cohort receives weight zero.

\subsection{Exact decomposition of the doubly robust estimator}
\label{sec:app_dr_decomp_proof}

The doubly robust estimator of Equation~\eqref{equation_DR_callaway} is not a regression
coefficient, so the projection argument above does not apply directly. Its weighting representation instead follows from the linearity of the outcome regression, as in Theorem~3 of \citet{chattopadhyay2023implied}.

Write $\Delta Y_{i} =\Delta^{t_y-t_1+\kappa+1}Y^{\obs}_{it_y}=Y^{\obs}_{it_y}-Y^{\obs}_{it_1-\kappa-1}$ for the within-unit difference used by the estimator, and let $\tilde\bX_i=(1,\bX_i^{\top})^{\top}$ denote the covariates with an intercept appended. Suppose the fitted outcome model is linear, $\widehat{m}_{t_1, t_y, \kappa} (\bX_i)=\tilde\bX_i^{\top}\hat\gamma$, with $\hat\gamma$ estimated by least squares among the never-treated,
\begin{equation}
\label{eq:app_dr_gamma}
    \hat\gamma = \bm S^{-1}\sum_{j:G_j=\infty}\tilde\bX_j \Delta Y_{j}, \qquad \bm S = \sum_{j:G_j=\infty}\tilde\bX_j\tilde\bX_j^{\top}.
\end{equation}
Write $\nu_i = n^{-1}\widehat{q}_{it_1}^{\text{never}}$ for the normalized propensity-score weights of the never-treated, which satisfy $\sum_{i:G_i=\infty}\nu_i=1$, and $w_i^{\text{treat}} = n^{-1}\widehat{q}_{it_1}^{\text{treat}} = \mathbbm{1}\{G_i=t_1\}/\sum_j \mathbbm{1}\{G_j=t_1\}$.

\begin{proposition}
\label{prop:app_dr_weights}
Under the linear outcome model \eqref{eq:app_dr_gamma}, the estimator \eqref{equation_DR_callaway} admits the representation \eqref{equation_DR_decomp} with never-treated weights
\begin{equation}
\label{eq:app_dr_w}
    w_i^{\text{never}} = \nu_i + \big(\bar{\tilde\bX}^{\text{treat}} - \bar{\tilde\bX}^{\nu}\big)^{\top}\bm S^{-1}\tilde\bX_i,
\qquad i:G_i=\infty,
\end{equation}
where $\bar{\tilde\bX}^{\text{treat}}=\sum_{i:G_i=t_1}w_i^{\text{treat}}\tilde\bX_i$ and $\bar{\tilde\bX}^{\nu}=\sum_{i:G_i=\infty}\nu_i\tilde\bX_i$. These weights satisfy
\begin{equation}
\label{eq:app_dr_balance}
\sum_{i:G_i=\infty} w_i^{\text{never}}\,\tilde\bX_i
= \sum_{i:G_i=t_1} w_i^{\text{treat}}\,\tilde\bX_i ,
\end{equation}
so both components have equal weighted covariate means, and both sets of weights sum to one.
\end{proposition}

\begin{proof}
Substituting the linear outcome model into \eqref{equation_DR_callaway},
\begin{equation*}
\begin{aligned}
    \widehat{\ate}^{\Scal_t}_{t_y}(t_1,\infty)
    &= \sum_{i:G_i=t_1} w_i^{\text{treat}}\big(\Delta Y_{i}-\tilde\bX_i^{\top}\hat\gamma\big)
    - \sum_{i:G_i=\infty} \nu_i\big(\Delta Y_{i}-\tilde\bX_i^{\top}\hat\gamma\big) \\ 
    & = \sum_{i:G_i=t_1} w_i^{\text{treat}}\Delta Y_{i}
    - \sum_{i:G_i=\infty} \nu_i\Delta Y_{i}
    - \big(\bar{\tilde\bX}^{\text{treat}} - \bar{\tilde\bX}^{\nu}\big)^{\top}\hat\gamma .
\end{aligned}
\end{equation*}
Replacing $\hat\gamma$ by \eqref{eq:app_dr_gamma} turns the last term into a linear
combination of the never-treated differences,
\begin{equation*}
    \big(\bar{\tilde\bX}^{\text{treat}} - \bar{\tilde\bX}^{\nu}\big)^{\top}\bm S^{-1}
    \sum_{i:G_i=\infty}\tilde\bX_i\Delta Y_{i}
    = \sum_{i:G_i=\infty}\Big[\big(\bar{\tilde\bX}^{\text{treat}} - \bar{\tilde\bX}^{\nu}\big)^{\top}\bm S^{-1}\tilde\bX_i\Big]\Delta Y_{i},
\end{equation*}
which combines with the second sum to give \eqref{eq:app_dr_w}. The treated weights are
unchanged.

For \eqref{eq:app_dr_balance}, multiply \eqref{eq:app_dr_w} by $\tilde\bX_i$ and sum:
\begin{equation*}
    \sum_{i:G_i=\infty} w_i^{\text{never}}\tilde\bX_i
    = \bar{\tilde\bX}^{\nu} + \bm S \bm S^{-1}\big(\bar{\tilde\bX}^{\text{treat}} - \bar{\tilde\bX}^{\nu}\big)
    = \bar{\tilde\bX}^{\text{treat}},
\end{equation*}
using $\sum_{i:G_i=\infty}\tilde\bX_i\tilde\bX_i^{\top}=\bm S$. Reading the first coordinate of \eqref{eq:app_dr_balance}, which corresponds to the intercept in $\tilde\bX_i$, gives $\sum_{i:G_i=\infty} w_i^{\text{never}} = \sum_{i:G_i=t_1} w_i^{\text{treat}} = 1$.
\end{proof}
Equation~\eqref{eq:app_dr_w} separates the never-treated weight into a propensity-score base weight $\nu_i$ and a correction that is linear in $\tilde\bX_i$. Unit-fixed effects are balanced by
construction through the within-unit differencing that produces $\Delta Y_i$, and time-fixed effects are balanced because both components use the same two periods, $t_y$ and $t_1-\kappa-1$. The correction is not sign-constrained, so $w_i^{\text{never}}$ can be negative even when every $\nu_i$ is positive. Moreover, $\bm S$ is formed from the never-treated units alone, so a cohort whose treated units sit far from the never-treated in covariate space can produce a large correction, with extreme weights on the never-treated units with the most extreme $\tilde\bX_i$. 

\subsection{TWFE estimator for the general estimand $\ate^\Pcal_{t_y}(t_1, t_{\mathbf{p}})$}
\label{sec:app_twfe_general_correction}

We now compare the dynamic TWFE coefficient with the estimand defined under a stochastic reference regime, $\ate^\Pcal_{t_y}(t_1, t_{\mathbf{p}})$. Let $l^*=t_y-t_1$. Suppose the dynamic TWFE specification correctly describes the mean potential outcomes in the target population. In particular, for every finite initiation time $g\in\Tcal$ relevant to the target estimand,
\begin{equation*}
    \E_{\P}\{Y_{it_y}(g)\} = \bar{\alpha}_{\Pcal} + \beta_{t_y} + \tau_{t_y-g},
\end{equation*}
where
$\bar{\alpha}_{\Pcal}=\E_{\P}(\alpha_i)$, while
\begin{equation*}
    \E_{\P}\{Y_{it_y}(\infty)\}
    = \bar{\alpha}_{\Pcal} + \beta_{t_y}.
\end{equation*}
We assume that the event time $t_y-g$ is included for every $g$ with $p_g>0$. We normalize the reference period as $\tau_{-1}=0$. Because $p_\infty+\sum_{g\in\Tcal}p_g=1$, the general estimand satisfies
\begin{equation}
\label{equation_TWFE_general_inconsistency}
\begin{aligned}[b]
    \ate^\Pcal_{t_y}(t_1,t_{\mathbf p}) 
    &= \E_{\P}\{Y_{it_y}(t_1)\} - p_\infty\E_{\P}\{Y_{it_y}(\infty)\} - \sum_{g\in\Tcal} p_g\E_{\P}\{Y_{it_y}(g)\} \\
    &= \bar{\alpha}_{\Pcal} + \beta_{t_y} + \tau_{t_y-t_1} - p_\infty \left( \bar{\alpha}_{\Pcal} + \beta_{t_y} \right) - \sum_{g\in\Tcal} p_g \left( \bar{\alpha}_{\Pcal} + \beta_{t_y} + \tau_{t_y-g} \right) \\ 
    &= \tau_{t_y-t_1} - \sum_{g\in\Tcal} p_g\tau_{t_y-g}.
\end{aligned}
\end{equation}
Thus, $\tau_{t_y-t_1}$ corresponds to $\ate^\Pcal_{t_y}(t_1,t_{\mathbf p})$ only when $\sum_{g\in\Tcal}p_g\tau_{t_y-g}=0$. When all relevant event-time coefficients are identified, the corresponding plug-in estimator is
\begin{equation*}
    \widehat{\tau}_{t_y-t_1} - \sum_{g\in\Tcal} p_g\widehat{\tau}_{t_y-g}.
\end{equation*}
As a special case, when $p_\infty=1$, Equation~\eqref{equation_TWFE_general_inconsistency} reduces to
\begin{equation}
\label{equation_TWFE_standard_consistency}
    \ate^\Pcal_{t_y}(t_1,\infty) = \tau_{t_y-t_1}.
\end{equation}
Therefore, under the correctly specified dynamic potential outcome
model, the event-time coefficient $\tau_{t_y-t_1}$ corresponds to the conventional never-treated estimand.

\subsection{The dynamic TWFE estimator as a robust weighting estimator}
\label{sec:app_twfe_rw_equivalence}

Section~\ref{sec3:weight_estimator} claims that our weighting (RW) approach recovers the dynamic TWFE regression estimator as a special case. We justify this claim in Proposition~\ref{prop:app_twfe_rw}. Specifically, the TWFE estimator can be considered as implicitly solving a balancing problem of the form \eqref{minimal_1}, with a particular information set, a particular set of balancing constraints, and the relaxed positivity constraint on weights.

\begin{proposition}
\label{prop:app_twfe_rw}
Consider the weighting problem
\begin{equation}
\label{eq:app_twfe_rw_problem}
    \min_{w \in \R^{N}} \ \sum_{(i,t) \in \Ical} w_{it}^{2} \quad \text{subject to} \quad \bm D^{\top}w = \bm{u}_{l^*}, \quad \bm F^{\top}w = \bm 0,
\end{equation}
in which the information set is the full panel $\Ical$, the balancing constraints impose exact balance on the unit and time indicators and any additional regressors, normalize the target event-time indicator to one, and set the weighted totals of the non-target event-time indicators to zero. The weights are unrestricted and may be negative. The unique solution of \eqref{eq:app_twfe_rw_problem} is the vector $w$ in \eqref{eq:app_decomp_w}, and hence $\sum_{(i,t)\in\Ical} w_{it}Y^{\obs}_{it} = \hat\tau_{l^*}$.
\end{proposition}

\begin{proof}
The constraints define an affine subspace of $\R^{N}$, and the objective is strictly convex, so the solution is unique. To identify the minimizer, write any feasible vector as $w=w_0+v$, where $w_0$ is the vector in \eqref{eq:app_decomp_w} and $v$ satisfies $\bm D^{\top}v=\bm 0$ and $\bm F^{\top}v=\bm 0$.
Thus, $v\perp\operatorname{col}([\,\bm F\ \bm D\,])$. By Proposition~\ref{prop:app_twfe_weights}, $w_0 = Mw_0 \in \operatorname{col}(\bm F)^{\perp}$ and $w_0 \in \operatorname{col}(\tilde{\bm D}) \subseteq \operatorname{col}([\,\bm F\ \ \bm D\,])$, so $w_0^{\top}v=0$ and $\|w\|^{2}=\|w_0\|^{2}+\|v\|^{2} \geq \|w_0\|^{2}$, with equality only at $v=\bm 0$.
\end{proof}

Thus, the RW estimator and the TWFE estimators minimize the same dispersion criterion. The two methods differ in three aspects. First, the information set: the TWFE estimator (Equation \eqref{eq:app_twfe_rw_problem}) uses the full panel, whereas the RW estimator (solving Equation \eqref{minimal_1} or \eqref{minimal_2}) uses only the observations that the assumptions justify. Second, the balancing constraints: the TWFE estimator additionally balances the non-target event-time indicators, which is what draws the delay and dissipation observations into the contrast, whereas RW approach balances the basis functions $\bm B(\cdot)$ in $\Mcal$ relative to the target sample, which does not include the non-target event time indicators. Third, the positivity restriction: the TWFE estimator leaves the weights unrestricted, so the resulting estimator is not sample-bounded, whereas the RW estimator can constrain the weights to be non-negative and produce the sample-bounded estimator. Thus, the TWFE estimator and the RW estimator differ in which observations they use (information set) and which variables they balance.

Proposition~\ref{prop:app_twfe_rw} is an algebraic identity, so it should hold exactly in any panel. Table~\ref{tab:supp_twfe_rw} confirms it in the Castle Doctrine data. We focus on $\ate^{\Scal_{t}}_{2008}(2006,\infty)$, each time solving problem~\eqref{eq:app_twfe_rw_problem} over a larger information set. The first row uses only the ideal contrast; the second adds the observations justified by invariance to time shifts; the third adds the not-yet-treated observations justified by no anticipation; the fourth adds those justified by delayed onset; and the fifth adds those justified by effect dissipation, at which point the information set is the entire panel. Balancing constraints are expanded with the information set. Specifically, the non-target event-time indicators are balanced once observations at those event times enter the comparison component. The positivity constraint on weights are relaxed throughout.

The final row reproduces the estimates from the dynamic TWFE regression exactly. The intermediate rows show that the additional observations that we include affect the point estimate. The ideal contrast is $0.064$, roughly half of the TWFE estimate. The change is also not monotone; adding the invariance group pushes the estimate above the TWFE estimate, whereas adding the not-yet-treated observations brings it below.

\begin{table}[htbp]
\centering
\caption{\label{tab:supp_twfe_rw}
The dynamic TWFE estimator as a robust weighting estimator. Each row reports $\widehat{\ate}^{\Scal_{t}}_{2008}(2006,\infty)$ from the minimum dispersion problem~\eqref{eq:app_twfe_rw_problem}, solved over the progressively larger information set and with the relaxed nonnegativity constraint on weights. Year indicators are balanced in every row, and the non-target event-time indicators are balanced once observations at those event times enter the estimation. $N_{\text{used}}$ counts the observations the estimator uses, out of $N_{\text{total}} = 550$. Standard errors in parentheses come from $500$ state-level bootstrap replications.}
\resizebox{.7\textwidth}{!}{%
\begin{tabular}{@{}l c c c c}
\toprule
 & Event-time & Observations & Balance & \\
Information set & balance & $N_{\text{used}}/N_{\text{total}}$ & constraints & Estimate \\
\midrule
Ideal contrast &  & $84/550$ & 45 & $0.064\ (0.079)$ \\
\quad $+$ Invariance &  & $243/550$ & 57 & $0.119\ (0.057)$ \\
\quad $+$ No anticipation & \checkmark & $475/550$ & 70 & $0.106\ (0.063)$ \\
\quad $+$ Delayed onset & \checkmark & $517/550$ & 72 & $0.115\ (0.066)$ \\
\quad $+$ Effect dissipation & \checkmark & $550/550$ & 75 & $0.115\ (0.066)$ \\
\midrule
Dynamic TWFE & \checkmark & $550/550$ & 75 & $0.115\ (0.066)$ \\
\bottomrule
\end{tabular}
}
\end{table}

%% file: supp4_diagnostics.tex

\section{Diagnostics}
\label{sec:app_diagnostics}
This section provides further details on the diagnostics introduced in Section~\ref{sec:diagnostic_metrics}. Section~\ref{sec:app_diag_info} discusses the information measures, Section~\ref{sec:app_diag_influence} the influence measure, and Section~\ref{sec:app_diag_balance} covariate balance, sign reversal, and the implied target profile. Throughout, $w_{it}$ denotes the implied weight assigned to observation $(i,t)$ by one of the estimators in Section~\ref{sec:app_decomp}. We use $\Gcal$ to denote a partition of the observations into groups. In the case study, these are the groups defined in Table~\ref{tab:app_groups}, but the same diagnostics can be applied to other groupings, such as by unit, calendar period, or treatment cohort. We write $g\in\Gcal$ for a generic group. In Equation~\eqref{equation_SA_PT}, $g$ indexes a treatment cohort.

\subsection{Weight dispersion and information}
\label{sec:app_diag_info}
For each group $g\in\Gcal$, define its absolute weight mass and squared weight mass as 
\begin{equation*} 
    S_g = \sum_{i=1}^{n}\sum_{t=1}^{T} |w_{it}|\,\mathbbm{1}\{(i,t)\in g\}, \qquad 
    Q_g = \sum_{i=1}^{n}\sum_{t=1}^{T} w_{it}^{2}\,\mathbbm{1}\{(i,t)\in g\}. 
\end{equation*} 
Let $S=\sum_{g\in\Gcal}S_g$ and $Q=\sum_{g\in\Gcal}Q_g$. The group-wise effective sample size and squared-weight share defined in Section~\ref{sec:diagnostic_metrics} can then be written as
$n_{\text{eff},g} = S_g^2/Q_g$ and the squared-weight share is $D_g = Q_g/Q$. If we apply the same formula to the full set of weights across the panel, regardless of any partition, the overall effective sample size is
\begin{equation}
\label{eq:app_ess_overall}
    n_{\text{eff}} = \frac{S^{2}}{Q} = \frac{\left(\sum_{i=1}^{n}\sum_{t=1}^{T} |w_{it}|\right)^{2}}{\sum_{i=1}^{n}\sum_{t=1}^{T} w_{it}^{2}},
\end{equation}
which measures how many equally weighted observations (effective samples) the estimator draws on in total.

The squared-weight shares are additive by construction, $\sum_{g \in \Gcal} D_g = 1$. The group-wise effective sample sizes are not, and in general $n_{\text{eff}}$ is smaller than $\sum_{g \in \Gcal} n_{\text{eff},g}$, with the two equal only when $S_g/Q_g$ is the same in every group. We define the information share as 
\begin{equation}
    p_{\mathrm{info},g} = \frac{n_{\mathrm{eff},g}} {\sum_{g'\in\Gcal}n_{\mathrm{eff},g'}}. 
\end{equation}
The information share is used to show how information is distributed across groups, whereas the squared-weight share summarizes the concentration of the estimator's weights. We also report the distribution of the absolute implied weights within each group, including the number of observations $N_g$, the minimum, quartiles, mean, upper percentiles, maximum, and total weight mass $S_g$. These summaries show features of the weight distribution that are not visible from $n_{\mathrm{eff},g}$ or $D_g$ alone. 

\subsection{Influence}
\label{sec:app_diag_influence}

Recall that $\operatorname{PE}_{it}= \hat{\tau}_{-(it)} - \hat{\tau}$, the change in the point estimate when observation $(i,t)$ is removed and the estimator is recomputed. For TWFE, we remove each non-reference state-year observation in turn and refit the full regression, including the state and year fixed effects and event-time indicators. The reference observations remain in the estimation sample. For the doubly robust and robust weighting estimators, removing an observation changes the weights and the weighted average of outcome measures. Therefore, we re-estimate the nuisance functions or re-solve the balancing problem, so that the change in the weights induced by the deletion is captured along with the change in the estimate. Observations with large implied weights are not necessarily influential, and the ranking in Figure~\ref{fig:twfe_influence_agg} does not reproduce the ranking of the weights. Removing a single observation isolates the contribution of one observation. Deleting a set of correlated cells, such as a unit's whole trajectory, requires the group-deletion diagnostics of \citet{banerjee1997influence} and \citet{zhu2012perturbation}, which we leave for future work.

\subsection{Covariate balance, sign reversal, and the implied target profile}
\label{sec:app_diag_balance}

\paragraph{Covariate balance.}
Each estimator of Section~\ref{sec:app_decomp} can be expressed as a weighted contrast between a treatment component and a control component. For a variable $x$ measured on the observations, we compare its weighted means across the two components with the standardized mean difference
\begin{equation}
\label{eq:app_smd}
    \text{SMD}(x)= \frac{ \bar{x}_{w,\text{treatment}}-\bar{x}_{w,\text{control}} }{\sqrt{(s_{\text{treatment}}^{2}+s_{\text{control}}^{2})/2}},
\end{equation}
where $\bar{x}_{w,\text{treatment}}$ and $\bar{x}_{w,\text{control}}$ are the weighted covariate means, and $s_{\text{treatment}}$ and $s_{\text{control}}$ are the respective unweighted sample standard deviations. The denominator is computed before weighting, so it does not depend on the weights. 

\paragraph{Sign reversal.}
The weights of the TWFE, interaction-weighted, and doubly robust estimators are unrestricted in sign, as shown in Sections~\ref{sec:app_twfe_decomp_proof} and~\ref{sec:app_dr_decomp_proof}. When an observation in the treatment component receives a negative weight, an increase in its outcome lowers the weighted mean of that component, so a larger treatment effect for that observation moves the estimate in the opposite direction \citep{small2017instrumental}. For each component, we report the number of observations with negative weights and the share of that component's total absolute weight assigned to those observations. The robust weighting estimator is not subject to sign reversal when the non-negativity constraint in \eqref{minimal_1} is applied.

\paragraph{Implied target profile.}
The weights also describe the population the estimate speaks to. Following \citet{chattopadhyay2020balancing}, we define the implied target profile of a component as its vector of weighted covariate means, and compare it with the covariate means of the study sample. For the TWFE estimator this profile need not resemble any recognizable population, since the weights are chosen to satisfy the balance conditions of Proposition~\ref{prop:app_twfe_weights} rather than to represent a target explicitly. For the robust weighting estimator, the profile is set by the investigator through the balancing constraints, which are defined relative to the study sample in \eqref{minimal_1} and relative to the target sample in \eqref{minimal_gen1}.

%% file: supp5_case_study.tex

\section{Additional case study results}
\label{sec_app_further}

This section provides additional weight-based diagnostics for event study estimators. We examine covariate balance, the populations represented by the implied weights, differences in group-wise ESS across estimators, and the information contributed by different observation groups. Section~\ref{sec_app_case_setup} describes the data and estimator specifications. Sections~\ref{sec_app_exact_weight}-~\ref{sec_app_case_profile} present results for the Castle Doctrine study. 

\subsection{Data and specifications}
\label{sec_app_case_setup}

The panel contains observations for 50 states from 2000 through 2010, for a total of 550 state-year observations. Twenty-one states enacted Castle Doctrine laws between 2005 and 2009. One state enacted a law in 2005, thirteen in 2006, four in 2007, two in 2008, and one in 2009. The remaining 29 states did not enact such a law during this period. The outcome is the log annual state-level homicide rate.

The baseline covariates $\bX_i$ are the log number of police officers, log income, and log number of prisoners. All three are measured at the beginning of follow-up and are constant over time. They are absorbed by the state fixed effects in the TWFE and IW specifications, as discussed in Section~\ref{sec:app_decomp_setup}. In the DR estimator, they enter the propensity score and outcome models. In the RW estimator, they enter the balancing constraints described in Section~\ref{sec:app_rw_implementation}.

All four estimators use the year immediately before law enactment as the reference period, so $\kappa=0$. Figure~\ref{fig:event_study_plot_estimators} reports estimates for event times from $-4$ through $4$. Standard errors are obtained from a state-level bootstrap with 500 replications.

\subsection{Cohort-specific estimates}
\label{sec_app_case_granular_estimation}

The main text focuses on the aggregate event study parameter. Here we report the estimates of cohort-specific building blocks estimands $\ate_{t_y}^{\Scal_t}(t_1,\infty)$ with $l = t_y - t_1 = 2$ in Table~\ref{tab:granular_estimation}. The estimates vary considerably across enactment cohorts. The IW estimate ranges from $0.064$ for the 2006 cohort to $0.271$ for the 2007 cohort, whereas TWFE produces a single event-time coefficient of $0.115$. Some of this variation may reflect the small number of states in certain cohorts, but it also shows the heterogeneity that an aggregate estimate can conceal. Without covariate adjustment, DR and RW reproduce the IW estimates exactly. The final two columns report DR and RW estimates adjusted for the baseline covariates described in Section~\ref{sec_app_case_setup}. Standard errors are obtained from 500 state-level bootstrap replications.

\begin{table}[h!tbp]
\caption{\label{tab:granular_estimation} Cohort-specific estimates two years after Castle Doctrine enactment. TWFE produces a single event-time coefficient, $\hat{\tau}_2$, which is repeated across rows for comparison. Standard errors from 500 state-level bootstrap replications are reported in parentheses. Cohorts with fewer than two states are omitted.}
\centering
\resizebox{.7\textwidth}{!}{%
\begin{tabular}{r cccccc}
\toprule
& \multicolumn{6}{c}{Estimator} \\
\cmidrule(lr){2-7}
& TWFE & IW & DR & RW & DR & RW \\
& & & $\Fcal = \emptyset$ & $\Fcal = \emptyset$ & $\Fcal = \bX_{\text{baseline}}$ & $\Fcal = \bX_{\text{baseline}}$ \\
\midrule
$t_1 = 2006$ & $0.115$ & $0.064$ & $0.064$ & $0.064$ & $0.143$ & $0.105$ \\
& $(0.066)$ & $(0.079)$ & $(0.079)$ & $(0.079)$ & $(0.208)$ & $(0.134)$ \\[4pt]
$t_1 = 2007$ & $0.115$ & $0.271$ & $0.271$ & $0.271$ & $0.270$ & $0.249$ \\
& $(0.066)$ & $(0.104)$ & $(0.104)$ & $(0.104)$ & $(0.137)$ & $(0.124)$ \\[4pt]
$t_1 = 2008$ & $0.115$ & $0.071$ & $0.071$ & $0.071$ & $0.220$ & $0.209$ \\
& $(0.066)$ & $(0.058)$ & $(0.058)$ & $(0.058)$ & $(0.173)$ & $(0.161)$ \\[4pt]
\bottomrule
\end{tabular}
}
\end{table}

\subsection{Exact weights}
\label{sec_app_exact_weight}

\begin{figure}[h!tbp]
    \centering
    \includegraphics[width=\textwidth]{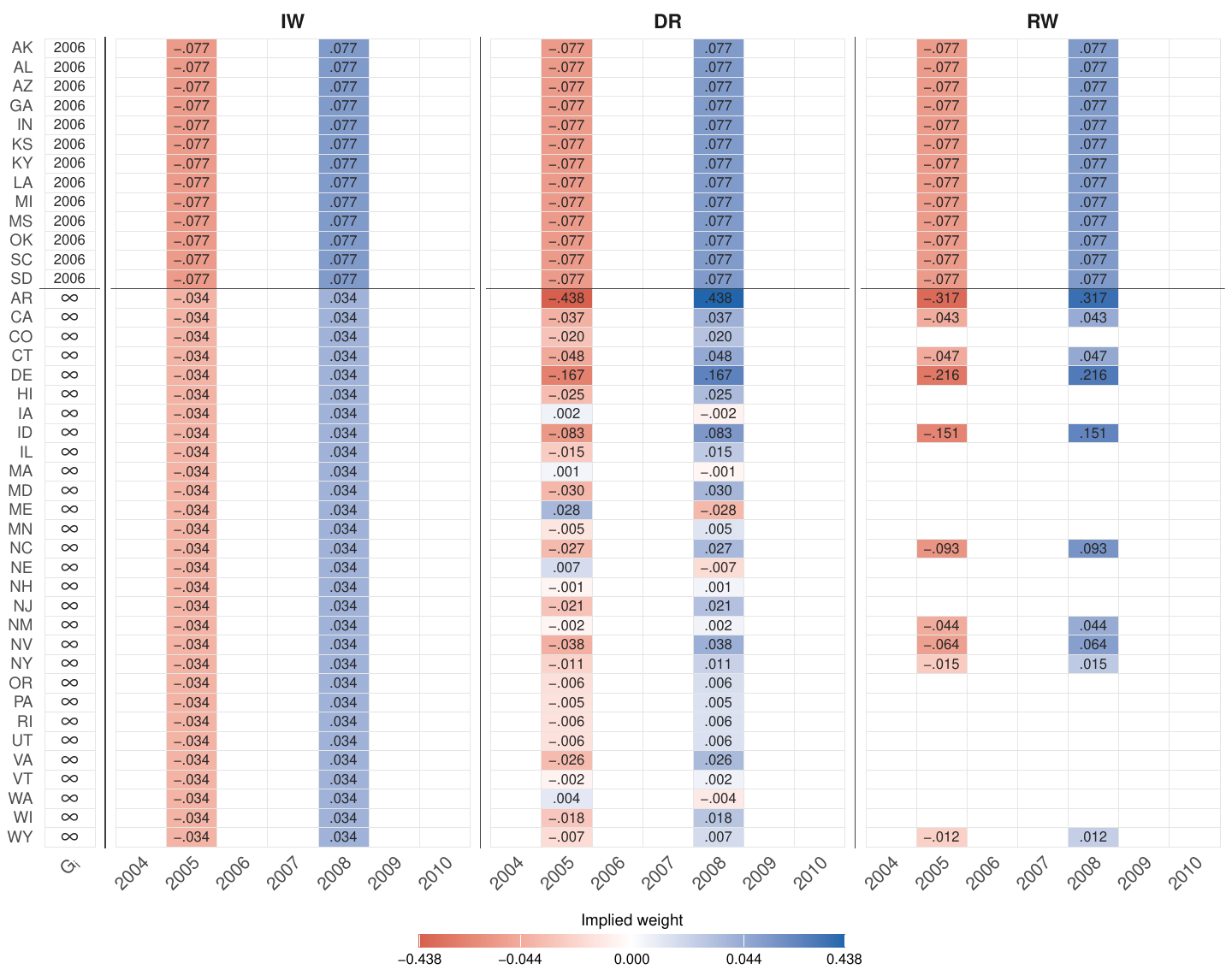}
    \caption{Implied weights assigned by the IW, DR, and RW estimators for the granular estimand $\ate^{\Scal_t}_{2008}(2006,\infty)$. The column $G_i$ reports each state's enactment year, and horizontal lines separate enactment cohorts. Blank cells have exactly zero weight; displayed values of $0.000$ are nonzero weights smaller than $0.0005$ in absolute value. }
    \label{app_fig:map_cell}
\end{figure}

Section~\ref{sec:case_study} reports the exact observation-level weights for the aggregate estimand. Figure~\ref{app_fig:map_cell} provides the corresponding weights for the cohort-specific estimand $\ate^{\Scal_t}_{2008}(2006,\infty)$ under IW, DR, and RW. These estimators use only observations from the target enactment cohort and never-treated states, with weights placed on the target outcome year and the reference year. All other observations receive zero weight and are omitted from the figure. For the cohort-specific estimand, 84 observations receive nonzero weight under IW and DR, compared with 46 under RW. For the aggregate estimand, the corresponding counts are 243 and 106.

IW assigns uniform weights to the never-treated states, whereas DR and RW reweight these states to adjust for differences in baseline covariates. RW sets more control weights to zero than DR, resulting in a sparser set of weights. It also constrains the weights within each component to be nonnegative and to sum to one, making the resulting estimate sample-bounded. 

\subsection{Covariate balance}
\label{sec_app_case_balance}

We assess balance using the standardized mean difference (SMD) defined in Section~\ref{sec:app_diag_balance} and Equation~\eqref{eq:app_smd}. The SMD is
the weighted mean in the treatment component minus the weighted mean in the control component, divided by the pooled unweighted standard deviation. 

Table~\ref{tab:comp_baseline} reports balance for the IW, DR, and RW estimators. For the granular estimand $\ate^{\Scal_t}_{2008}(2006,\infty)$, the treatment component contains within-state outcome changes from 2005 to 2008 for the 13 states that enacted a law in 2006. The control component contains the corresponding changes for the 29 never-treated states. Before weighting, the SMDs for log police officers, log income, and log prisoners are $0.228$, $-0.935$, and $1.190$, respectively. IW assigns uniform weights, so these differences are unchanged after weighting. DR achieves exact balance, while RW reduces all three absolute SMDs to $0.011$ or less. 

For the aggregate lag-2 estimand, the treatment component pools 20 states from the 2005--2008 enactment cohorts, observed two years after enactment. The corresponding baseline-covariate SMDs are $0.195$, $-1.016$, and $0.800$. Relative to the granular estimand, pooling cohorts reduces the differences in police officers and prisoners but increases the difference in income. IW again leaves these differences unchanged, DR reduces them to zero, and RW reduces each absolute SMD to $0.010$. The aggregate comparison covers outcome years 2007--2010. Before weighting, the treatment component is unevenly distributed across these years because the four cohorts contain 1, 13, 4, and 2 states, whereas the never-treated control observations are distributed equally across the four years. This produces SMDs of $-0.578$, $0.864$, $-0.118$, and $-0.398$. After weighting, IW, DR, and RW give the control component the same year distribution as the treatment component, reducing every calendar-year SMD to zero.

Table~\ref{tab:balance_twfe} summarizes balance under the dynamic TWFE implied weights. Before weighting, the largest absolute SMD is $0.552$ among the three baseline covariates, $0.206$ among the 50 state indicators, $1.487$ among the 11 calendar-year indicators, and $0.287$ among the 13 event-time indicators. The implied TWFE weights reduce all four maxima to zero because they exactly balance the variables in the regression. They also balance the baseline covariates because these covariates are time-invariant, so balancing state indicators will automatically balance the baseline covariates. These results have a different interpretation. The TWFE treatment and control components draw on observations throughout the panel, and the same treated state can appear in both components. Thus, the zero SMDs describe balance between the two components of the TWFE decomposition, rather than balance in the clean comparison between states that enacted a law and never-enacted states in Table~\ref{tab:comp_baseline}.

\begin{table}[h!tbp]
\centering
\caption{\label{tab:comp_baseline}
Standardized mean differences (SMDs) between the treatment and control components before and after weighting, by target estimand.}
\renewcommand{\arraystretch}{1.2}

\begin{subtable}{\textwidth}
\centering
\caption{\label{tab:comp_baseline_granular}
Target estimand: $\ate^{\Scal_t}_{2008}(2006, \infty)$.}
\resizebox{.6\textwidth}{!}{%
\begin{tabular}{lcccccc}
\toprule
 & \multicolumn{2}{c}{IW} & \multicolumn{2}{c}{DR} & \multicolumn{2}{c}{RW} \\
\cmidrule(lr){2-3} \cmidrule(lr){4-5} \cmidrule(lr){6-7}
Baseline covariate & Before & After & Before & After & Before & After \\
\midrule
l\_police & \phantom{-}0.228 & \phantom{-}0.228 & \phantom{-}0.228 & \phantom{-}0.000 & \phantom{-}0.228 & -0.011 \\
l\_income & -0.935 & -0.935 & -0.935 & \phantom{-}0.000 & -0.935 & -0.010 \\
l\_prisoner & \phantom{-}1.190 & \phantom{-}1.190 & \phantom{-}1.190 & \phantom{-}0.000 & \phantom{-}1.190 & \phantom{-}0.011 \\
\bottomrule
\end{tabular}%
}
\end{subtable}

\vspace{1em}

\begin{subtable}{\textwidth}
\centering
\caption{\label{tab:comp_baseline_aggregate}
Target estimand: $\ate^{\Scal_t}_{\mathrm{agg}}(l=2)$.}
\resizebox{.55\textwidth}{!}{%
\begin{tabular}{lcccccc}
\toprule
 & \multicolumn{2}{c}{IW} & \multicolumn{2}{c}{DR} & \multicolumn{2}{c}{RW} \\
\cmidrule(lr){2-3} \cmidrule(lr){4-5} \cmidrule(lr){6-7} 
Variable & Before & After & Before & After & Before & After \\
\midrule
\multicolumn{7}{l}{Baseline covariates} \\
\addlinespace[1pt]
l\_police & \phantom{-}0.195 & \phantom{-}0.195 & \phantom{-}0.195 & \phantom{-}0.000 & \phantom{-}0.195 & \phantom{-}0.010 \\
l\_income & -1.016 & -1.016 & -1.016 & \phantom{-}0.000 & -1.016 & -0.010 \\
l\_prisoner & \phantom{-}0.800 & \phantom{-}0.800 & \phantom{-}0.800 & \phantom{-}0.000 & \phantom{-}0.800 & \phantom{-}0.010 \\
\addlinespace[3pt]
\midrule
\multicolumn{7}{l}{Calendar year indicators} \\
\addlinespace[1pt]
2007 &  -0.578 & \phantom{-}0.000 & -0.578 & \phantom{-}0.000 & -0.578 & \phantom{-}0.000 \\
2008 & \phantom{-}0.864 & \phantom{-}0.000 & \phantom{-}0.864 & \phantom{-}0.000 & \phantom{-}0.864 & \phantom{-}0.000 \\
2009 & -0.118 & \phantom{-}0.000 & -0.118 & \phantom{-}0.000 & -0.118 & \phantom{-}0.000 \\
2010 & -0.398 & \phantom{-}0.000 & -0.398 & \phantom{-}0.000 & -0.398 & \phantom{-}0.000 \\
\bottomrule
\end{tabular}%
}
\end{subtable}
\vspace{3pt}
\begin{minipage}{.92\textwidth}
\footnotesize
l\_police: log police officers; l\_income: log income; l\_prisoner: log prisoners. All three covariates are measured at baseline and treated as time invariant.
\end{minipage}
\end{table}

\begin{table}[h!tbp]
\centering
\caption{\label{tab:balance_twfe} Balance of the dynamic TWFE implied weights. The two components partition the panel, so every regressor in the specification varies within both and has a well-defined standardized mean difference. Each row reports the number of variables in the block and the largest absolute difference across them. The target event-time indicator and the omitted reference period ($l=-1$) are excluded from the balance diagnostics.}
\renewcommand{\arraystretch}{1.2}
\resizebox{.72\textwidth}{!}{%
\begin{threeparttable}
\begin{tabular}{lccc}
\toprule
 & & \multicolumn{2}{c}{Largest absolute SMD} \\
\cmidrule(lr){3-4}
Variable block & Variables & Before weighting & After weighting \\
\midrule
Baseline covariates & 3 & \phantom{-}0.552 & \phantom{-}0.000 \\
State indicators & 50 & \phantom{-}0.206 & \phantom{-}0.000 \\
Calendar year indicators & 11 & \phantom{-}1.487 & \phantom{-}0.000 \\
Non-target event-time indicators & 13 & \phantom{-}0.287 & \phantom{-}0.000 \\
\bottomrule
\end{tabular}
\end{threeparttable}
}
\end{table}

\subsection{Weight dispersion}
\label{sec_app_case_dispersion}

Tables~\ref{tab:weight_dispersion_granular} and~\ref{tab:weight_dispersion_aggregate} report summary statistics of the absolute implied weights by observation group, for the granular and the aggregate lag-2 estimands.

For TWFE, the event-time-two observations receive the largest individual weights. The weights on the delay and dissipation groups are much smaller, but these 75 observations together carry $0.223$ in absolute weight. The 290 never-treated observations likewise have small individual weights but carry $1.560$ in total. Thus, many small TWFE weights add up to substantial weight outside the event-time-two observations. IW, DR, and RW weight the treated observations uniformly and differ in how they weight the controls. IW is uniform for the granular estimand and within each cohort comparison for the aggregate estimand. DR is the most concentrated, with maximum absolute control weights of $0.438$ and $0.285$ for the granular and aggregate estimands. The corresponding RW maxima are smaller, at $0.317$ and $0.114$, and many RW control weights are zero.

\begin{table}[h!tbp]
\caption{\label{tab:weight_dispersion_granular} Summary statistics of absolute implied weights by observation group for estimating $\ate^{\Scal_t}_{2008}(2006, \infty)$.}
\centering
\resizebox{.95\textwidth}{!}{%
\begin{tabular}{lccccccccc}
\toprule
 & $N$ & Min & \multicolumn{1}{p{1.4cm}}{\centering 25th\\Pctl.} & Median & Mean & \multicolumn{1}{p{1.4cm}}{\centering 75th\\Pctl.} & \multicolumn{1}{p{1.4cm}}{\centering 95th\\Pctl.} & Max. & $\sum|w|$ \\
\midrule
\multicolumn{10}{l}{\textbf{TWFE}} \\[3pt]
$l = 2$, $G_i = 2006$ (ideal) & 13 & 0.0462 & 0.0462 & 0.0462 & 0.0462 & 0.0462 & 0.0462 & 0.0462 & 0.6010 \\
$l = 2$, $G_i \neq 2006$ (invariance) & 7 & 0.0551 & 0.0551 & 0.0551 & 0.0570 & 0.0585 & 0.0606 & 0.0615 & 0.3990 \\
$l < -1$ (no anticipation) & 115 & 0.0000 & 0.0007 & 0.0017 & 0.0025 & 0.0026 & 0.0103 & 0.0215 & 0.2853 \\
$0 \leq l < 2$ (delay) & 42 & 0.0019 & 0.0024 & 0.0031 & 0.0046 & 0.0031 & 0.0121 & 0.0167 & 0.1937 \\
$l > 2$ (dissipate) & 33 & 0.0000 & 0.0002 & 0.0002 & 0.0009 & 0.0002 & 0.0029 & 0.0090 & 0.0288 \\
$G_i = \infty$ (never treated) & 290 & 0.0004 & 0.0006 & 0.0038 & 0.0054 & 0.0071 & 0.0157 & 0.0157 & 1.5596 \\
\hline
\multicolumn{10}{l}{\textbf{IW}} \\[3pt]
$l = 2$, $G_i = 2006$ (ideal) & 13 & 0.0769 & 0.0769 & 0.0769 & 0.0769 & 0.0769 & 0.0769 & 0.0769 & 1.0000 \\
$l = 2$, $G_i \neq 2006$ (invariance) & 0 & \multicolumn{8}{c}{---} \\
$l < -1$ (no anticipation) & 0 & \multicolumn{8}{c}{---} \\
$0 \leq l < 2$ (delay) & 0 & \multicolumn{8}{c}{---} \\
$l > 2$ (dissipate) & 0 & \multicolumn{8}{c}{---} \\
$G_i = \infty$ (never treated) & 29 & 0.0345 & 0.0345 & 0.0345 & 0.0345 & 0.0345 & 0.0345 & 0.0345 & 1.0000 \\
\hline
\multicolumn{10}{l}{\textbf{DR}} \\[3pt]
$l = 2$, $G_i = 2006$ (ideal) & 13 & 0.0769 & 0.0769 & 0.0769 & 0.0769 & 0.0769 & 0.0769 & 0.0769 & 1.0000 \\
$l = 2$, $G_i \neq 2006$ (invariance) & 0 & \multicolumn{8}{c}{---} \\
$l < -1$ (pre-treatment) & 0 & \multicolumn{8}{c}{---} \\
$0 \leq l < 2$ (delay) & 0 & \multicolumn{8}{c}{---} \\
$l > 2$ (dissipate) & 0 & \multicolumn{8}{c}{---} \\
$G_i = \infty$ (never treated) & 29 & 0.0013 & 0.0055 & 0.0149 & 0.0374 & 0.0280 & 0.1332 & 0.4376 & 1.0835 \\
\hline
\multicolumn{10}{l}{\textbf{RW}} \\[3pt]
$l = 2$, $G_i = 2006$ (ideal) & 13 & 0.0769 & 0.0769 & 0.0769 & 0.0769 & 0.0769 & 0.0769 & 0.0769 & 1.0000 \\
$l = 2$, $G_i \neq 2006$ (invariance) & 0 & \multicolumn{8}{c}{---} \\
$l < -1$ (no anticipation) & 0 & \multicolumn{8}{c}{---} \\
$0 \leq l < 2$ (delay) & 0 & \multicolumn{8}{c}{---} \\
$l > 2$ (dissipate) & 0 & \multicolumn{8}{c}{---} \\
$G_i = \infty$ (never treated) & 29 & 0.0000 & 0.0000 & 0.0000 & 0.0345 & 0.0428 & 0.1895 & 0.3166 & 1.0000 \\
\bottomrule
\end{tabular}
}
\end{table}

\begin{table}[h!tbp]
\caption{\label{tab:weight_dispersion_aggregate} Summary statistics of absolute implied weights by observation group for estimating the aggregate lag-2 estimand $\ate^{\Scal_t}_{\text{agg}}(l=2)$.}
\centering
\resizebox{.95\textwidth}{!}{%
\begin{tabular}{rcccccccccc}
\toprule
 & $N$ & Min & \multicolumn{1}{p{1.4cm}}{\centering 25th\\Pctl.} & Median & Mean & \multicolumn{1}{p{1.4cm}}{\centering 75th\\Pctl.} & \multicolumn{1}{p{1.4cm}}{\centering 95th\\Pctl.} & Max. & $\sum|w|$ \\
\midrule
\multicolumn{10}{l}{\textbf{TWFE}} \\[3pt]
$l = 2$ (ideal) & 20 & 0.0462 & 0.0462 & 0.0462 & 0.0500 & 0.0551 & 0.0587 & 0.0615 & 1.0000 \\
$l < -1$ (no anticipation) & 115 & 0.0000 & 0.0007 & 0.0017 & 0.0025 & 0.0026 & 0.0103 & 0.0215 & 0.2853 \\
$0 \leq l < 2$ (delay) & 42 & 0.0019 & 0.0024 & 0.0031 & 0.0046 & 0.0031 & 0.0121 & 0.0167 & 0.1937 \\
$l > 2$ (dissipate) & 33 & 0.0000 & 0.0002 & 0.0002 & 0.0009 & 0.0002 & 0.0029 & 0.0090 & 0.0288 \\
$G_i = \infty$ (never treated) & 290 & 0.0004 & 0.0006 & 0.0038 & 0.0054 & 0.0071 & 0.0157 & 0.0157 & 1.5596 \\
\hline
\multicolumn{10}{l}{\textbf{IW}} \\[3pt]
$l = 2$ (ideal) & 20 & 0.0500 & 0.0500 & 0.0500 & 0.0500 & 0.0500 & 0.0500 & 0.0500 & 1.0000 \\
$l < -1$ (no anticipation) & 0 & \multicolumn{8}{c}{---} \\
$0 \leq l < 2$ (delay) & 0 & \multicolumn{8}{c}{---} \\
$l > 2$ (dissipate) & 0 & \multicolumn{8}{c}{---} \\
$G_i = \infty$ (never treated) & 116 & 0.0017 & 0.0030 & 0.0052 & 0.0086 & 0.0108 & 0.0224 & 0.0224 & 1.0000 \\
\hline
\multicolumn{10}{l}{\textbf{DR}} \\[3pt]
$l = 2$ (ideal) & 20 & 0.0500 & 0.0500 & 0.0500 & 0.0500 & 0.0500 & 0.0500 & 0.0500 & 1.0000 \\
$l < -1$ (no anticipation) & 0 & \multicolumn{8}{c}{---} \\
$0 \leq l < 2$ (delay) & 0 & \multicolumn{8}{c}{---} \\
$l > 2$ (dissipate) & 0 & \multicolumn{8}{c}{---} \\
$G_i = \infty$ (never treated) & 116 & 0.0000 & 0.0004 & 0.0028 & 0.0093 & 0.0073 & 0.0306 & 0.2845 & 1.0774 \\
\hline
\multicolumn{10}{l}{\textbf{RW}} \\[3pt]
$l = 2$ (ideal) & 20 & 0.0500 & 0.0500 & 0.0500 & 0.0500 & 0.0500 & 0.0500 & 0.0500 & 1.0000 \\
$l < -1$ (no anticipation) & 0 & \multicolumn{8}{c}{---} \\
$0 \leq l < 2$ (delay) & 0 & \multicolumn{8}{c}{---} \\
$l > 2$ (dissipate) & 0 & \multicolumn{8}{c}{---} \\
$G_i = \infty$ (never treated) & 116 & 0.0000 & 0.0000 & 0.0000 & 0.0086 & 0.0053 & 0.0479 & 0.1135 & 1.0000 \\
\bottomrule
\end{tabular}
}
\end{table}

\subsection{Effective sample sizes and information shares}
\label{sec_app_case_ess}

Table~\ref{tab:ess_aggregate} in the main text reports the group-wise effective sample sizes, information shares, and squared-weight shares for the aggregated estimand. Table~\ref{tab:ess_granular} reports the same quantities for the cohort-specific estimand $\ate^{\Scal_t}_{2008}(2006, \infty)$.

The results follow the similar general pattern as the aggregated results. For TWFE, the ideal group accounts for only $5.6\%$ of the information share, while the delay and dissipation groups together account for $12.9\%$. IW, DR, and RW differ only in how they weight the controls. IW assigns equal weights within each cohort comparison and has a control effective sample size of $29.00$. Covariate adjustment makes the control weights variable, where the effective sample size falls to $4.96$ under DR and $5.31$ under RW. The corresponding squared-weight shares are $0.755$ and $0.710$. Thus, the RW control weights are less concentrated than the DR control weights, consistent with the dispersion statistics in Table~\ref{tab:weight_dispersion_granular}.

\begin{table}[!t]
\centering
\caption{\label{tab:ess_granular} Group-wise effective sample sizes (ESS), information shares, and squared-weight share for estimating $\ate^{\Scal_t}_{2008}(2006, \infty)$. $N_g$ counts the differenced observation; each state has one period removed as the reference period.}
\renewcommand{\arraystretch}{1.2}
\resizebox{\textwidth}{!}{%
\begin{tabular}{@{}rcccc@{\qquad}rcccc@{}}
\toprule
\multicolumn{5}{c}{\textbf{TWFE}} & \multicolumn{5}{c}{\textbf{IW}} \\
\cmidrule(r){1-5} \cmidrule(l){6-10}
Observation group ($g$) & $N_g$ & $n_{\text{eff}, g}$ & $p_{\text{info}, g}$ & $D_g$ &
Observation group ($g$) & $N_g$ & $n_{\text{eff}, g}$ & $p_{\text{info}, g}$ & $D_g$ \\
\midrule
$l = 2$, $G_i = 2006$ (ideal) & 13 & 13.00 & 0.056 & 0.389 &
$l = 2$, $G_i = 2006$ (ideal) & 13 & 13.00 & 0.310 & 0.690 \\
$l = 2$, $G_i \neq 2006$ (invariance) & 7 & 6.99 & 0.030 & 0.319 &
$l = 2$, $G_i \neq 2006$ (invariance) & 0 & \multicolumn{3}{c}{---} \\
$l < -1$ (no anticipation) & 115 & 41.97 & 0.180 & 0.027 &
$l < -1$ (no anticipation) & 0 & \multicolumn{3}{c}{---} \\
$0 \leq l < 2$ (delay) & 42 & 23.34 & 0.100 & 0.022 &
$0 \leq l < 2$ (delay) & 0 & \multicolumn{3}{c}{---} \\
$l > 2$ (dissipate) & 33 & 6.71 & 0.029 & 0.002 &
$l > 2$ (dissipate) & 0 & \multicolumn{3}{c}{---} \\
$G_i = \infty$ (never treated) & 290 & 141.25 & 0.606 & 0.241 &
$G_i = \infty$ (never treated) & 29 & 29.00 & 0.690 & 0.310 \\
Total & 500 & 233.26 & 1.000 & 1.000 &
Total & 42 & 42.00 & 1.000 & 1.000 \\
\midrule
\multicolumn{5}{c}{\textbf{DR}} & \multicolumn{5}{c}{\textbf{RW}} \\
\cmidrule(r){1-5} \cmidrule(l){6-10}
Observation group ($g$) & $N_g$ & $n_{\text{eff}, g}$ & $p_{\text{info}, g}$ & $D_g$ &
Observation group ($g$) & $N_g$ & $n_{\text{eff}, g}$ & $p_{\text{info}, g}$ & $D_g$ \\
\midrule
$l = 2$, $G_i = 2006$ (ideal) & 13 & 13.00 & 0.724 & 0.245 &
$l = 2$, $G_i = 2006$ (ideal) & 13 & 13.00 & 0.710 & 0.290 \\
$l = 2$, $G_i \neq 2006$ (invariance) & 0 & \multicolumn{3}{c}{---} &
$l = 2$, $G_i \neq 2006$ (invariance) & 0 & \multicolumn{3}{c}{---} \\
$l < -1$ (no anticipation) & 0 & \multicolumn{3}{c}{---} &
$l < -1$ (no anticipation) & 0 & \multicolumn{3}{c}{---} \\
$0 \leq l < 2$ (delay) & 0 & \multicolumn{3}{c}{---} &
$0 \leq l < 2$ (delay) & 0 & \multicolumn{3}{c}{---} \\
$l > 2$ (dissipate) & 0 & \multicolumn{3}{c}{---} &
$l > 2$ (dissipate) & 0 & \multicolumn{3}{c}{---} \\
$G_i = \infty$ (never treated) & 29 & 4.96 & 0.276 & 0.755 &
$G_i = \infty$ (never treated) & 29 & 5.31 & 0.290 & 0.710 \\
Total & 42 & 17.96 & 1.000 & 1.000 &
Total & 42 & 18.31 & 1.000 & 1.000 \\
\bottomrule
\end{tabular}%
}
\end{table}
\subsection{Influence}
\label{sec_app_case_influence}

\begin{figure}[h!tbp]
    \centering

    \begin{subfigure}{\textwidth}
        \centering
        \includegraphics[width=.8\textwidth]
        {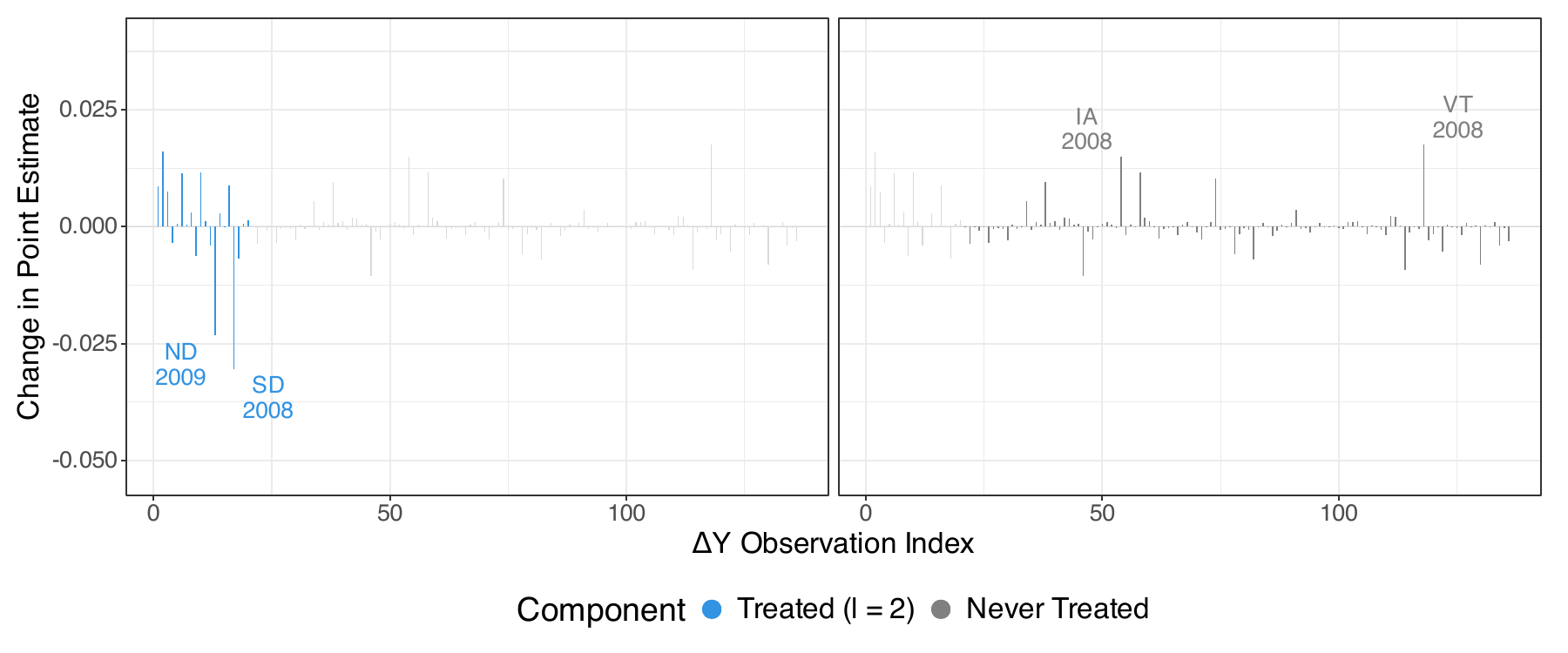}
        \caption{IW estimator.}
        \label{app_fig:influence_iw}
    \end{subfigure}

    \vspace{.5em}

    \begin{subfigure}{\textwidth}
        \centering
        \includegraphics[width=.8\textwidth]
        {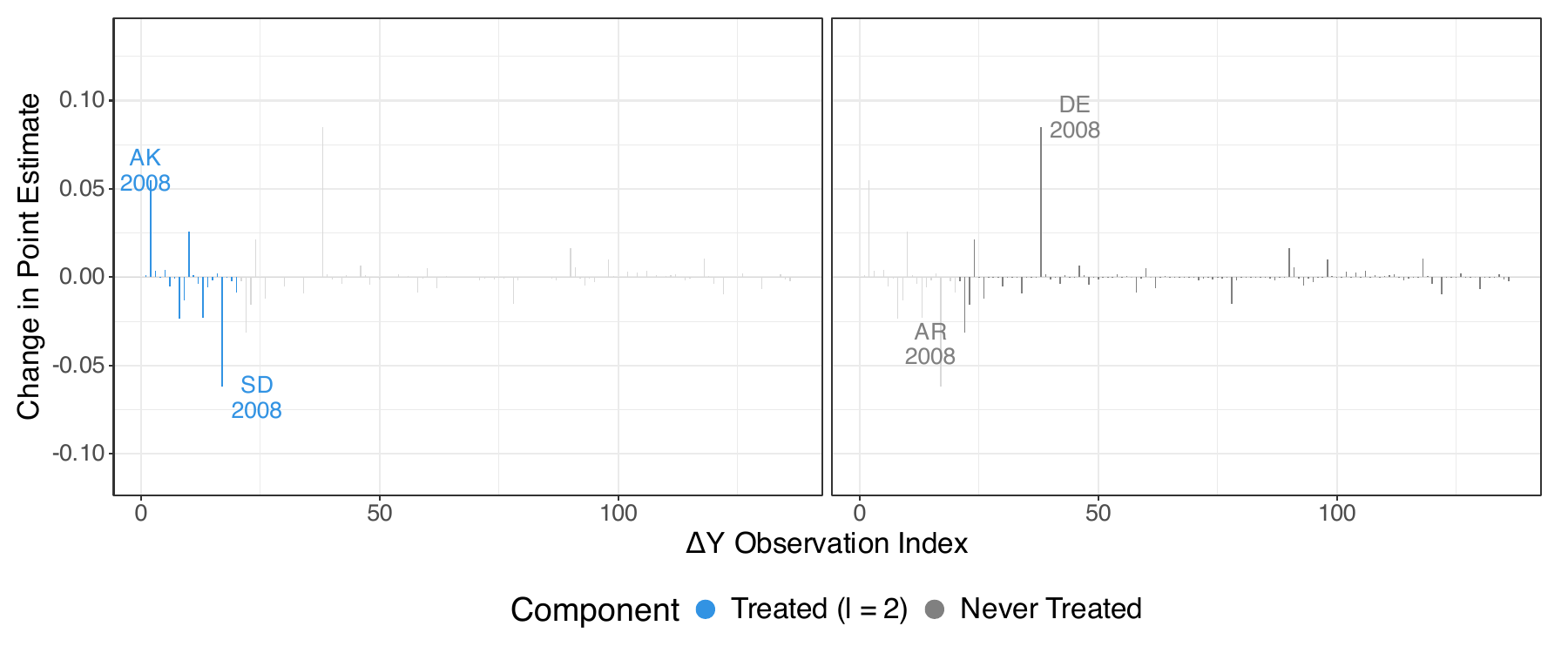}
        \caption{DR estimator.}
        \label{app_fig:influence_dr}
    \end{subfigure}

    \vspace{.5em}

    \begin{subfigure}{\textwidth}
        \centering
        \includegraphics[width=.8\textwidth]
        {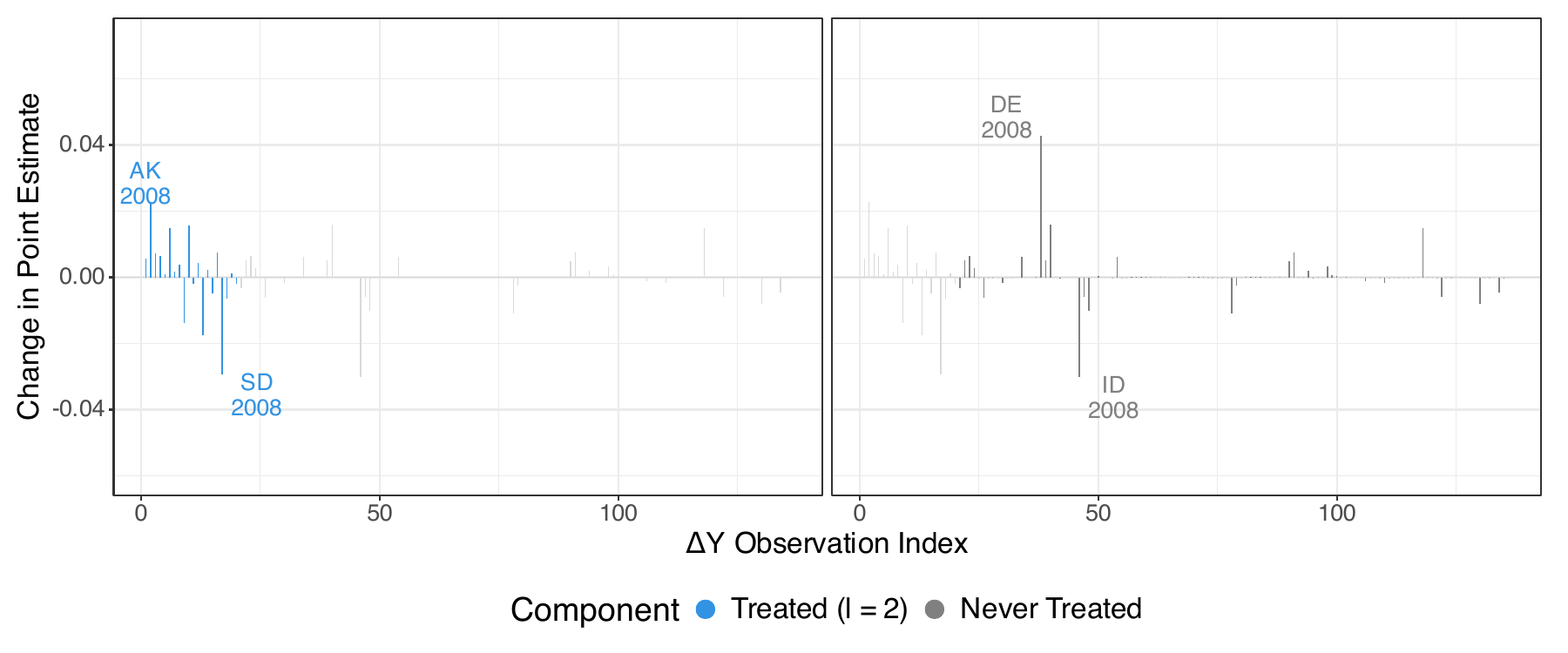}
        \caption{RW estimator.}
        \label{app_fig:influence_rw}
    \end{subfigure}

    \caption{\label{app_fig:influence_aggregate}
    Leave-one-observation-out influence for the IW, DR, and RW estimators at the aggregate lag-2 estimand. Each segment shows the change in the point estimate from removing one observation. Observations are colored by observation groups. Within each observation group, the most influential observations are labeled by state and year.}
\end{figure}

Figure~\ref{fig:twfe_influence_agg} in the main text shows the change in the dynamic TWFE estimate when each $\Delta Y^{\obs}$ observation is removed in turn. Figure~\ref{app_fig:influence_aggregate} reports the same diagnostic for IW, DR, and RW at the aggregate lag-2 estimand. Each estimator has two panels, one for treated observations at $l=2$ and one for never-treated controls, since these are the only observation groups used by the three estimators. Each segment represents one observation and shows the change in the estimate when it is removed. Within each observation group, the observations producing the largest changes are labeled by state and year.

For IW, the largest changes come from treated states. Removing South Dakota produces the largest change, $-0.030$, which is about $27\%$ of the IW estimate of $0.112$. For DR, the largest change comes from the never-treated control group. Removing Delaware changes the DR estimate by $0.085$, corresponding to $48\%$ of its estimate. For RW, the largest change also comes from the never-treated control group. Removing Delaware changes the RW estimate by $0.043$, corresponding to $29\%$ of its estimate. These large changes partly reflect the limited effective sample sizes, which make the estimates more sensitive to the removal of individual observations.


\subsection{Implied target profile}
\label{sec_app_case_profile}

\begin{figure}[h!tbp]
    \centering
    \includegraphics[width=.85\textwidth]{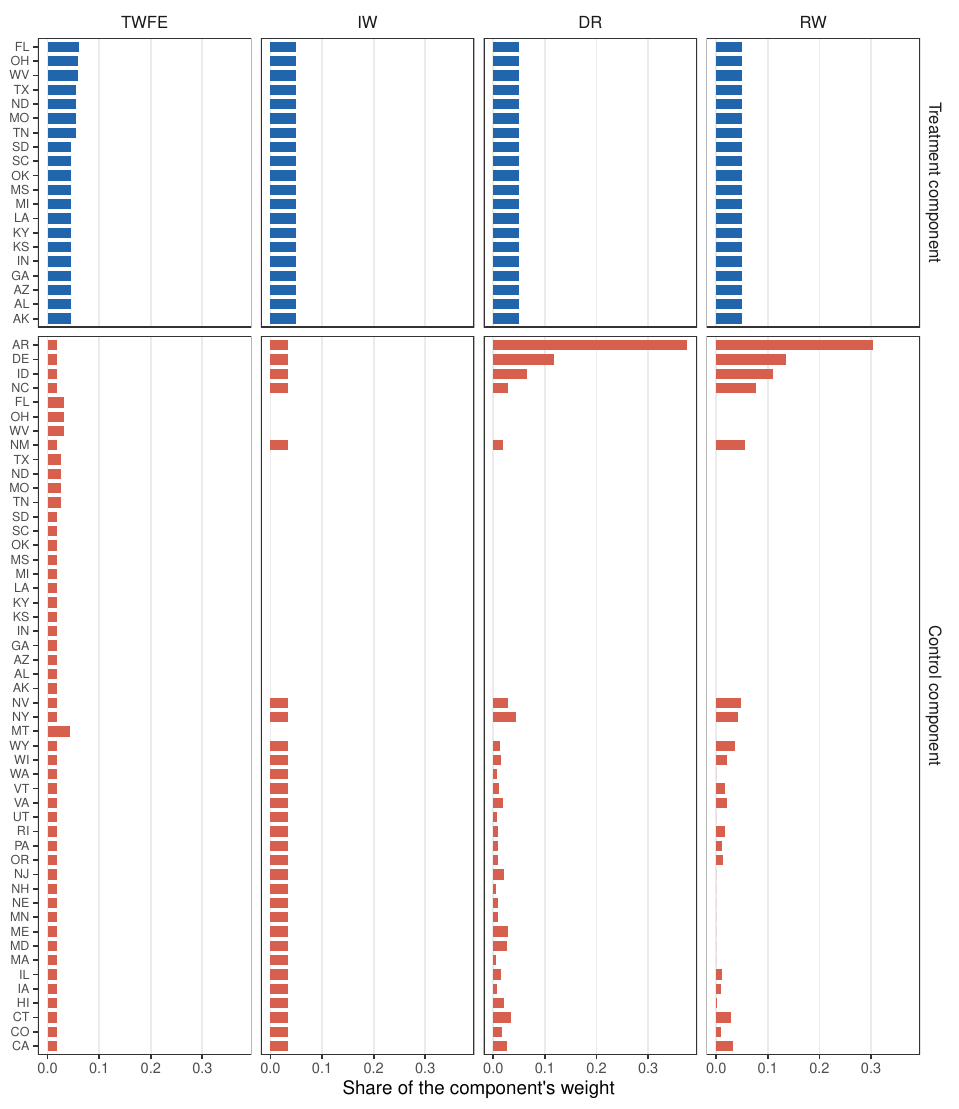}
    \caption{Distribution of implied weights across states for the aggregate lag-2 estimand $\ate^{\Scal_t}_{\mathrm{agg}}(l=2)$ in the Castle Doctrine study. Columns show the TWFE, IW, DR, and RW estimators. The upper and lower panels show the treatment and control components, respectively. Each bar shows a state's share of the component's total absolute weight, summing absolute observation weights across calendar years. Shares sum to one within each panel.}
    \label{app_fig:profile_state}
\end{figure}

\begin{figure}[h!tbp]
    \centering
    \includegraphics[width=.75\textwidth]{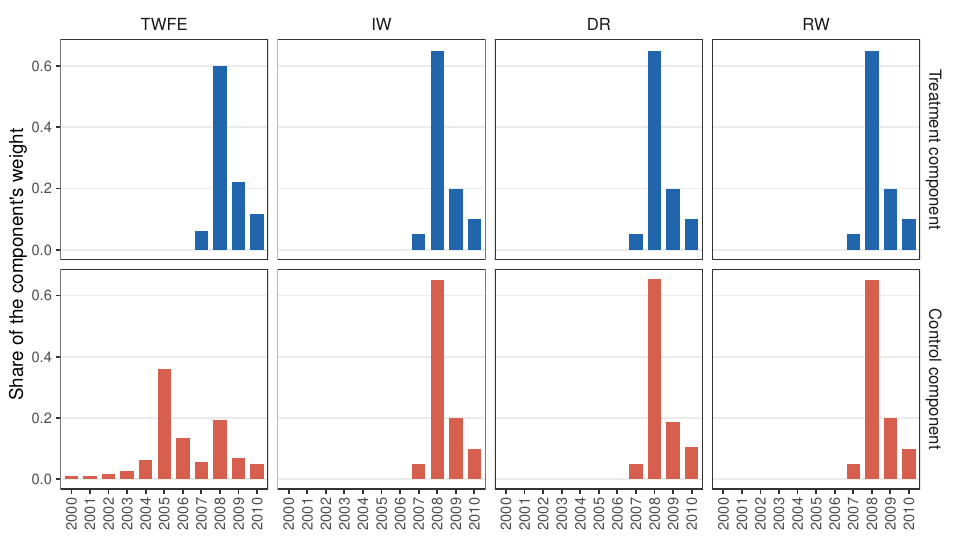}
    \caption{Distribution of implied weights across calendar years for the aggregate lag-2 estimand $\ate^{\Scal_t}_{\mathrm{agg}}(l=2)$ in the Castle Doctrine study. Columns show the TWFE, IW, DR, and RW estimators. The upper and lower panels show the treatment and control components, respectively. Each bar shows a calendar year's share of the component's total absolute weight, summing absolute observation weights across states. Shares sum to one within each panel.}
    \label{app_fig:profile_year}
\end{figure}

Table~\ref{tab:target_profiles} shows the baseline covariate profiles of the treatment and control components before and after weighting. Panel~(a) reports the granular estimand, and panel~(b) reports the aggregate estimand. Figures~\ref{app_fig:profile_state} and~\ref{app_fig:profile_year} show how the weighted profiles are composed across states and calendar years.

The after-weighting columns show the baseline covariate profile represented by each component. Under IW, weighting does not change either profile, so the differences between states that enacted the law and states that did not remain. DR matches the control profile to the treatment profile exactly, while RW brings the two profiles almost same to each other. For IW, DR, and RW, the treatment profile is the simple average of the 20 states that enacted the law. Under IW, the control profile is likewise the simple average of all 29 states that did not enact the law. DR and RW instead construct the control profile using unequal contributions from these states. Arkansas contributes the most, accounting for $38\%$ of the DR control profile and $30\%$ of the RW control profile. The RW profile uses only 20 of the 29 control states and is sparser. The TWFE control profile
draws on all 50 states, including states that enacted the law, and places substantial weight on calendar years outside 2007--2010, which are the years represented in its treatment profile.

\begin{table}[h!tbp]
\centering
\caption{\label{tab:target_profiles}
Implied target profiles for the granular and aggregate lag-2 estimands. For each estimator, the table reports the unweighted study sample mean and the means of the treatment and control components before and after weighting. The after-weighting means are the implied target profiles defined in
Section~\ref{sec:app_diag_balance}.}
\renewcommand{\arraystretch}{1.15}
\begin{subtable}{.9\textwidth}
\centering
\caption{\label{tab:profile_granular}
Target estimand: $\ate^{\Scal_t}_{2008}(2006,\infty)$.}
\resizebox{.65\textwidth}{!}{%
\begin{tabular}{lccccc}
\toprule
 & Study & \multicolumn{2}{c}{Treatment component}
 & \multicolumn{2}{c}{Control component} \\
\cmidrule(lr){3-4} \cmidrule(lr){5-6}
Variable & sample & Before & After & Before & After \\
\midrule
\multicolumn{6}{l}{\textbf{IW}} \\[3pt]
l\_police
 & \phantom{-}5.726 & \phantom{-}5.758 & \phantom{-}5.758
 & \phantom{-}5.711 & \phantom{-}5.711 \\
l\_income
 & \phantom{-}10.864 & \phantom{-}10.794 & \phantom{-}10.794
 & \phantom{-}10.926 & \phantom{-}10.926 \\
l\_prisoner
 & \phantom{-}5.950 & \phantom{-}6.257 & \phantom{-}6.257
 & \phantom{-}5.817 & \phantom{-}5.817 \\
\midrule
\multicolumn{6}{l}{\textbf{DR}} \\[3pt]
l\_police
 & \phantom{-}5.726 & \phantom{-}5.758 & \phantom{-}5.758
 & \phantom{-}5.711 & \phantom{-}5.758 \\
l\_income
 & \phantom{-}10.864 & \phantom{-}10.794 & \phantom{-}10.794
 & \phantom{-}10.926 & \phantom{-}10.794 \\
l\_prisoner
 & \phantom{-}5.950 & \phantom{-}6.257 & \phantom{-}6.257
 & \phantom{-}5.817 & \phantom{-}6.257 \\
\midrule
\multicolumn{6}{l}{\textbf{RW}} \\[3pt]
l\_police
 & \phantom{-}5.726 & \phantom{-}5.758 & \phantom{-}5.758
 & \phantom{-}5.711 & \phantom{-}5.760 \\
l\_income
 & \phantom{-}10.864 & \phantom{-}10.794 & \phantom{-}10.794
 & \phantom{-}10.926 & \phantom{-}10.795 \\
l\_prisoner
 & \phantom{-}5.950 & \phantom{-}6.257 & \phantom{-}6.257
 & \phantom{-}5.817 & \phantom{-}6.252 \\
\bottomrule
\end{tabular}%
}
\end{subtable}

\vspace{1em}

\begin{subtable}{.9\textwidth}
\centering
\caption{\label{tab:profile_aggregate}
Target estimand: $\ate^{\Scal_t}_{\mathrm{agg}}(l=2)$.}
\resizebox{.65\textwidth}{!}{%
\begin{tabular}{lccccc}
\toprule
 & Study & \multicolumn{2}{c}{Treatment component}
 & \multicolumn{2}{c}{Control component} \\
\cmidrule(lr){3-4} \cmidrule(lr){5-6}
Variable & sample & Before & After & Before & After \\
\midrule
\multicolumn{6}{l}{\textbf{TWFE}} \\[3pt]
l\_police
 & \phantom{-}5.726 & \phantom{-}5.752 & \phantom{-}5.751
 & \phantom{-}5.725 & \phantom{-}5.751 \\
l\_income
 & \phantom{-}10.864 & \phantom{-}10.785 & \phantom{-}10.784
 & \phantom{-}10.867 & \phantom{-}10.784 \\
l\_prisoner
 & \phantom{-}5.950 & \phantom{-}6.148 & \phantom{-}6.132
 & \phantom{-}5.943 & \phantom{-}6.132 \\
\midrule
\multicolumn{6}{l}{\textbf{IW}} \\[3pt]
l\_police
 & \phantom{-}5.726 & \phantom{-}5.752 & \phantom{-}5.752
 & \phantom{-}5.711 & \phantom{-}5.711 \\
l\_income
 & \phantom{-}10.864 & \phantom{-}10.785 & \phantom{-}10.785
 & \phantom{-}10.926 & \phantom{-}10.926 \\
l\_prisoner
 & \phantom{-}5.950 & \phantom{-}6.148 & \phantom{-}6.148
 & \phantom{-}5.817 & \phantom{-}5.817 \\
\midrule
\multicolumn{6}{l}{\textbf{DR}} \\[3pt]
l\_police
 & \phantom{-}5.726 & \phantom{-}5.752 & \phantom{-}5.752
 & \phantom{-}5.711 & \phantom{-}5.752 \\
l\_income
 & \phantom{-}10.864 & \phantom{-}10.785 & \phantom{-}10.785
 & \phantom{-}10.926 & \phantom{-}10.785 \\
l\_prisoner
 & \phantom{-}5.950 & \phantom{-}6.148 & \phantom{-}6.148
 & \phantom{-}5.817 & \phantom{-}6.148 \\
\midrule
\multicolumn{6}{l}{\textbf{RW}} \\[3pt]
l\_police
 & \phantom{-}5.726 & \phantom{-}5.752 & \phantom{-}5.752
 & \phantom{-}5.711 & \phantom{-}5.749 \\
l\_income
 & \phantom{-}10.864 & \phantom{-}10.785 & \phantom{-}10.785
 & \phantom{-}10.926 & \phantom{-}10.786 \\
l\_prisoner
 & \phantom{-}5.950 & \phantom{-}6.148 & \phantom{-}6.148
 & \phantom{-}5.817 & \phantom{-}6.144 \\
\bottomrule
\end{tabular}%
}
\end{subtable}
\end{table}

\subsection{Sign reversal}
\label{sec_app_sign_reversal}

Section~\ref{sec:app_diag_balance} defines sign reversal as a negative weight on a $\Delta Y^{\obs}$ observation. We report the number of such observations and their share of the component's absolute weight. No estimator assigns negative weights in the treatment component. Among the 480 $\Delta Y^{\obs}$ observations in the TWFE control component, 270 receive negative weights, accounting for $49.5\%$ of the component's absolute weight. For DR, negative weights appear on 5 of 29 control observations for the granular estimand and 28 of 116 for the aggregate estimand, accounting for $3.9\%$ and $3.6\%$ of absolute control weight. IW and RW have no negative weights: IW weights eligible controls uniformly, while RW imposes nonnegativity in~\eqref{minimal_1}. 

Figures~\ref{fig:map_agg} and~\ref{app_fig:map_cell} display the exact weights assigned by different estimators to the panel observations. For the granular DR estimator, negative control weights are assigned to Iowa, Maine, Massachusetts, Nebraska, and Washington.

%% file: supp6_simulation.tex

\section{Simulation study}
\label{sec:app_simulation}
This simulation examines two sources of bias. First, we examine how the dynamic TWFE estimator behaves when treatment effects differ across treatment initiation cohorts. Second, we examine the bias that results when the control component is expanded under incorrect assumptions about delayed treatment onset or treatment effect dissipation. Parallel trends and no anticipation hold by construction throughout. Thus, this simulation does not study misspecification of $\kappa$.

\subsection{Simulation design}
\label{sec:app_sim_design}

The simulated panel contains 210 units observed from 2000 through 2010. Treatment initiation is staggered: 5, 15, 35, 50, 30, and 15 units initiate treatment in 2004 through 2009, respectively, while 60 units remain untreated throughout the panel.

\begin{figure}[h!tbp]
\centering
\includegraphics[width=\textwidth]
{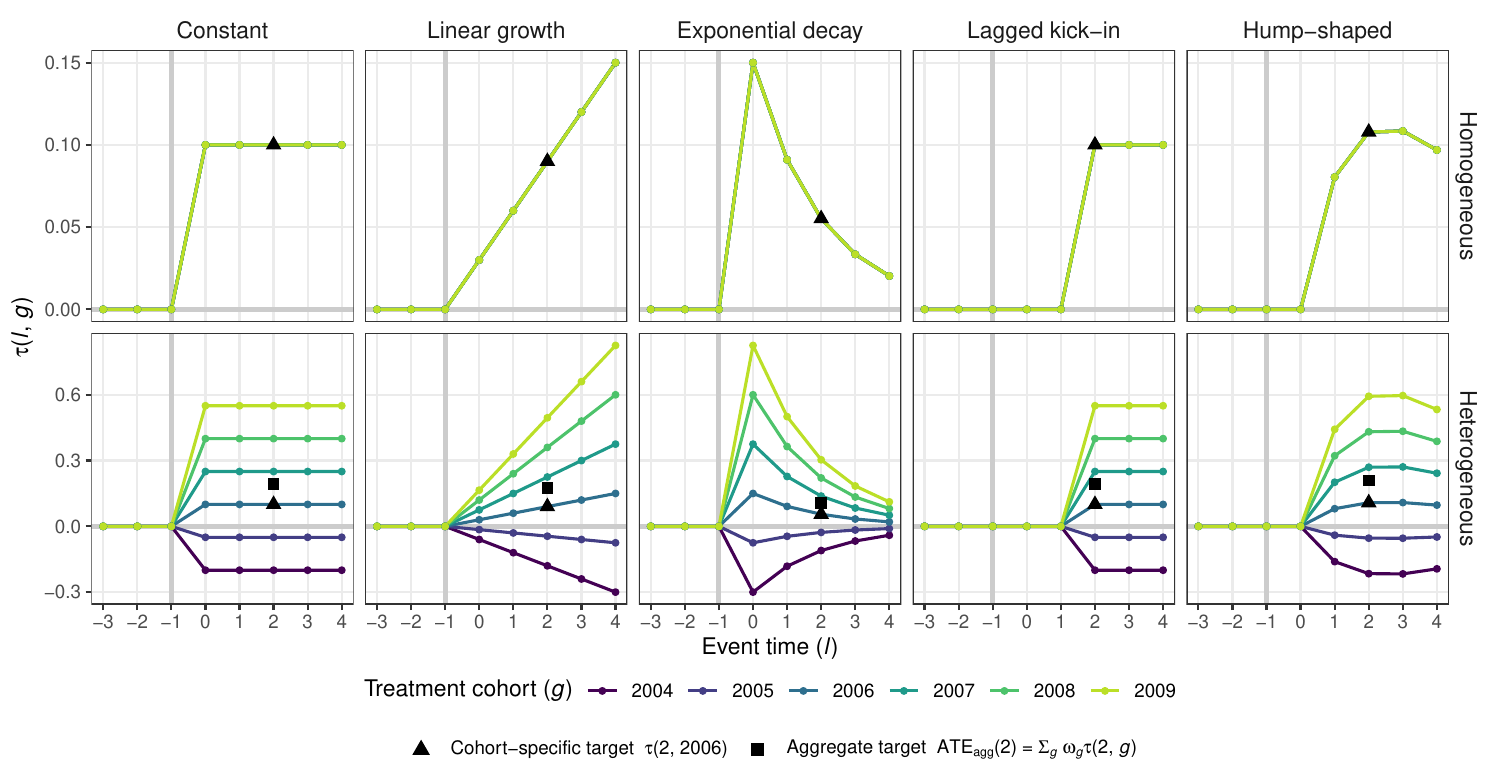}
\caption{\label{fig:app_sim_dgp}
Treatment effect paths in the ten simulation designs. Columns correspond to the five effect paths, and rows correspond to homogeneous and heterogeneous cohort effects. Colors distinguish treatment initiation cohorts. The triangle marks the target effect $\tau(2,2006)$, the rectangle marks the aggregated effect $\ate^{\Scal_t}_{\mathrm{agg}}(l=2) = \sum_{g=2004}^{2008}\omega_g\,\tau(2,g)$, and the vertical gray line marks the omitted reference period, $l=-1$. Because $m(g)=1$ under cohort homogeneity, the two target effects are equal and their markers coincide, so only the triangle is shown in the upper row.
}
\end{figure}

Outcomes are generated as
\begin{equation}
\label{eq:app_sim_dgp}
Y_{it} =
\alpha_i+\delta_t
+\mathbbm{1}\{G_i<\infty\}\tau(t-G_i,G_i)
+\varepsilon_{it},
\end{equation}
where $\alpha_i\sim N(0,\sigma_\alpha^2)$,
$\delta_t\sim N(0,\sigma_\delta^2)$, and
$\varepsilon_{it}\sim N(0,\sigma_\varepsilon^2)$, with
$\sigma_\alpha=1$, $\sigma_\delta=0.5$, and
$\sigma_\varepsilon=0.3$. The additive unit and period effects ensure that parallel trends holds by construction. We set $\tau(l,g)=0$ for $l<0$, so there is no anticipation. For treated units, the treatment effect is
\begin{equation*}
\tau(l,g)=f(l)m(g),
\end{equation*}
where $f(l)$ determines the effect path over event time and $m(g)$ allows the path to vary across treatment initiation cohorts. We consider five effect paths: constant, linear growth, exponential decay, lagged kick-in, and hump-shaped. Each is evaluated under two cohort regimes. Under homogeneity, $m(g)=1$ for every cohort. Under heterogeneity,
\begin{equation*}
m(g)=1+1.5(g-2006),
\end{equation*}
giving multipliers $-2.0$, $-0.5$, $1.0$, $2.5$, and $4.0$ for the 2004--2008 cohorts. Cohort 2009 is not observed at $l=2$ within a panel that ends in 2010, so it appears in the effect paths but not in the aggregate estimand. Because $m(2006)=1$, the effect for the 2006 cohort is unchanged across the two regimes. Figure~\ref{fig:app_sim_dgp} shows the resulting treatment effect paths for the five paths under the homogeneous and heterogeneous cohort regimes.

We consider the same two target estimands as in the case study. The first is the effect for the 2006 cohort in 2008,
\begin{equation*}
\ate^{\Scal_t}_{2008}(2006,\infty) = \tau(2,2006).
\end{equation*}
The second is the aggregate lag-2 estimand defined in Equation~\eqref{eq:app_agg_estimand},
\begin{equation*}
\ate^{\Scal_t}_{\mathrm{agg}}(l=2)
= \sum_{g=2004}^{2008}
\omega_g\,\ate^{\Scal_g}_{g+2}(g,\infty)
= \sum_{g=2004}^{2008}\omega_g\,\tau(2,g).
\end{equation*}
The 2004--2008 cohorts contain 5, 15, 35, 50, and 30 units, giving weights of $0.037$, $0.111$, $0.259$, $0.370$, and $0.222$, respectively. Under cohort homogeneity, the two target estimands are equal.
We generate 500 replications for each design. Within each replication, the same unit effects, period effects, and idiosyncratic errors are used across all ten designs. This makes comparisons across designs less sensitive to simulation noise.

\subsection{Cohort heterogeneity and dynamic TWFE}
\label{sec:app_sim_twfe}

\begin{figure}[h!tbp]
\centering

\begin{subfigure}[t]{\textwidth}
    \centering
    \includegraphics[width=\textwidth]
    {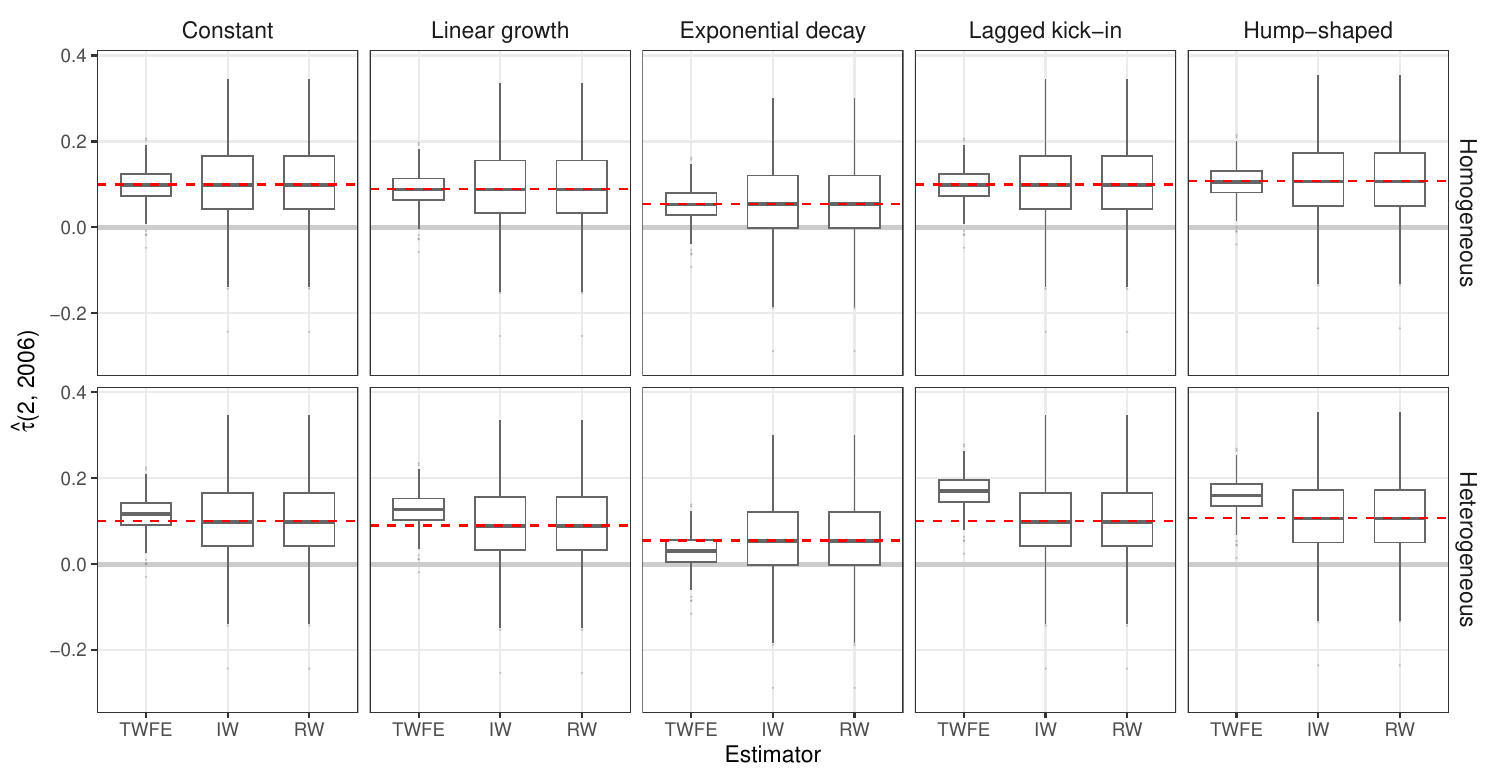}
    \caption{Cohort-specific target,
    $\ate^{\Scal_t}_{2008}(2006,\infty)=\tau(2,2006)$.}
    \label{fig:app_sim_boxplot}
\end{subfigure}

\vspace{0.75em}

\begin{subfigure}[t]{\textwidth}
    \centering
    \includegraphics[width=\textwidth]
    {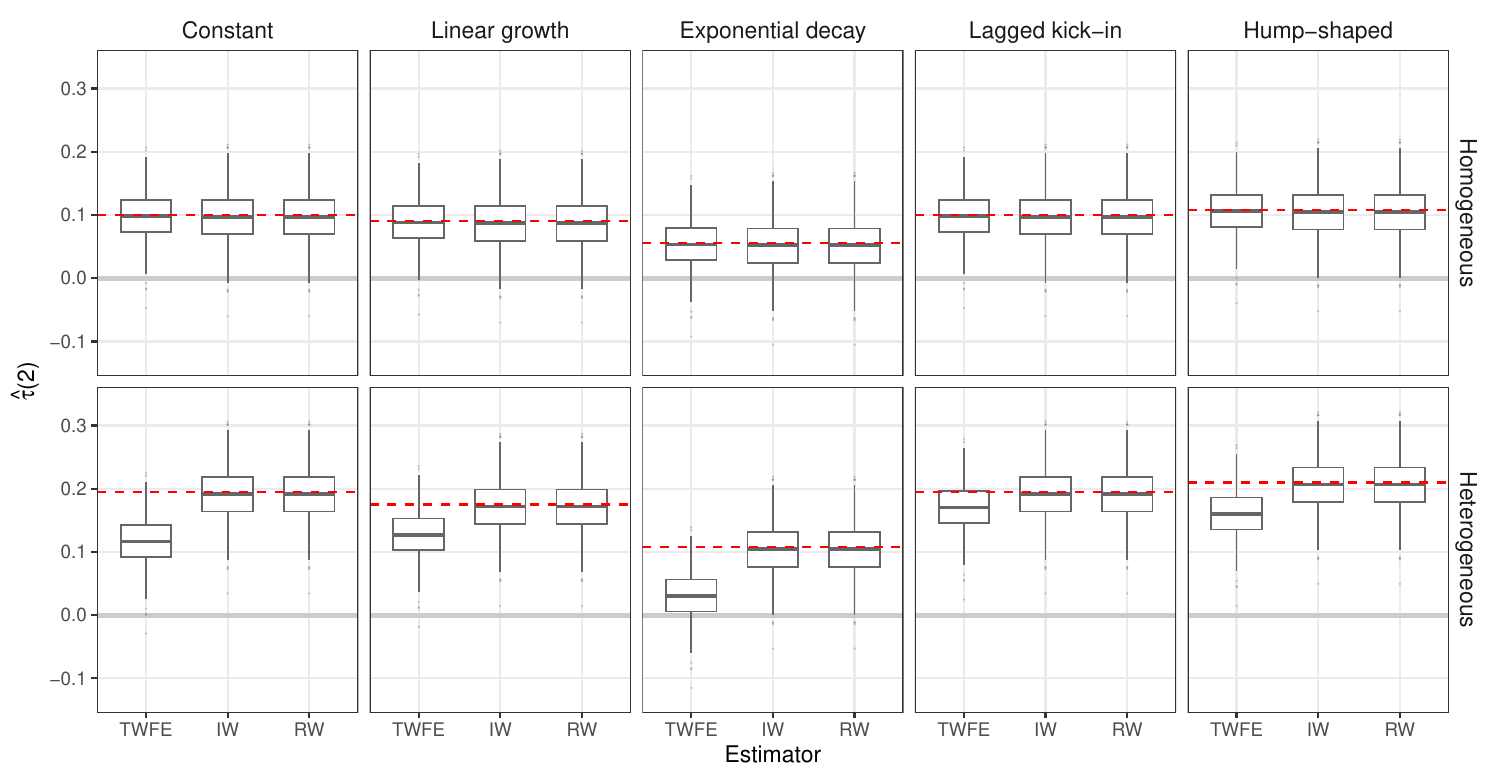}
    \caption{Aggregate lag-2 target,
    $\ate^{\Scal_t}_{\mathrm{agg}}(l=2)$.}
    \label{fig:app_sim_boxplot_pooled}
\end{subfigure}

\caption{\label{fig:app_sim_estimator_boxplots}
Sampling distributions of the dynamic TWFE, IW, and RW estimators over
$500$ replications. Columns show the five treatment effect paths, and
rows show the homogeneous and heterogeneous cohort regimes. The red
dashed line marks the true target effect in each design.}
\end{figure}

Figure~\ref{fig:app_sim_estimator_boxplots} compares the three estimators for the cohort-specific target in panel~(a) and the aggregate lag-2 target in panel~(b). For the cohort-specific target estimand, although the dynamic TWFE coefficient is not cohort specific, we compare it with this target to mirror the comparison in the case study. The IW and RW estimators use only the ideal experiment observations. DR is omitted because there are no covariates in the simulation, in which case it coincides with IW as RW does.

Under cohort homogeneity, all three estimators are centered near the true effect. Under cohort heterogeneity, IW and RW remain centered near the target, while dynamic TWFE is biased in all five designs. Its bias ranges from $0.017$ to $0.070$; in the lagged kick-in design, the bias is about $70\%$ of the target effect. The difference arises because dynamic TWFE combines information across treatment cohorts. When effects are common across cohorts, contributions from the non-target cohorts cancel because their implied weights sum to zero within each non-target event-time bin, as shown in Proposition~\ref{prop:app_twfe_weights}. This cancellation no longer holds when effects vary across cohorts.

For the aggregate lag-2 target, the same pattern remains. All three estimators are centered near the target under cohort homogeneity, while under heterogeneity IW and RW remain centered near the aggregate target and dynamic TWFE does not.

\subsection{Information borrowing with the RW estimator}
\label{sec:app_sim_borrowing}

We next use the RW estimator to examine how expanding the control component affects bias and accuracy across the ten simulation settings. The target throughout is the aggregate lag-2 estimand, $\ate^{\Scal_t}_{\mathrm{agg}}(l=2)$.

We begin with the ideal contrast, which uses never-treated units as controls, and then add not-yet-treated units under $\kappa=0$. From this common baseline, we consider delayed onset and effect dissipation separately. The delayed-onset specifications add observations at $l=0$ under $\phi=0$ and then at $l\in \{0,1\}$ under $\phi=1$. The effect-dissipation specifications instead add observations at $l\geq4$ under $\xi=4$ and then at $l\geq3$ under $\xi=3$. The final specification includes all of these groups. It draws on the same broad set of observation groups that can enter dynamic TWFE, although RW and TWFE assign them different weights. Panel A of Table~\ref{tab:app_sim_borrowing} reports bias, and Panel B reports root mean squared error. Green cells indicate specifications whose borrowing assumptions are satisfied by the true data-generating process and treatment effect path. Red cells indicate specifications for which at least one of the invoked assumptions is false.

\begin{table}[h!tbp]
\centering
\caption{\label{tab:app_sim_borrowing} Bias (Panel A) and root mean squared error (Panel B) of the robust weighting estimator under seven information sets, based on $500$ replications. Bias is the average estimate minus the true aggregate lag-2 effect. The delayed-onset and effect-dissipation specifications are evaluated separately from the common ideal-plus-not-yet-treated baseline, while the final column combines all observation groups. Green cells indicate specifications whose borrowing assumptions are satisfied by the true data-generating process; red cells indicate specifications for which at least one invoked assumption is false.}
\resizebox{0.9\textwidth}{!}{%
\begin{tabular}{@{}llcccccc@{\hspace{1.1em}}c@{}}
\toprule
 & & & & \multicolumn{2}{c}{Delayed onset} & \multicolumn{2}{c}{Effect dissipation} & \\
\cmidrule(lr){5-6} \cmidrule(lr){7-8}
True effect & Cohort & Ideal & $+$ Not yet & $+\, l = 0$ & $+\, l \in \{0,1\}$ & $+\, l \ge 4$ & $+\, l \ge 3$ & All \\
paths & regime & contrast & treated & ($\phi = 0$) & ($\phi = 1$) & ($\xi = 4$) & ($\xi = 3$) & information \\
\midrule
\multicolumn{9}{@{}l}{Panel A: Bias} \\[2pt]
Constant & Homogeneous & \cellcolor{safegreen}$-0.003$ & \cellcolor{safegreen}$-0.003$ & \cellcolor{biasred}$-0.022$ & \cellcolor{biasred}$-0.042$ & \cellcolor{biasred}$-0.003$ & \cellcolor{biasred}$-0.002$ & \cellcolor{biasred}$-0.032$ \\
         & Heterogeneous & \cellcolor{safegreen}$-0.003$ & \cellcolor{safegreen}$-0.003$ & \cellcolor{biasred}$-0.083$ & \cellcolor{biasred}$-0.150$ & \cellcolor{biasred}$-0.003$ & \cellcolor{biasred}$-0.002$ & \cellcolor{biasred}$-0.105$ \\
Linear growth & Homogeneous & \cellcolor{safegreen}$-0.003$ & \cellcolor{safegreen}$-0.003$ & \cellcolor{biasred}$-0.009$ & \cellcolor{biasred}$-0.022$ & \cellcolor{biasred}$-0.022$ & \cellcolor{biasred}$-0.036$ & \cellcolor{biasred}$-0.043$ \\
              & Heterogeneous & \cellcolor{safegreen}$-0.003$ & \cellcolor{safegreen}$-0.003$ & \cellcolor{biasred}$-0.027$ & \cellcolor{biasred}$-0.074$ & \cellcolor{biasred}$\phantom{-}0.004$ & \cellcolor{biasred}$-0.019$ & \cellcolor{biasred}$-0.067$ \\
Exponential decay & Homogeneous & \cellcolor{safegreen}$-0.003$ & \cellcolor{safegreen}$-0.003$ & \cellcolor{biasred}$-0.032$ & \cellcolor{biasred}$-0.047$ & \cellcolor{biasred}$\phantom{-}0.011$ & \cellcolor{biasred}$\phantom{-}0.034$ & \cellcolor{biasred}$-0.009$ \\
                  & Heterogeneous & \cellcolor{safegreen}$-0.003$ & \cellcolor{safegreen}$-0.003$ & \cellcolor{biasred}$-0.123$ & \cellcolor{biasred}$-0.170$ & \cellcolor{biasred}$-0.006$ & \cellcolor{biasred}$\phantom{-}0.025$ & \cellcolor{biasred}$-0.098$ \\
Lagged kick-in & Homogeneous & \cellcolor{safegreen}$-0.003$ & \cellcolor{safegreen}$-0.003$ & \cellcolor{safegreen}$-0.003$ & \cellcolor{safegreen}$-0.003$ & \cellcolor{biasred}$-0.018$ & \cellcolor{biasred}$-0.036$ & \cellcolor{biasred}$-0.028$ \\
              & Heterogeneous & \cellcolor{safegreen}$-0.003$ & \cellcolor{safegreen}$-0.003$ & \cellcolor{safegreen}$-0.003$ & \cellcolor{safegreen}$-0.003$ & \cellcolor{biasred}$-0.003$ & \cellcolor{biasred}$-0.027$ & \cellcolor{biasred}$-0.025$ \\
Hump-shaped & Homogeneous & \cellcolor{safegreen}$-0.003$ & \cellcolor{safegreen}$-0.003$ & \cellcolor{safegreen}$-0.003$ & \cellcolor{biasred}$-0.024$ & \cellcolor{biasred}$-0.003$ & \cellcolor{biasred}$-0.027$ & \cellcolor{biasred}$-0.036$ \\
            & Heterogeneous & \cellcolor{safegreen}$-0.003$ & \cellcolor{safegreen}$-0.003$ & \cellcolor{safegreen}$-0.003$ & \cellcolor{biasred}$-0.075$ & \cellcolor{biasred}$-0.005$ & \cellcolor{biasred}$-0.031$ & \cellcolor{biasred}$-0.074$ \\
\midrule
\multicolumn{9}{@{}l}{Panel B: Root mean squared error} \\[2pt]
Constant & Homogeneous & \cellcolor{safegreen}$\phantom{-}0.043$ & \cellcolor{safegreen}$\phantom{-}0.044$ & \cellcolor{biasred}$\phantom{-}0.048$ & \cellcolor{biasred}$\phantom{-}0.058$ & \cellcolor{biasred}$\phantom{-}0.042$ & \cellcolor{biasred}$\phantom{-}0.042$ & \cellcolor{biasred}$\phantom{-}0.051$ \\
         & Heterogeneous & \cellcolor{safegreen}$\phantom{-}0.043$ & \cellcolor{safegreen}$\phantom{-}0.044$ & \cellcolor{biasred}$\phantom{-}0.093$ & \cellcolor{biasred}$\phantom{-}0.155$ & \cellcolor{biasred}$\phantom{-}0.042$ & \cellcolor{biasred}$\phantom{-}0.042$ & \cellcolor{biasred}$\phantom{-}0.112$ \\
Linear growth & Homogeneous & \cellcolor{safegreen}$\phantom{-}0.043$ & \cellcolor{safegreen}$\phantom{-}0.044$ & \cellcolor{biasred}$\phantom{-}0.043$ & \cellcolor{biasred}$\phantom{-}0.046$ & \cellcolor{biasred}$\phantom{-}0.048$ & \cellcolor{biasred}$\phantom{-}0.055$ & \cellcolor{biasred}$\phantom{-}0.058$ \\
             & Heterogeneous & \cellcolor{safegreen}$\phantom{-}0.043$ & \cellcolor{safegreen}$\phantom{-}0.044$ & \cellcolor{biasred}$\phantom{-}0.050$ & \cellcolor{biasred}$\phantom{-}0.084$ & \cellcolor{biasred}$\phantom{-}0.043$ & \cellcolor{biasred}$\phantom{-}0.046$ & \cellcolor{biasred}$\phantom{-}0.078$ \\
Exponential decay & Homogeneous & \cellcolor{safegreen}$\phantom{-}0.043$ & \cellcolor{safegreen}$\phantom{-}0.044$ & \cellcolor{biasred}$\phantom{-}0.053$ & \cellcolor{biasred}$\phantom{-}0.062$ & \cellcolor{biasred}$\phantom{-}0.044$ & \cellcolor{biasred}$\phantom{-}0.054$ & \cellcolor{biasred}$\phantom{-}0.040$ \\
                  & Heterogeneous & \cellcolor{safegreen}$\phantom{-}0.043$ & \cellcolor{safegreen}$\phantom{-}0.044$ & \cellcolor{biasred}$\phantom{-}0.130$ & \cellcolor{biasred}$\phantom{-}0.175$ & \cellcolor{biasred}$\phantom{-}0.043$ & \cellcolor{biasred}$\phantom{-}0.048$ & \cellcolor{biasred}$\phantom{-}0.105$ \\
Lagged kick-in & Homogeneous & \cellcolor{safegreen}$\phantom{-}0.043$ & \cellcolor{safegreen}$\phantom{-}0.044$ & \cellcolor{safegreen}$\phantom{-}0.043$ & \cellcolor{safegreen}$\phantom{-}0.040$ & \cellcolor{biasred}$\phantom{-}0.046$ & \cellcolor{biasred}$\phantom{-}0.055$ & \cellcolor{biasred}$\phantom{-}0.048$ \\
               & Heterogeneous & \cellcolor{safegreen}$\phantom{-}0.043$ & \cellcolor{safegreen}$\phantom{-}0.044$ & \cellcolor{safegreen}$\phantom{-}0.043$ & \cellcolor{safegreen}$\phantom{-}0.040$ & \cellcolor{biasred}$\phantom{-}0.042$ & \cellcolor{biasred}$\phantom{-}0.049$ & \cellcolor{biasred}$\phantom{-}0.047$ \\
Hump-shaped & Homogeneous & \cellcolor{safegreen}$\phantom{-}0.043$ & \cellcolor{safegreen}$\phantom{-}0.044$ & \cellcolor{safegreen}$\phantom{-}0.043$ & \cellcolor{biasred}$\phantom{-}0.046$ & \cellcolor{biasred}$\phantom{-}0.042$ & \cellcolor{biasred}$\phantom{-}0.050$ & \cellcolor{biasred}$\phantom{-}0.053$ \\
            & Heterogeneous & \cellcolor{safegreen}$\phantom{-}0.043$ & \cellcolor{safegreen}$\phantom{-}0.044$ & \cellcolor{safegreen}$\phantom{-}0.043$ & \cellcolor{biasred}$\phantom{-}0.085$ & \cellcolor{biasred}$\phantom{-}0.043$ & \cellcolor{biasred}$\phantom{-}0.052$ & \cellcolor{biasred}$\phantom{-}0.084$ \\
\bottomrule
\end{tabular}%
}
\end{table}

The ideal contrast and the specification that adds not-yet-treated units have negligible bias in every scenario. Their common bias is $-0.003$, and their RMSEs are nearly identical at $0.043$ and $0.044$. Thus, including not-yet-treated observations under the correctly specified no-anticipation assumption does not introduce meaningful bias.

When the delayed-onset restriction is correct, borrowing can modestly improve accuracy as measured by RMSE. In the lagged kick-in scenario, effects are zero at both $l=0$ and $l=1$. Adding these observations leaves the bias at $-0.003$ and lowers RMSE from $0.043$ to $0.040$. When the restriction is incorrect, however, the cost can be large. Under heterogeneous cohort effects, adding observations through $l=1$ produces a bias of $-0.150$ in the constant-effect scenario and $-0.170$ in the exponential-decay scenario, with RMSEs of $0.155$ and $0.175$.

One caveat is that a borrowing assumption that is not satisfied does not always produce large bias or RMSE in some settings. Several red effect-dissipation specifications perform reasonably well because treatment effects happen to cancel within outcome differences or across cohorts. This is a feature of the particular data-generating processes considered here, not evidence that the corresponding assumptions are valid. In applications, where the true treatment effect dynamics are unknown, investigators should generally restrict the contrast to observations justified by assumptions they are willing to defend, expanding the information set only when there is strong substantive justification or when feasibility or precision is a serious concern.

The final specification combines all observation groups and illustrates the overall tradeoff. Despite using the widest information set, its RMSE exceeds that of the ideal contrast in nine of the ten settings because of the resulting bias. The advantage of RW is not that borrowing more observations necessarily improves estimation. Rather, it allows the investigator to specify which observations may enter the contrast.